\documentclass[superscriptaddress,11pt]{article}%
\usepackage{authblk}
\usepackage{array}
\usepackage{type1cm}
\usepackage{lettrine}
\usepackage{graphicx}
\usepackage{geometry}
\usepackage{psfrag}
\usepackage{amsmath}
\usepackage{amsfonts}
\usepackage{appendix}
\usepackage[margin=10pt,font=footnotesize,labelfont=sc,format=hang,labelformat=parens]%
{caption}
\usepackage{amssymb}%
\providecommand{\U}[1]{\protect\rule{.1in}{.1in}}
\numberwithin{equation}{section}
\begin{document}

\title{Towards an Anomaly-Free Quantum Dynamics for a Weak Coupling Limit of
Euclidean Gravity}
%\author{Casey Tomlin
%\and Madhavan Varadarajan}
\author[a,b]{Casey Tomlin}
\author[b]{Madhavan Varadarajan}
\affil[a]{Institute for Gravitation and the Cosmos\\Pennsylvania State University, University Park, PA 16802-6300, U.S.A}
\affil[b]{Raman Research Institute\\Bangalore-560 080, India}

\maketitle

\begin{abstract}
The $G_{\mathrm{Newton}}\rightarrow0$ limit of Euclidean gravity introduced by
Smolin is described by a generally covariant $\mathrm{U}(1)^{3}$ gauge theory.
The Poisson bracket algebra of its Hamiltonian and diffeomorphism constraints
is isomorphic to that of gravity. Motivated by recent results in Parameterized
Field Theory and by the search for an anomaly-free quantum dynamics for Loop
Quantum Gravity (LQG), the quantum Hamiltonian constraint of density weight
$4/3$ for this $\mathrm{U}(1)^{3}$ theory is constructed so as to produce a
non-trivial LQG-type representation of its Poisson brackets through the
following steps. First, the constraint at finite triangulation, as well as the
commutator between a pair of such constraints, are constructed as operators on
the `charge' network basis. Next, the continuum limit of the commutator is
evaluated with respect to an operator topology defined by a certain space of
`vertex smooth' distributions. Finally, the operator corresponding to the
Poisson bracket between a pair of Hamiltonian constraints is constructed at
finite triangulation in such a way as to generate a `generalised'
diffeomorphism and its continuum limit is shown to agree with that of the
commutator between a pair of finite triangulation Hamiltonian constraints. Our
results in conjunction with the recent work of Henderson, Laddha and Tomlin in
a 2+1-dimensional context, constitute the necessary first steps toward a
satisfactory treatment of the quantum dynamics of this model.

\end{abstract}

\thispagestyle{empty}
\let\oldthefootnote\thefootnote\renewcommand{\thefootnote}{\fnsymbol{footnote}}
\footnotetext{Email: casey@gravity.psu.edu, madhavan@rri.res.in}
\let\thefootnote\oldthefootnote

\section{Introduction}

\label{introduction}

%\lettrine[lines=2]
A key open issue in canonical LQG relates to the definition of the Hamiltonian
constraint operator. This operator is constructed as the continuum limit of
its finite triangulation approximant \cite{qsd,aajurekreview}. The latter is
the quantum correspondent of a classical approximant which is uniquely defined
only up to terms which vanish in the classical continuum limit wherein the
triangulation of the spatial manifold is taken to be infinitely fine. In
contrast to the classical continuum limit, the continuum limit of the quantum
operator is not independent of the choice of finite triangulation approximant
thus resulting in an infinitely manifold choice in the definition of the
quantum dynamics of LQG. On the other hand, a necessary condition for the very
consistency of the quantum theory is an anomaly free representation of the
constraint algebra. Therefore, one possible way to restrict the choice of
quantum dynamics is to demand that the ensuing algebra of quantum constraints
is free from anomalies. Unfortunately, irrespective of the specific choice of
quantum dynamics made in the current state of art in LQG, the quantum
constraint algebra trivializes i.e. the commutator of a pair of Hamiltonian
constraints as well as the operator corresponding to their classical Poisson
bracket vanish in the continuum limit \cite{ttbook,habitat1,habitat2}. While
it is remarkable that no obvious inconsistency arises,
%with the current set of choices which go into the definition
%of the quantum dynamics,
we believe that the situation is unsatisfactory for reasons we now elaborate.

We refer to the commutator between two Hamiltonian constraints as the Left
Hand Side (LHS) and the operator corresponding to their Poisson bracket as the
Right Hand Side (RHS). While the LHS and the RHS both vanish in the continuum
limit, they do so for very different reasons. The LHS vanishes because the
second Hamiltonian constraint acts trivially on spin network deformations
produced by the action of the first Hamiltonian constraint
\cite{ttbook,habitat1}. In contrast, the RHS vanishes because there are too
many powers of the parameter $\delta$ in its expression at finite
triangulation, the continuum limit being defined by $\delta\rightarrow0$. More
in detail, the finite triangulation approximant to the RHS is built out of the
basic operators of LQG as follows. The curvature is approximated by a small
loop holonomy (divided by its area $\sim\delta^{2}$), the densitized triad by
the electric flux through a small surface (divided by its coordinate area
$\sim\delta^{2}$), and, powers of $\sqrt{q}$ by small region volumes (divided
by $\delta^{3}$ since $\sqrt{q}\delta^{3}\sim$ volume operator). The lower the
density of the Hamiltonian constraints in the LHS, the lower is the power of
$\sqrt{q}$ in the RHS, and hence, the higher the overall power of $\delta$ in
the RHS. For Hamiltonian constraints of density weight one, it is
straightforward to see that one obtains an overall power of $\delta$ in the
RHS which then kills the RHS as $\delta\rightarrow0$ \emph{irrespective of its
finer details}.

Thus one may expect that the consideration of higher density weight
Hamiltonian constraints would yield a non-vanishing RHS with an LHS which
still vanishes because of the independence of the successive actions of the
Hamiltonian constraint alluded to above. Hence, it could well be the case that
the current definitions of the Hamiltonian constraint are anomalous, the
anomaly being hidden by the low density weight.\footnote{A hint that something
may be wrong is already seen in the `scaling by hand' calculations of
Lewandowski and Marolf \cite{habitat1}.}

Our view that the current set of choices for the quantum dynamics of LQG may
be physically incorrect, and that the consideration of higher density
constraints is vital to obtain a non-trivial constraint algebra, is supported
by recent work on Parameterized Field Theory (PFT) \cite{ppftham} and the
Husain-Kucha{\v{r}} model \cite{mvaldiffeo}. In these works the physically
correct finite triangulation approximants to the constraints involve choices
which are qualitatively different from those currently used. Indeed the
approximants bear a qualitative similarity with the physically appropriate
ones used in `improved' LQC \cite{apslqc}. Moreover, the non-triviality of the
quantum constraint algebra in these works is seen to be directly tied to the
kinematically singular nature of the constraint operators which in turn are a
consequence of the higher density nature of the constraints
\cite{ppftham,mvaldiffeo}.

Given this situation, our aim is to use the insights gained from the study of
PFT and the Husain-Kucha{\v{r}} model to construct higher density weight
constraint operators for LQG which yield a non-trivial anomaly-free
representation of the classical constraint algebra. While PFT and the
Husain-Kucha{\v{r}} model have proven to be immensely useful, they suffer from
one structural oversimplification vis a vis gravity: Their constraint algebras
are Lie algebras, unlike the gravitational constraint algebra, which has
structure functions. Therefore, before attempting LQG with all its
complications, it is advisable to tackle a simpler system whose constraint
algebra bears more of a structural similarity with gravity. Just such a system
has been proposed recently by Laddha and its quantum dynamics studied in a
2+1-dimensional context in \cite{hendersonal,alokrecent}. The system is obtained by
replacing, in the phase space description of Euclidean gravity in terms of
triads and connections, the triad rotation group $\mathrm{SU}(2)$ by the group
$\mathrm{U}(1)^{3}$. The $\mathrm{U}(1)^{3}$ model (in 3+1 dimensions) has
three Gauss Law constraints, three spatial diffeomorphism constraints and a
Hamiltonian constraint. The constraint algebra for the Hamiltonian and
diffeomorphism constraints is isomorphic to that of gravity. In fact, it turns
out that this system is exactly the $G_{\mathrm{N}}\rightarrow0$ limit of
Euclidean gravity studied by Smolin in \cite{leeg=0}.\footnote{We thank Miguel
Campiglia for pointing this out to us.}

In this work we initiate the investigation of the quantum dynamics of this
$\mathrm{U}(1)^{3}$ model in 3+1 dimensions with a view to obtaining a
non-trivial representation of the Poisson bracket between a pair of
Hamiltonian constraints. The work entails many new techniques and
constructions and for simplicity we shall \emph{ignore issues of spatial
covariance.} Modifications to our constructions which incorporate spatial
covariance will be discussed in a future publication \cite{meinprep}, this
work serving as a necessary precursor to that one.

The layout of the paper is as follows. Section \ref{u1model} describes the
classical Hamiltonian formulation of the $\mathrm{U}(1)^{3}$ model and
provides a brief review of the $\mathrm{U}(1)^{3}$ `charge' network
representation which comprises its LQG-type quantum kinematics. In Section
\ref{logicsketch} we describe the main steps in our considerations so as to
provide the reader with overall the logical structure of our work. In Section
\ref{hatfinitet} we motivate and define the action of the Hamiltonian
constraint at finite triangulation and compute the action of its commutator
(at finite triangulation) on the charge network basis.

In the last part of Section \ref{hatfinitet}, we compute the continuum limit
of this finite-triangulation commutator. The notion of continuum limits in LQG
is a delicate one. In the literature two different definitions of the
continuum limit exist, one through the specification of Thiemann's Uniform
Rovelli-Smolin (URS) topology \cite{ttbook}, and one through the specification
of the Lewandowski-Marolf habitat \cite{habitat1,ttbook}. The continuum limit
we use is, roughly speaking, an intermediate between the two, and can best be
described in analogy to the case of the URS topology. The URS topology is a
topology on the space of operators on the kinematic Hilbert space (the
finite-triangulation constraint operators belong to this space) which is
defined by a family of seminorms which, in turn, are specified by
diffeomorphism-invariant distributions. These distributions do not lie in the
kinematic Hilbert space but in the algebraic dual space.\footnote{Recall that
an element of algebraic dual consists of linear mappings from the finite span
of charge network states to the complex numbers.} The continuum limit is then
specified in terms of Cauchy sequences of finite-triangulation operators in
this topology. In the present work as well the continuum limit is specified in
term of Cauchy sequences of finite triangulation operators. However, the
operator topology is defined by a different subspace of the algebraic dual. As
we shall see, examples of elements of this subspace are provided by rough
analogs of the Lewandowski-Marolf habitat states
\cite{habitat1,ppftham,mvaldiffeo} which we call `vertex smooth algebraic'
states (VSA states).\footnote{Just as a diffeomorphism-invariant distribution
can be thought of as a kinematically non-normalizable sum over all
diffeomorphically related spin (or `charge') network bras, a VSA state is
constructed as a weighted, kinematically non-normalizable sum over a certain
set of charge network bras where the weights are provided by the evaluation of
smooth complex valued functions on the spatial manifold at certain vertices of
the bra. The set of bras is closed under diffeomorphisms but contains
diffeomorphically distinct bras in contrast to the Lewandowski-Marolf habitat
states.} In Section \ref{hatfinitet}, we obtain the continuum limit of the
finite-triangulation commutator in the `VSA' topology under certain
assumptions about the space of VSA states.

In Section \ref{rhs} we construct the finite-triangulation operator which
corresponds to the RHS. The construction is based on a remarkable classical
identity which we derive in Section \ref{rhsid}. As shown in Appendix
\ref{su2rhsid}, the identity extends to the case of internal group
$\mathrm{SU}(2)$ i.e. to the case of gravity and, hence, is of interest in its
own right. To our knowledge this identity has not been noticed before. As in
Section \ref{hatfinitet}, we evaluate the continuum limit of the
finite-triangulation operator for the RHS under certain assumptions on the
space of VSA states. Section \ref{vsasection} is devoted to a proof that there
exists a large space of VSA states subject to the assumptions of Sections
\ref{hatfinitet} and \ref{rhs}.

The final conclusion of our work in Sections \ref{hatfinitet} and \ref{rhs} is
that the continuum limits of the LHS and RHS agree in the VSA topology induced
by the space of VSA states constructed in Section \ref{vsasection}. \emph{This
agreement is what we mean by an anomaly-free representation of the Poisson
bracket between a pair of Hamiltonian constraints.}

Section \ref{discussion} is devoted to a discussion of our results as well as
an elaboration of open issues, the two key open issues being: (i) an
improvement of our considerations so as to incorporate diffeomorphism
covariance; (ii) the promotion of our VSA topology-based calculations to the
context of a genuine habitat.

%BEG
We work with the semianalytic category in this paper so that the Cauchy slice
$\Sigma$, coordinate charts thereon, its diffeomorphisms and the graphs
embedded in it are semianalytic and $C^{k}$, $k\gg1$.
%END

\section{The U$(1)^{3}$ model}

\label{u1model}In Section \ref{u1modelham} we obtain the Hamiltonian
formulation of the $\mathrm{U}(1)^{3}$ model from that of Euclidean gravity
through Smolin's $G_{\mathrm{N}}\rightarrow0$ limit \cite{leeg=0}. In Section
\ref{quantumkinematics} we briefly review its quantum kinematics in the
polymer representation.

\subsection{The Hamiltonian Formulation}

\label{u1modelham}Recall that Euclidean gravity is described, in its
Hamiltonian formulation, by the action:
\begin{equation}
S[E,\mathcal{A}]=\frac{1}{G_{\mathrm{N}}}\int\mathrm{d}t\int_{\Sigma
}\mathrm{d}^{3}x~\left(  E_{i}^{a}\dot{\mathcal{A}}_{a}^{i}-\Lambda
^{i}\mathcal{D}_{a}E_{i}^{a}-N^{a}(E_{i}^{b}\mathcal{F}_{ab}^{i}%
-\mathcal{A}_{a}^{i}\mathcal{D}_{b}E_{i}^{b})-N\epsilon^{ijk}E_{i}^{a}%
E_{j}^{b}\mathcal{F}_{ab}^{k}\right)  . \label{hamactionsu2}%
\end{equation}
Here $E_{i}^{a},\mathcal{A}_{a}^{i}$ are the canonically conjugate densitized
triad and $\mathrm{SU}(2)$ connection. The curvature of the connection is
$\mathcal{F}_{ab}^{i}:=\partial_{a}\mathcal{A}_{b}^{i}-\partial_{b}%
\mathcal{A}_{a}^{i}+\epsilon_{jk}^{i}\mathcal{A}_{a}^{i}\mathcal{A}_{b}^{j}$
and $\mathcal{D}_{a}$ is the gauge covariant derivative so that $\mathcal{D}%
_{a}E_{i}^{a}={\partial}_{a}E_{i}^{a}+\epsilon_{{i}j{k}}\mathcal{A}_{a}%
^{j}E_{k}^{a}$. $N,N^{a},\Lambda^{i}$ are the (appropriately densitized)
lapse, shift and internal gauge Lagrange multipliers.

We have set the speed of light to be unity so that $G_{\mathrm{N}}$ has
dimensions $[$length$][$mass$]^{-1}$, $\mathcal{A}_{a}^{i},\Lambda^{i}$ have
dimensions $[$length$]^{-1}$ and the triad, lapse, and shift are dimensionless
so that Equation (\ref{hamactionsu2}) acquires the dimensions of action.
Following Smolin, we define the rescaled connection $A_{a}^{i}:=G_{\mathrm{N}%
}^{-1}\mathcal{A}_{a}^{i}$ so that the curvature takes the form $\mathcal{F}%
_{ab}^{i}=G_{\mathrm{N}}(\partial_{a}{A}_{b}^{i}-\partial_{b}{A}_{a}%
^{i}+G_{\mathrm{N}}\epsilon_{jk}^{i}{A}_{a}^{i}{A}_{b}^{j})$ and
$\mathcal{D}_{a}E_{i}^{a}={\partial}_{a}E_{i}^{a}+G_{\mathrm{N}}\epsilon
_{ijk}{A}_{a}^{j}E_{k}^{a}$.

Rewriting the action in terms of the scaled connection and then setting
$G_{\mathrm{N}}=0$, it is easy to obtain:
%BEG
%\footnote{In fact, one is rescaling
%all four spacetime components of the self-dual connection, in which case the
%Lagrange multiplier $\Lambda^{i}\propto\mathcal{A}_{0}^{i}$ is also
%rescaled.}
%END%
\begin{equation}
S[E,{A}]=\int\mathrm{d}t\left(  \int\mathrm{d}^{3}x~E_{i}^{a}\dot{{A}}_{a}%
^{i}-G[\Lambda]-D[\vec{N}]-H[N]\right)  , \label{hamactionu13}%
\end{equation}
where
\begin{align}
G[\Lambda]  &  =\int\mathrm{d}^{3}x~\Lambda^{i}\partial_{a}E_{i}^{a}\\
D[\vec{N}]  &  =\int\mathrm{d}^{3}x~N^{a}\left(  E_{i}^{b}F_{ab}^{i}-A_{a}%
^{i}\partial_{b}E_{i}^{b}\right) \\
H[N]  &  =\tfrac{1}{2}\int\mathrm{d}^{3}x~{N}\epsilon^{ijk}E_{i}^{a}E_{j}%
^{b}F_{ab}^{k},
\end{align}
are the Gauss law, diffeomorphism, and Hamiltonian constraints of the theory,
and where $F_{ab}^{i}:=\partial_{a}{A}_{b}^{i}-\partial_{b}{A}_{a}^{i}$. Note
that the Gauss law constraints generate three independent $\mathrm{U}(1)^{3}$
gauge transformations on the connections $A_{a}^{i},i=1,2,3$ with
gauge-invariant curvature $F_{ab}^{i}$ and that the three electric fields
$E_{i}^{a},i=1,2,3$ are gauge-invariant. Thus, the action (\ref{hamactionu13})
describes a $\mathrm{U}(1)^{3}$ theory as claimed.

The constraints $G[\Lambda],D[\vec{N}],H[N]$ are first class. Their Poisson
bracket algebra is
\begin{align}
\{G[\Lambda],G[\Lambda^{\prime}]\}  &  =\{G[\Lambda],H[N]\}=0\\
\{D[\vec{N}],G[\Lambda]\}  &  =G[\pounds _{\vec{N}}\Lambda]\\
\{D[\vec{N}],D[\vec{M}]\}  &  =D[\pounds _{\vec{N}}\vec{M}]\\
\{D[\vec{N}],H[N]\}  &  =H[\pounds _{\vec{N}}N]\\
\{H[N],H[M]\}  &  =D[\vec{\omega}]+G[A\cdot\vec{\omega}],\qquad\omega
^{a}:=E_{i}^{a}E_{i}^{b}\left(  M\partial_{b}N-N\partial_{b}M\right)
\end{align}

The last Poisson bracket (between the Hamiltonian constraints) exhibits
structure functions just as in gravity. Working towards a representation of
this last Poisson bracket in quantum theory will occupy the rest of this work.

\subsection{Quantum Kinematics}

\label{quantumkinematics}

\subsubsection{The Holonomy-Flux Algebra}

\label{hfalg}Let $e$ be a $C^{k},k\gg1$ semianalytic, embedded edge
$e:[0,1]\rightarrow\Sigma$. An edge holonomy in the $j^{\text{th}}$ copy of
U$(1)$ is denoted by $h_{e,q^{j}}$ with
\begin{equation}
h_{e,q^{j}}=\mathrm{e}^{\mathrm{i}\kappa\gamma q^{j}\int_{e_{I}}A_{a}%
^{j}\mathrm{d}x^{a}}.
\end{equation}

Here $q^{j}$ is an integer, $\kappa$ is a constant of dimension $[$%
length$][$mass$]^{-1}$ and $\gamma$ is a positive real number. For fixed
$\kappa,\gamma$, the edge holonomies for all edges and all values of the
`charges' $q^{j}$ form a complete set of functions of the connection
$A_{a}^{j}$; i.e., the knowledge of all these holonomies allows the
reconstruction of $A_{a}^{j}$. We fix $\kappa$ once and for all. We shall see
below that $\gamma$ is a Barbero-Immirizi-like parameter of the theory which
labels inequivalent quantum representations.\footnote{We could have chosen
three different parameters $\gamma_{i}$ and obtained a 3-parameter family of
inequivalent representations. For simplicity and to maintain similarity with
the case of gravity where there is a single Barbero-Immirzi parameter, we set
$\gamma_{i}=\gamma,i=1,2,3$.} The edge holonomy $h_{e,{\vec{q}}}$ valued in
$\mathrm{U}(1)^{3}$ is defined to be the product of edge holonomies over the
three copies of U$(1)$:
\begin{equation}
h_{e,{\vec{q}}}=\mathrm{e}^{\mathrm{i}\kappa\gamma\sum_{j=1}^{3}q^{j}%
\int_{e_{I}}A_{a}^{j}\mathrm{d}x^{a}}.
\end{equation}
Given a closed, oriented graph $\alpha$ with $N$ edges, the graph holonomy
$h_{\alpha,\{{\vec{q}}\}}:=h_{\alpha,\{{\vec{q}}_{I}|I=1,\dots,N\}}$ is just
the product of the edge holonomies over the edges of the graph, so that
\begin{equation}
h_{\alpha,\{{\vec{q}}\}}:=\prod_{I=1}^{N}h_{e_{I},{\vec{q}}_{I}}
\label{graphhol}%
\end{equation}

It is easily verified that the graph holonomy $h_{\alpha,\{{\vec{q}}\}}$ is
invariant under U$(1)^{3}$ gauge transformations if and only if, for every
vertex $v$ of the graph $\alpha$ and for each $i$,
\begin{equation}
\sum_{I_{v}}\tau({I_{v}})q_{I_{v}}^{i}=0. \label{ggeinv}%
\end{equation}
where $e_{I_{v}}$ ranges over the edges incident at $v$ and $\tau({I_{v}})$ is
$+1$ if the edge is outgoing at $v$ and $-1$ if ingoing. The labels
$\alpha,\{{\vec{q}}_{I}|I=1,\dots,N\}$ define a colored graph which we refer
to as a \emph{charge network}. A charge network $c=c(\alpha,\{{\vec{q}}%
_{I}|I=1,\dots,N\})$ is closed oriented graph whose edges are `colored' by
representation labels of U$(1)^{3}$; i.e., each edge $e_{I}$ is colored with
the triple of charges $(q_{I}^{1},q_{I}^{2},q_{I}^{3}):={\vec{q}}_{I}$. If the
charges satisfy Equation (\ref{ggeinv}), we shall say that the charge network
is gauge-invariant.\footnote{More precisely, charge networks are associated
with equivalence classes of colored oriented closed graphs; colored graphs
which yield the same graph holonomy via (\ref{graphhol}) define such an
equivalence class.} Thus, graph holonomies are labelled by charge networks and
we may write $h_{\alpha,\{{\vec{q}}\}}:=h_{c}$. For future purposes it is
useful to write the graph holonomy $h_{c}$ in the form
\begin{equation}
h_{c}=\exp\left(  \int\mathrm{d}^{3}x~c_{i}^{a}A_{a}^{i}\right)  \label{hcdef}%
\end{equation}
where
\begin{equation}
c_{i}^{a}(x)=c_{i}^{a}(x;\{e_{I}\},\{q_{I}\})=\sum_{I=1}^{M}\mathrm{i}%
\gamma\kappa q_{I}^{i}\int\mathrm{d}t_{I}~\delta^{(3)}(e_{I}(t_{I}),x)\dot
{e}_{I}^{a}(t_{I}). \label{chrgcoord}%
\end{equation}
Here $t_{I}$ is a parameter which runs along the edge $e_{I}$. Adapting the
old terminology of Gambini and Pullin \cite{loopcoord}, shall refer to
$c_{i}^{a}(x)$ as a \emph{charge network coordinate}.

The gauge-invariant electric flux $E_{i}(S)$ through a two-dimensional
oriented surface $S$ is given by integrating the 2-form $\eta_{abc}{E}_{i}%
^{a}$ over $S$ so that
\begin{equation}
{E}_{i}(S):=\int_{S}\eta_{abc}{E}_{i}^{a}.
\end{equation}

The only non-trivial Poisson bracket amongst the holonomy-flux variables is
$\{h_{c},{E}_{i}(S)\}$, which is readily computed:
\begin{equation}
\{h_{c},{E}_{i}(S)\}=\mathrm{i}\frac{\gamma\kappa}{2}\sum\epsilon({e_{I}%
,S})q_{I}^{i}h_{c}.
\end{equation}
Here the graph $\alpha(c)$ underlying $c$ is chosen to be fine enough that
isolated intersection points of the graph with $S$ are at its vertices and the
integer $\epsilon({e_{I},S})$ vanishes unless $e_{I}$ intersects $S$
transversely in which case $\epsilon({e_{I},S})=1$ if $e_{I}$ is outgoing from
and above $S$ or incoming to and below $S$ and $-1$ otherwise. Unless
indicated explicitly below, we will always assume that charge network edges
are outgoing at vertices or relevant interior edge points.

\subsubsection{The Polymer Representation}

\label{polyrep}An orthonormal basis for the kinematic Hilbert space is
provided by `charge network' states. To every distinct charge network label
$c$ we assign the unit norm charge network state $|c\rangle\equiv
|\gamma,\{{\vec{q}}_{I}\}\rangle$. Two charge network states are orthogonal if
and only if their charge network labels differ; i.e., if the colored graphs
which label them are inequivalent. We denote this inner product between charge
network states by
\begin{equation}
\langle c^{\prime}|c\rangle=\delta_{c^{\prime},c} \label{polyip}%
\end{equation}
where the Kronecker delta $\delta_{c^{\prime},c}$ vanishes unless there is a
choice of colored graph underlying $c$ which is identical to a choice of
colored graph underlying $c^{\prime}$ in which case $c=c^{\prime}$ and
$\delta_{c,c^{\prime}}=1$.

Let the finite span of the charge network states be $\mathcal{D}$. The Cauchy
completion of $\mathcal{D}$ in the inner product (\ref{polyip}) yields the
kinematic Hilbert space $\mathcal{H}_{\mathrm{kin}}$.

The holonomy operators act as follows:
\begin{equation}
\hat{h}_{c}|c^{\prime}\rangle=|c+c^{\prime}\rangle
\end{equation}
The charge network $c+c^{\prime}$ is defined as follows: Let $\alpha$ be a
fine enough closed, oriented graph which underlies both $c$ and $c^{\prime}$.
Add the charge labels of $c,c^{\prime}$ edgewise to obtain to new charge
labels for $\alpha$. This newly colored graph specifies the charge network
$c+c^{\prime}$. The flux operators act as follows:
\begin{equation}
\hat{E}_{i}(S)|c\rangle=\frac{\hbar\gamma\kappa}{2}\sum\epsilon({e_{I}%
,S})q_{I}^{i}|c\rangle
\end{equation}
It can be verified that the above operator actions provide a representation of
the holonomy-flux Poisson bracket algebra on $\mathcal{H}_{\mathrm{kin}}$.
Finally note that, as in LQG, we may derive these operator actions by
thinking, heuristically, of the charge network states as wave functions which
depend on smooth connections via $|c\rangle\sim c(A)=h_{c}(A)$ and by seeking
to represent the holonomy operators by multiplication and the electric field
operators by functional differentiation.

\section{Sketch of Overall Logical Structure}

\label{logicsketch}Our purpose in this section is to give the reader a rough
global view of the logical structure of our considerations. In Section
\ref{logicsketchsteps} we provide a brief sketch of the main steps in our
work. Section \ref{topologynote} contains a precise definition of the
continuum limit in terms of a topology on the space of operators and indicates
the sense in which the implementation of the steps of Section
\ref{logicsketchsteps} establishes the existence of a non-trivial anomaly-free
representation of the constraint algebra. In Section \ref{choices} we briefly
describe the various choices made in order to implement the steps of Section
\ref{logicsketchsteps}. To avoid unnecessary clutter we shall not worry about
overall factors, both dimensional and numerical (only in this section!).

As in LQG, we are faced with a tension between the local nature of the
constraints of the model (most importantly the dependence on $F_{ab}^{i}$) and
the non-local and discontinuous nature of some of the basic operators of the
quantum theory (namely the holonomy operators). Since there is no way to
extract a connection (or curvature) operator out of the holonomy operators due
to their discontinuous action with respect to any shrinking procedure applied
to the loops which label them, one proceeds in close analogy to Thiemann's
seminal work \cite{qsd}. We fix a one-parameter family of triangulations
$T_{\delta}$ of the spatial manifold $\Sigma$ where $\delta$ labels the
fineness of the triangulation, with $\delta\rightarrow0$ being the continuum
limit of infinite refinement, construct finite triangulation approximants to
the classical constraints, construct the corresponding operators and then take
an appropriate continuum limit, the hope being that while individual operators
may not possess a continuum limit, the conglomeration of operators which
combine to form the constraint does possess a continuum limit.

\subsection{Steps}

\label{logicsketchsteps}\noindent\emph{Step 1. The finite-triangulation
Hamiltonian constraint and its continuum limit:} Let the Hamiltonian
constraint at finite triangulation $T_{\delta}$ be $C_{\delta}[N]$.
$C_{\delta}[N]$ is a discrete approximant to the Hamiltonian constraint $C[N]$
(see, however, the remark after Step 4 below) so that $\lim_{\delta
\rightarrow0}C_{\delta}[N]=C[N]$. Let the corresponding operator ${\hat{C}%
}_{\delta}[N]$ be such that ${\hat{C}}_{\delta}[N]:\mathcal{D}\rightarrow
\mathcal{D}$ where $\mathcal{D}$ is the finite span of charge network states.
Let $\mathcal{D}^{\ast}$ be the algebraic dual to $\mathcal{D}$ so that every
$\Psi\in\mathcal{D}^{\ast}$ is a linear map from $\mathcal{D}$ to $%
%TCIMACRO{\U{2102} }%
%BeginExpansion
\mathbb{C}
%EndExpansion
$. Let $|c\rangle$ be a charge network state. Then for every pair
$(\Psi,|c\rangle)$ we compute the one-parameter family of complex numbers
$\Psi({\hat{C}}_{\delta}[N]|c\rangle)$. The continuum limit action of
${\hat{C}}_{\delta}[N]$ is defined to be
\begin{equation}
\lim_{\delta\rightarrow0}\Psi({\hat{C}}_{\delta}[N]|c\rangle)
\label{contcndef}%
\end{equation}

\bigskip

\noindent\emph{Step 2. Finite triangulation commutator and its continuum
limit:} Let $T_{\delta^{\prime}}$ be a refinement of $T_{\delta}$ so that
$\delta^{\prime}<\delta$. Define a discrete approximant to $C[N]C[M]$ by
$C[N]_{\delta^{\prime}}C[M]_{\delta}$. The corresponding operator product is
${\hat{C}}[N]_{\delta^{\prime}}{\hat{C}}[M]_{\delta}$. The commutator at
finite triangulation is then\textbf{ }${\hat{C}}[N]_{\delta^{\prime}}{\hat{C}%
}[M]_{\delta}-{\hat{C}}[M]_{\delta^{\prime}}{\hat{C}}[N]_{\delta}$ and its
continuum limit action is
\begin{equation}
\lim_{\delta\rightarrow0}\lim_{\delta^{\prime}\rightarrow0}\Psi([{\hat{C}%
}[N]_{\delta^{\prime}}{\hat{C}}[M]_{\delta}-{\hat{C}}[M]_{\delta^{\prime}%
}{\hat{C}}[N]_{\delta}]|c\rangle) \label{contlhsdef}%
\end{equation}

\bigskip

\noindent\emph{Step 3. RHS at finite triangulation and its continuum limit:}
Recall that the RHS, $D[{\vec{\omega}}]$, is just the diffeomorphism
constraint smeared with a metric-dependent shift. One could define it at
finite triangulation by some discrete approximant $D_{\delta}[{\vec{\omega}}%
]$. Note that the LHS at finite triangulation, by virtue of the quadratic
dependence of the commutator on the constraint, depends on the \emph{pair} of
parameters $\delta,\delta^{\prime}$. Clearly, a better comparison of the LHS
and RHS would result if the RHS could also naturally accommodate a commutator
description. Remarkably, it so happens that the classical expression for the
RHS, \emph{can} be written as the Poisson bracket between a pair of
\emph{diffeomorphism} constraints with triad dependent shifts. Specifically,
we have that $D[{\vec{\omega}}]=\sum_{i=1}^{3}\{D[N_{i}],D[M_{i}]\}$ where
$D(N_{i})$ is the diffeomorphism constraint smeared with the shift $N_{i}^{a}$
which is constructed out of the lapse $N$ and the electric field variable (see
Section \ref{rhsid}). Let $D_{\delta}[N_{i}]$ be a finite triangulation
approximant to $D[N_{i}]$. Then the finite-triangulation RHS operator can be
written as $\sum_{i}{\hat{D}}[N_{i}]_{\delta^{\prime}}{\hat{D}}[M_{i}%
]_{\delta}-{\hat{D}}[M_{i}]_{\delta^{\prime}}{\hat{D}}[N_{i}]_{\delta}$ and
its continuum limit action is defined to be
\begin{equation}
\lim_{\delta\rightarrow0}\lim_{\delta^{\prime}\rightarrow0}\sum_{i=1}^{3}%
\Psi([{\hat{D}}[N_{i}]_{\delta^{\prime}}{\hat{D}}[M_{i}]_{\delta}-{\hat{D}%
}[M_{i}]_{\delta^{\prime}}{\hat{D}}[N_{i}]_{\delta}]|c\rangle)
\label{contrhsdef}%
\end{equation}

\bigskip

\noindent\emph{Step 4. Existence of the continuum limit for suitable algebraic
dual states:} We look for a large (infinite-dimensional) subspace
$\mathcal{D}_{\mathrm{cont}}^{\ast}\subset\mathcal{D}^{\ast}$ such that for
every $\Psi\in\mathcal{D}_{\mathrm{cont}}^{\ast}$ and every charge network
state $|c\rangle$, the limits (\ref{contcndef}), (\ref{contlhsdef}), and
(\ref{contrhsdef}) exist with (\ref{contlhsdef}$)=($\ref{contrhsdef}). Further
we require that (\ref{contlhsdef}), and (\ref{contrhsdef}) do not vanish
identically for every pair $(\Psi,|c\rangle)$.
%This is what we mean by a nontrivial  anomaly-free representation of the
%Poisson bracket between a pair of Hamiltonian constraints.

\bigskip

\noindent\emph{Remark:} In accordance with Step 1 above we should first find a
classical approximant to the classical constraints such that the approximant
is built out of small edge holonomies and small surface fluxes (where the
notion of smallness is defined by the finite triangulation parameter $\delta
$). We should then replace the classical phase space functions by their
quantum counterparts to obtain the constraint operator at finite
triangulation. Instead, in Section 4 we \emph{directly} motivate, through
heuristic considerations, finite-triangulation quantum constraint operators.
It is desirable that it be shown that these operators correspond to the
quantization of classical finite triangulation approximants. Based on our
experience with PFT and the HK model, we are fairly sure that this should be
easy to do. However since this is one of the first attempts at obtaining a
nontrivial representation of the constraint algebra we choose to press on and
leave loose ends such as this to be tied up by future work.

\subsection{A Note on the `Topology' Interpretation of the Continuum Limit}

\label{topologynote}Given any operator ${\hat{O}}:\mathcal{D}\rightarrow
\mathcal{D}$ and a pair $(\Psi,|c\rangle)$ with $\Psi\in\mathcal{D}%
_{\mathrm{cont}}^{\ast}$ and $|c\rangle$ being a charge network state, we may
define the \emph{seminorm} of the operator ${\hat{O}}$ to be $||{\hat{O}%
}||_{\Psi,c}=|\Psi({\hat{O}}|c\rangle)|$. The family of seminorms
$||\;\;||_{\Psi,c}$ for every pair $(\Psi,|c\rangle)$ defines a topology on
the vector space of operators from $\mathcal{D}$ to itself. It is
straightforward to check that the sequences of operators (indexed by $\delta$,
$\delta^{\prime}$) defined in the previous section can be interpreted as
sequences which are Cauchy in this topology. Of course there is no guarantee
that the limit of such a Cauchy sequence is also an operator from
$\mathcal{D}$ to itself. Indeed, we shall see that the limit is interpretable
as an operator from $\mathcal{D}_{\mathrm{cont}}^{\ast}$ into $\mathcal{D}%
^{\ast}$; this follows straightforwardly from the fact that every operator
${\hat{O}}$ from $\mathcal{D}$ to itself defines an operator $(\hat{O}^{\dag
})^{\prime}$ on $\mathcal{D}^{\ast}$ by dual action.

It is straightforward to see that the successful implementation of Step 4
implies that:

\begin{enumerate}
\item[(i)] The sequence\footnote{Strictly speaking, the statement applies to
any appropriately defined countably-infinite subset of the 1 (or 2) parameter
set of operators under consideration.} of finite-triangulation Hamiltonian
constraint operators is Cauchy and converges to a non-trivial operator from
$\mathcal{D}_{\mathrm{cont}}^{\ast}$ into $\mathcal{D}^{\ast}$.

\item[(ii)] Likewise for the sequences of finite-triangulation LHS
approximants and finite-triangulation RHS approximants.

\item[(iii)] The difference between the RHS and LHS operators at finite
triangulation also form a Cauchy sequence. This sequence converges to zero.
\end{enumerate}

The statements (i)-(iii) constitute a precise definition of what we mean by a
nontrivial anomaly-free representation of the Poisson bracket between a pair
of Hamiltonian constraints. These statements hold in PFT and the
Husain-Kucha{\v{r}} model. However, there, one has the stronger statement that
the finite-triangulation operators as well as their limits are operators from
$\mathcal{D}_{\mathrm{cont}}^{\ast}$ to itself; the linear vector space
$\mathcal{D}_{\mathrm{cont}}^{\ast}$ then acts as a linear representation
space which supports a representation of the constraint algebra. Following
Lewandowski and Marolf \cite{habitat1}, such a representation space is called
a \emph{habitat}.

We are optimistic that our considerations here admit a generalization to a
habitat-based representation. Indeed, as we shall see briefly in Section
\ref{choices} and in detail later, our choice of $\mathcal{D}_{\mathrm{cont}%
}^{\ast}$ closely mimics that of the habitats of PFT \cite{ppftham} and the
Husain-Kucha{\v{r}} model \cite{mvaldiffeo}.

\subsection{Choices}

\label{choices}\noindent\emph{1. The action of the finite-triangulation
Hamiltonian constraint operator:} As in LQG \cite{qsd}, the Hamiltonian
constraint acts only at charge network vertices. Recall, from Section
\ref{introduction}, that the reason the LHS trivializes in LQG can be traced
to the fact that the second Hamiltonian constraint does not act on graph
deformations generated by the first. As argued in \cite{habitat1} this is
because the Hamiltonian constraint does not move the vertex it acts on. Here
we define the action of ${\hat{C}}_{\delta}[N]$ after a careful study of the
Hamiltonian vector field of $C[N]$. This study motivates an operator action
which \emph{does} move the vertices it acts upon. This is the reason we get a
non-trivial LHS with the desired dependence on derivatives of the lapse (see
Equation (\ref{lhsfinal})); the derivative is born of the fact that the second
Hamiltonian constraint acts at the closely displaced vertex created by the
first Hamiltonian constraint.\newline

\noindent\emph{2. $\mathcal{D}_{\mathrm{cont}}^{\ast}$, vertex smooth
functions, and density weight:} The choice of $\mathcal{D}_{\mathrm{cont}%
}^{\ast}$ for PFT and the Husain-Kucha{\v{r}} model is characterized by vertex
smooth functions (see Footnote 4). An element $\Psi_{f}$ of $\mathcal{D}%
_{\mathrm{cont}}^{\ast}$ is obtained by summing over an uncountably infinite
set of charge network bras with weights which correspond to the evaluation of
a smooth function $f$ (from copies of the spatial manifold to $%
%TCIMACRO{\U{2102} }%
%BeginExpansion
\mathbb{C}
%EndExpansion
$) at points on the spatial manifold given by the vertices of the bra. In our
notation, with $|c\rangle$ being an appropriate spin/charge network state, one
typically obtains $\Psi_{f}({\hat{C}}_{\delta}[N]|c\rangle)$ to be the
difference of the evaluation of the function at points on the manifold which
are `$\delta$' apart divided by an overall power of $\delta$. In the continuum
limit this translates to a derivative of $f$. If the overall factor of
$\delta$ was absent, one would get a trivial result by virtue of the
smoothness of $f$. As discussed in Section \ref{introduction}, the overall
factor of $\delta$ is tied to the choice of density weight of the constraint.
As has been known for a long time, density weight one objects constructed
solely out of the phase space variables when integrated with scalar smearing
functions typically lead to LQG operators with no overall factors of $\delta$.
This is what would happen if we used the density weight one constraint. Hence
in order to get an overall factor of $\delta^{-1}$, we need to multiply the
density weight one constraint by $\sqrt{q}^{1/3}$ (recall that $\sqrt{\hat{q}%
}\delta^{3}\sim$ volume operator) i.e. we need to consider a Hamiltonian
constraint of weight $4/3$. It is then straightforward to check that the RHS
also acquires an overall factor of $\delta^{-1}$ which, as we shall see, also
goes into producing a derivative of $f$ in the continuum limit. Thus the
higher density weight allows on one hand the moving of vertices caused by the
Hamiltonian constraint to manifest nontrivially, thereby giving rise to a
nontrivial LHS, and on the other, compensates for the (hitherto) `too many
factors of $\delta$' in the RHS, thereby leaving an overall factor of
$\delta^{-1}$ which is responsible for \emph{its} non-triviality.

\section{The Hamiltonian Constraint Operator at Finite Triangulation and the
Continuum Limit of its Commutator}

\label{hatfinitet}The Hamiltonian constraint of density weight $4/3$ smeared
with a lapse $N$ (of density weight $-1/3$) is:
\begin{equation}
H[N]=\tfrac{1}{2}\int_{\Sigma}\mathrm{d}^{3}x~\epsilon^{ijk}F_{ab}^{k}%
E_{j}^{b}(NE_{i}^{a}q^{-\frac{1}{3}}). \label{ham4/3}%
\end{equation}
Note that the last piece of the above expression,
\begin{equation}
N_{i}^{a}:=NE_{i}^{a}q^{-\frac{1}{3}} \label{electricshift}%
\end{equation}
defines an electric field-dependent vector field for each $i$. For reasons
which will become clear shortly, we shall refer to $N_{i}^{a}$ as the
\emph{electric shift}. We refer to its quantum correspondent as the
\emph{quantum shift}.

In Section \ref{regulatingstructures} we detail our choice of regulating
structures. In Section \ref{quantumshift} we construct the quantum shift
operator. Since its phase space dependence is solely on the electric field,
the operator is diagonalized in the charge network basis. Moreover, due to its
dependence on the inverse metric, its action is non-trivial only at vertices.

In Section \ref{heuristic} we provide heuristic motivation for the action of
the constraint operator at finite triangulation. Motivated by previous work in
PFT, the Husain-Kucha{\v{r}} model, and LQC \cite{ppftham,mvaldiffeo,apslqc},
as well as by the requirement that constraint move the vertex on which it
acts, we assign a key role to the quantum shift in this action. Specifically,
using the key classical identity,
\begin{equation}
N_{i}^{a}F_{ab}^{k}=\pounds _{\vec{N}_{i}}A_{b}^{k}- \partial_{b}(N_{i}%
^{c}A_{c}^{i}) \label{nfidentity}%
\end{equation}
as motivation, the quantum shift is used to deform the graph underlying the
charge network.
%Since the quantum shift only acts at vertices of the charge
%network, the Hamiltonian constraint (as in LQG) also acts only on vertices.
While the classical electric shift is smooth, by virtue of the discrete
`quantum geometry', the quantum shift is not a smooth vector field and the
choice of the deformations it defines is made on the basis of intuition gained
by the study of PFT and the Husain-Kucha{\v{r}} model. We detail this choice
in Section \ref{deformations} and conclude with the evaluation of the action
of the Hamiltonian constraint operator at finite triangulation on the charge
network basis. Note that since the quantum shift only acts at vertices of the
charge network, the Hamiltonian constraint (as in LQG) also acts only on vertices.

In Section \ref{2ndh}, we evaluate the commutator of two Hamiltonian
constraints at finite triangulation on the charge network basis, and in
Section \ref{continuum} we compute the continuum limit.

\subsection{Choice of Triangulation and Regulating Structures}

\label{regulatingstructures}Scalar densities of non-trivial weight need
coordinate systems (more precisely $n$-forms in $n$ dimensions) for their
evaluation. Since the lapse is no longer a scalar, it turns out that we need
to fix regulating coordinate systems to define the finite-triangulation
Hamiltonian constraint. Accordingly, once and for all, around every
$p\in\Sigma$ we fix an open neighborhood $U_{p}$ with coordinate system
$\{x\}_{p}$ such that $p$ is at the origin of $\{x\}_{p}$. When there is no
confusion we shall drop the label $p$ and refer to the coordinate patch as
$\{x\}$.

We shall use the regulating coordinate patches to specify the fineness of the
triangulation below, to define the quantum shift in Section \ref{quantumshift}
and to specify the detailed graph deformations generated by the Hamiltonian
constraint in Section \ref{deformations}. An immediate concern is the
interaction of this choice of coordinate patches with the spatial covariance
of the Hamiltonian constraint. While we shall comment on this issue towards
the end of this paper, we shall (as mentioned in Section \ref{introduction})
defer a comprehensive treatment of the issue to Reference \cite{meinprep}.

The one parameter family of triangulations $T_{\delta}$ are adapted to the
charge network on which the finite triangulation approximants act.
Specifically, we require that $T_{\delta}$ (for sufficiently small $\delta$)
be such that every vertex $v$ of the coarsest graph underlying the charge
network is contained in the interior of a cell $\triangle_{\delta(v)}\in
T_{\delta},$ and every cell of $T_{\delta}$ contains at most one such vertex.
The size of $\triangle_{\delta(v)}$ is restricted to be of $O(\delta^{3})$
\emph{as measured in the coordinate system $\{x\}_{v}$}.

\subsection{The Quantum Shift}

\label{quantumshift}Let $\hat{q}^{-1/3}$ act non-trivially at a vertex $v$ of
the charge network $c$. We shall refer to such vertices as non-degenerate. Let
$\{x\}$ denote the coordinate patch at $v$.
%We evaluate the action of the quantum shift operator at $v$.
Fix a coordinate ball $B_{\tau}(v)$ of radius $\tau$ centered at $v$, and
restrict attention to small enough $\tau$ in the following manipulations so
that all constructions happen within the domain of $\{x\}$. Let $\hat{q}%
_{\tau}^{-1/3}$ denote the regularization of $\hat{q}^{-1/3}$ using this
coordinate ball. From Appendix \ref{invq} (and from our general arguments in
and prior to Section \ref{choices}), the eigenvalue of $\hat{q}_{\tau}^{-1/3}$
for the eigenstate $|c\rangle$ takes the form $\tau^{2}(\hbar\kappa
\gamma)^{-1}\nu^{-2/3}$ where $\nu$ is a number constructed out of the charges
which label the edges of $c$ at $v$.

Treating $\hat{E}_{i}^{a}$ as a functional derivative and $\vert c\rangle$ as
a function of the connection, the action of $\hat{E}_{i}^{a}$ at the point $v$
naturally decomposes into a sum of contributions per edge \cite{aajurekarea}
$\hat{E}_{i}^{a}=\sum_{I}\hat{E}_{i}^{aI}$ with
\begin{equation}
\hat{E}_{i}^{aI}(x(v))|c\rangle=\hbar\kappa\gamma q_{I}^{i}\int_{0}%
^{1}\mathrm{d}t~\dot{e}_{I}^{a}(t)\delta^{(3)}(e_{I}(t),x(v))|c\rangle.
\label{sumei}%
\end{equation}

Next, we define the regulated quantum shift, ${\hat{N}}_{\tau}^{a_{i}}$,
evaluated at the point $v$ by
\begin{equation}
{\hat{N}}_{\tau}^{a_{i}}:=N(x(v))\hat{q}_{\tau}^{-1/3}\frac{1}{\frac{4\pi
\tau^{3}}{3}}\int_{B_{\tau}(v)}\mathrm{d}^{3}x~\hat{E}_{i}^{a}(x)
\end{equation}
From Equation (\ref{sumei}) and the form of the eigenvalue of $\hat{q}_{\tau
}^{-1/3}$, we obtain
\begin{equation}
{\hat{N}}_{\tau}^{a_{i}}|c\rangle=\sum_{I}N_{i}^{aI}(v)_{\{x\},\tau}|c\rangle
\end{equation}
with
\begin{align}
N_{i}^{aI}(v)_{\{x\},\tau}  &  =\hbar\kappa\gamma N(x(v))\tau^{2}(\hbar
\kappa\gamma)^{-1}\nu^{-2/3}q_{I}^{i}\frac{1}{\tfrac{4}{3}\pi\tau^{3}}%
\int_{B_{\tau}(v)}\mathrm{d}^{3}x\int_{0}^{1}\mathrm{d}t~\dot{e}_{I}%
^{a}(t)\delta^{(3)}(e_{I}(t),x)\nonumber\\
&  =N(x(v))\nu^{-2/3}q_{I}^{i}\frac{1}{\tfrac{4}{3}\pi\tau}\int_{B_{\tau
}(v)\cap e_{I}}\mathrm{d}e_{I}^{a}\nonumber\\
&  =\frac{3}{4\pi}N(x(v))\nu^{-2/3}q_{I}^{i}\hat{e}_{I\tau}^{a}
\label{evaluenI}%
\end{align}
where $\hat{e}_{I\tau}^{a}$ is a unit vector which pierces $B_{\tau}(v)$ at
the point where $e_{I}$ intersects it$.$ That is, the point $\partial B_{\tau
}(v)\cap e_{I}$ has coordinates $\tau\hat{e}_{I\tau}^{a}$ in the coordinate
system $\{x\}$. The appearance of $\{x\},\tau$ remind us that these values
refer to a particular choice of coordinates $\{x\}$ and a parameter $\tau$
defining the size of $B_{\tau}(v)$.

We may now take the regulating parameter $\tau\rightarrow0$ to obtain
\begin{equation}
{\hat N^{a}_{i}}(v)\vert c\rangle:= N^{a}_{i} (v)\vert c\rangle=\sum_{I}%
N_{i}^{aI}(v)_{\{x\}} \vert c\rangle\label{hatnc=nc}%
\end{equation}
with
\begin{equation}
N_{i}^{aI}(v)_{\{x\}}:= \lim_{\tau\rightarrow0}N_{i}^{aI}(v)_{\{x\},\tau} =
\frac{3}{4\pi}N(x(v))\nu^{-2/3}q_{I}^{i}\hat{e}_{I}^{a} \label{niievalue}%
\end{equation}
where $\hat{e}_{I}^{a}$ is the unit tangent vector at $v$ along the edge
$e_{I}$ in the coordinate system $\{ x\}$.

\subsection{Heuristic Operator Action}

\label{heuristic}We motivate a definition for a finite-triangulation
Hamiltonian constraint through the following heuristic arguments. Using
Equation (\ref{nfidentity}) and by parts integration, the Hamiltonian
constraint (\ref{ham4/3}) can be written, modulo terms proportional to the
Gauss constraints (recall that these constraints are $G_{i}=\partial_{a}%
E_{i}^{a}$), as:%

\begin{equation}
C[N]=-\tfrac{1}{2}\int_{\Sigma}\mathrm{d}^{3}x~\epsilon^{ijk}(\pounds _{\vec
{N}_{i}}A_{a}^{j})E_{k}^{a},\qquad N_{i}^{a}:=Nq^{-1/3}E_{i}^{a}%
\end{equation}
where $N_{i}^{a}$ is the electric shift (\ref{electricshift}).

Next, we add a classically-vanishing term which leads to the modified
expression:%
\begin{equation}
C^{\prime}[N]:=C[N]+\tfrac{1}{2}\int_{\Sigma}\mathrm{d}^{3}x~N_{i}^{a}%
F_{ab}^{i}E_{i}^{b}=\tfrac{1}{2}\int_{\Sigma}\mathrm{d}^{3}x~\left(
-\epsilon^{ijk}(\pounds _{\vec{N}_{j}}A_{b}^{k})E_{i}^{b}+%
%TCIMACRO{\tsum \nolimits_{i}}%
%BeginExpansion
{\textstyle\sum\nolimits_{i}}
%EndExpansion
(\pounds _{\vec{N}_{i}}A_{b}^{i})E_{i}^{b}\right)  \label{cprimen}%
\end{equation}
While classically trivial, we shall see in Sections \ref{deformations} and
\ref{2ndh} that this term ensures that in the quantum theory the second
Hamiltonian constraint acts on a vertex displaced by the first one; this is
why we add it above.

We shall think of gauge-invariant charge network states $|c\rangle$ as wave
functions $c(A$) of the connection $A_{a}^{i}$. We write $c(A)$ in the form of
a gauge-invariant graph holonomy (see Section \ref{hfalg}):
\begin{equation}
c(A)=\exp\left(  \int\mathrm{d}^{3}x~c_{i}^{a}A_{a}^{i}\right)
\end{equation}
where we recall that the charge network coordinate $c_{i}^{a}(x)$ is given by
\begin{equation}
c_{i}^{a}(x)=c_{i}^{a}(x;\{e_{I}\},\{q_{I}\})=\sum_{I=1}^{M}\mathrm{i}%
\gamma\kappa q_{I}^{i}\int\mathrm{d}t_{I}~\delta^{(3)}(e_{I}(t_{I}),x)\dot
{e}_{I}^{a}(t_{I}),
\end{equation}

We now seek the action of the quantum correspondent of $C^{\prime}[N]$ on
$c(A)$. Accordingly, we replace the electric shift in Equation (\ref{cprimen})
by the eigenvalue of the quantum shift operator (\ref{hatnc=nc}). The
eigenvalue is no longer a smooth field but, as part of our heuristics, in what
follows below, we shall treat it as a smooth field which is supported only in
the cells $\triangle_{\delta(v)}$ which contain the vertices $v$ of $c$. Next,
we shall think of the remaining electric field operator (corresponding to the
right most term in Equation (\ref{cprimen}) as $\frac{\hbar}{\mathrm{i}}%
\frac{\delta}{\delta A_{b}^{j}}$. We are lead to the following a heuristic
operator action:%
\begin{align}
\hat{C}^{\prime}[N]c(A)  &  =c(A)\int_{\Sigma}\mathrm{d}^{3}x~c_{i}^{a}%
(x)\hat{C}^{\prime}[N]A_{a}^{i}(x)\nonumber\\
&  =\frac{\hbar}{2\mathrm{i}}c(A)\int\mathrm{d}^{3}x~c_{i}^{a}(x)\int
\mathrm{d}^{3}y~\left(  \epsilon^{ljk}(\pounds _{\vec{N}_{l}}A_{b}^{k}%
)\frac{\delta A_{a}^{i}(x)}{\delta A_{b}^{j}(y)}+(\pounds _{\vec{N}_{j}}%
A_{b}^{j})\frac{\delta A_{a}^{i}(x)}{\delta A_{b}^{j}(y)}\right) \nonumber\\
&  =\frac{\hbar}{2\mathrm{i}}c(A)\int_{\Sigma}\mathrm{d}^{3}x~\left(
\epsilon^{ijk}c_{i}^{a}\pounds _{\vec{N}_{k}}A_{a}^{j}+c_{i}^{a}%
\pounds _{\vec{N}_{i}}A_{a}^{i}\right) \nonumber\\
&  =-\frac{\hbar}{2\mathrm{i}}c(A)\int_{\Sigma}\mathrm{d}^{3}x~A_{a}%
^{i}\left(  \epsilon^{ijk}\pounds _{\vec{N}_{j}}c_{k}^{a}+\pounds _{\vec
{N}_{i}}c_{i}^{a}\right)  \label{heurcn}%
\end{align}
Expanding,%
\begin{align}
\hat{C}^{\prime}[N]c(A)  &  =-\frac{\hbar}{2\mathrm{i}}c(A)\int_{\Sigma
}\mathrm{d}^{3}x~\nonumber\\
&  \left(  (\pounds _{\vec{N}_{1}}c_{2}^{a})A_{a}^{3}+(\pounds _{\vec{N}_{1}%
}\bar{c}_{3}^{a})A_{a}^{2}+(\pounds _{\vec{N}_{1}}c_{1}^{a})A_{a}%
^{1}+(\pounds _{\vec{N}_{2}}c_{3}^{a})A_{a}^{1}\right.  \label{cprimehatsigma}%
\\
&  +\left.  (\pounds _{\vec{N}_{2}}\bar{c}_{1}^{a})A_{a}^{3}+(\pounds _{\vec
{N}_{2}}c_{2}^{a})A_{a}^{2}+(\pounds _{\vec{N}_{3}}c_{1}^{a})A_{a}%
^{2}+(\pounds _{\vec{N}_{3}}\bar{c}_{2}^{a})A_{a}^{1}+(\pounds _{\vec{N}_{3}%
}c_{3}^{a})A_{a}^{3}\right)  ,\nonumber
\end{align}
where we have written $\bar{c}_{i}^{a}\equiv-c_{i}^{a}.$
%Because of the way we will quantize $q^{-\alpha}$, and the sequence in which
%it is regarded to act in $\hat{C}^{\prime}$ (right-most), the action of
%$\hat{C}^{\prime}$ is nontrivial only at the vertices of charge networks. So
%that this feature is reflected in the present heuristic computation, we assume
Since the quantum shifts $\vec{N}_{i}$ have support only within the cells
$\triangle_{\delta(v)}$ which contain vertices of the charge network, the
integral in (\ref{cprimehatsigma}) gets contributions only from such cells. If
we further decompose the quantum shift $N_{i}^{a}$ into its edge contributions
$N_{i}^{aI_{v}}$ (see Equation (\ref{hatnc=nc}); $I_{v}$ signifies that the
edges emanate from $v$) at each vertex $v$ and think of each of these
contributions as being of compact support in $\triangle_{\delta(v)}$, the
expression (\ref{cprimehatsigma}) of the Lie derivative with respect to
$N_{i}^{a}$ splits into a sum over edge contributions in each cell
$\triangle_{\delta(v)}$. We obtain
\begin{equation}
\hat{C}^{\prime}[N]c(A)=\sum_{v\in V(c)}\sum_{I_{v}}^{\mathrm{val}(v)}\hat
{C}_{v}^{\prime}[N^{I_{v}}]c(A)
\end{equation}
where val$(v)$ is the valence of $v,$ and
\begin{equation}
\hat{C}_{v}^{\prime}[N^{I_{v}}]c(A)=-\frac{\hbar}{2\mathrm{i}}c(A)\int
_{\triangle_{\delta(v)}}\mathrm{d}^{3}x~A_{a}^{i}\left(  \epsilon
^{ijk}\pounds _{\vec{N}_{j}^{I_{v}}}c_{k}^{a}+\pounds _{\vec{N}_{i}^{I_{v}}%
}c_{i}^{a}\right)
\end{equation}

Since the kinematics of LQG supports the action of finite diffeomorphisms
rather than infinitesimal ones, we approximate the Lie derivative with respect
to $N_{i}^{aI_{v}}$ by small, finite diffeomorphisms, $\varphi(\vec{N}_{i}%
^{I},\delta)$, generated by $N_{i}^{aI_{v}}$:
\begin{equation}
(\pounds _{\vec{N}_{i}^{I}}c_{j}^{a})A_{a}^{k}=-\frac{\varphi(\vec{N}_{i}%
^{I},\delta)^{\ast}c_{j}^{a}A_{a}^{k}-c_{j}^{a}A_{a}^{k}}{\delta}+O(\delta).
\label{lien}%
\end{equation}
Hence
\begin{equation}
\hat{C}_{v}^{\prime}[N^{I_{v}}]c(A)=\frac{1}{\delta}\frac{\hbar}{2\mathrm{i}%
}c(A)\int_{\triangle_{\delta}(v)}\mathrm{d}^{3}x~\left[  \cdots\right]
^{I}+O(\delta) \label{postlien1}%
\end{equation}
where the integrand $\left[  \cdots\right]  ^{I}$ is given by%
\begin{align}
\left[  \cdots\right]  ^{I}  &  =\left[  (\varphi_{1}c_{2}^{a})A_{a}^{3}%
-c_{2}^{a}A_{a}^{3}\right]  +\left[  (\varphi_{1}\bar{c}_{3}^{a})A_{a}%
^{2}-\bar{c}_{3}^{a}A_{a}^{2}\right]  +\left[  (\varphi_{1}c_{1}^{a})A_{a}%
^{1}-c_{1}^{a}A_{a}^{1}\right] \nonumber\\
&  +\left[  (\varphi_{2}c_{3}^{a})A_{a}^{1}-c_{3}^{a}A_{a}^{1}\right]
+\left[  (\varphi_{2}\bar{c}_{1}^{a})A_{a}^{3}-\bar{c}_{1}^{a}A_{a}%
^{3}\right]  +\left[  (\varphi_{2}c_{2}^{a})A_{a}^{2}-c_{2}^{a}A_{a}%
^{2}\right] \nonumber\\
&  +\left[  (\varphi_{3}c_{1}^{a})A_{a}^{2}-c_{1}^{a}A_{a}^{2}\right]
+\left[  (\varphi_{3}\bar{c}_{2}^{a})A_{a}^{1}-\bar{c}_{2}^{a}A_{a}%
^{1}\right]  +\left[  (\varphi_{3}c_{3}^{a})A_{a}^{3}-c_{3}^{a}A_{a}%
^{3}\right]  \label{postlien2}%
\end{align}
We have used the shorthand $\varphi_{i}c_{j}^{a}\equiv\varphi(\vec{N}_{i}%
^{I},\delta)^{\ast}c_{j}^{a}$ and dropped the common $I$. In the above
expression, each line consists of terms which are deformed along a single
shift minus the undeformed quantity; note also that each square-bracketed pair
of terms is $O(\delta)$. Making all sums explicit, we have, in obvious
notation,
\begin{equation}
\hat{C}^{\prime}[N]c(A)=\sum_{v\in V(c)}\sum_{I_{v}}\hat{C}_{v}^{\prime
}[N^{I_{v}}]c(A)=\frac{\hbar}{2\mathrm{i}}c(A)\sum_{v\in V(c)}\sum_{I_{v}}%
\sum_{i}\frac{1}{\delta}\int_{\triangle_{\delta}(v)}\left[  \cdots\right]
_{N_{i}^{I_{v}}}+O(\delta)
\end{equation}
Since the square bracketed terms are $O(\delta)$, we may write%
\begin{equation}
\hat{C}^{\prime}[N]c(A)=\frac{\hbar}{2\mathrm{i}}c(A)\sum_{v\in V(c)}%
\sum_{I_{v},i}\frac{\mathrm{e}^{\int_{\triangle_{\delta}(v)}\left[
\cdots\right]  _{N_{i}^{I_{v}}}}-1}{\delta}+O(\delta). \label{heuristic1}%
\end{equation}
The reason we exponentiate the square bracket is that each summand (to the
right of the summation signs) is proportional to a graph holonomy (minus the
identity) so that the right hand side of the above equation defines a linear
combination of charge network states. For instance (suppressing some of the
$v$ dependence),
\begin{equation}
\mathrm{e}^{\int_{\triangle_{\delta}(v)}\left[  \cdots\right]  _{N_{1}^{I}}%
}=\mathrm{e}^{\int_{\triangle_{\delta}(v)}\left[  (\varphi_{1}^{I}c_{2}%
^{a})A_{a}^{3}-c_{2}^{a}A_{a}^{3}\right]  +\left[  (\varphi_{1}^{I}\bar{c}%
_{3}^{a})A_{a}^{2}-\bar{c}_{3}^{a}A_{a}^{2}\right]  +\left[  (\varphi_{1}%
^{I}c_{1}^{a})A_{a}^{1}-c_{1}^{a}A_{a}^{1}\right]  }%
\end{equation}
describes a graph holonomy which lives on a graph deformation of the original
graph underlying $c$ multiplied by a graph holonomy which lives in the
undeformed vicinity of $v$. The deformation is confined to the vicinity of the
vertex $v$, moves the vertex $v$ along the $I^{\mathrm{th}}$ edge direction
and \textquotedblleft flips\textquotedblright\ the charges on all edges in the
vicinity of deformation by the replacements $q^{2}\rightarrow-q^{3}$,
$q^{3}\rightarrow q^{2}$, $q^{1}\rightarrow q^{1}$, and the undeformed piece
has charges with an inverse flip (see Section \ref{deformations} below).

So far all of these manipulations have been formal and we only use the result
to motivate our definition of the constraint operator. In the next section we
shall discuss these graph deformations at length as they lie at the heart of
our proposed action of the Hamiltonian constraint.

\subsection{Deformations}

\label{deformations}In the previous section we persisted in the fiction that
the quantum shift eigenvalue was a smooth function on $\Sigma$. In actuality,
due to the discrete `quantum geometry' (in this case the discrete electric
lines of force along graphs), the quantum shift vanishes almost everywhere.
This contrast between discrete quantum structures and their smooth classical
correspondents is a characteristic feature of LQG and the appropriate
replacement of the latter by the former in the quantum theory is more of an
art than a deductive exercise. Accordingly, we view the manipulations of the
last section as motivational heuristics; the precise graph deformations
generated by the quantum shift are arrived at by the usual `physicist mixture' of
intuition and mathematical precision. While the details of our choices may
suffer from non-uniqueness, we believe that there is a certain robustness to
their main features. As a final remark, we note that our considerations are
guided by the view that there must be imprints of the graph deformations which
survive the action of diffeomorphisms and the possibility that the chosen
deformations have analogs in the SU$(2)$ case of gravity.

Before turning to the precise form of the deformations we are proposing, we
modify the heuristic starting point in two important ways:\bigskip

\noindent(i) As mentioned above, despite the quantum shift being supported
only at isolated points, we have imagined extending its support smoothly to
$\triangle_{\delta(v)}$ the idea being that as $\delta\rightarrow0$, the `1
point' support at $v$ is formally recovered. We choose to extend the quantum
shift to $\triangle_{\delta(v)}$ by keeping $\frac{3}{4\pi}N(x(v))\nu
_{v}^{-2/3}q_{I_{v}}^{i}$ as an overall factor and extending the edge tangent
${\hat{e}}_{I_{v}}$ at $v$ to $\triangle_{\delta(v)}$ in some smooth,
compactly supported way. This allows us to pull out the factor $\frac{3}{4\pi
}N(x(v))\nu^{-2/3}q_{I}^{i}$ in Equation (\ref{lien}) to obtain
\begin{equation}
(\pounds _{\vec{N}_{i}^{I}}c_{j}^{a})A_{a}^{k}=-\frac{3}{4\pi}N(x(v))\nu
_{v}^{-2/3}q_{I_{v}}^{i}\frac{\varphi(\vec{{\hat{e}}}_{I},\delta)^{\ast}%
c_{j}^{a}A_{a}^{k}-c_{j}^{a}A_{a}^{k}}{\delta}+O(\delta). \label{liee}%
\end{equation}
so that (\ref{postlien1}) is modified to
\begin{equation}
\hat{C}_{v}^{\prime}[N^{I_{v}}]c(A)=\frac{1}{\delta}\frac{\hbar}{2\mathrm{i}%
}c(A)\frac{3}{4\pi}N(x(v))\nu_{v}^{-2/3}q_{I_{v}}^{i}\int_{\Sigma}%
\mathrm{d}^{3}x~\left[  \cdots\right]  _{\delta}^{I_{v},i}+O(\delta)
\end{equation}
where the integrand $\left[  \cdots\right]  _{\delta}^{I_{v},i}$ is given by
\begin{align}
\left[  \cdots\right]  _{\delta}^{I_{v},1}  &  =\left[  (\varphi c_{2}%
^{a})A_{a}^{3}-c_{2}^{a}A_{a}^{3}\right]  +\left[  (\varphi\bar{c}_{3}%
^{a})A_{a}^{2}-\bar{c}_{3}^{a}A_{a}^{2}\right]  +\left[  (\varphi c_{1}%
^{a})A_{a}^{1}-c_{1}^{a}A_{a}^{1}\right] \nonumber\\
\left[  \cdots\right]  _{\delta}^{I_{v},2}  &  =\left[  (\varphi c_{3}%
^{a})A_{a}^{1}-c_{3}^{a}A_{a}^{1}\right]  +\left[  (\varphi\bar{c}_{1}%
^{a})A_{a}^{3}-\bar{c}_{1}^{a}A_{a}^{3}\right]  +\left[  (\varphi c_{2}%
^{a})A_{a}^{2}-c_{2}^{a}A_{a}^{2}\right] \nonumber\\
\left[  \cdots\right]  _{\delta}^{I_{v},3}  &  =\left[  (\varphi c_{1}%
^{a})A_{a}^{2}-c_{1}^{a}A_{a}^{2}\right]  +\left[  (\varphi\bar{c}_{2}%
^{a})A_{a}^{1}-\bar{c}_{2}^{a}A_{a}^{1}\right]  +\left[  (\varphi c_{3}%
^{a})A_{a}^{3}-c_{3}^{a}A_{a}^{3}\right]  , \label{postlien3}%
\end{align}
where $\varphi c_{j}^{a}\equiv\varphi(\vec{{\hat{e}}}_{I_{v}},\delta)^{\ast
}c_{j}^{a}$, and where we have replaced the region of integration
$\triangle_{\delta}(v)$ by $\Sigma$ by virtue of the compact support of
$\vec{{\hat{e}}}_{I_{v}}(x)$. Following a similar line of argument as before,
we are lead to the expression
\begin{align}
\hat{C}^{\prime}[N]c(A)  &  =\sum_{v\in V(c)}\sum_{I_{v}}\hat{C}_{v}^{\prime
}[N^{I_{v}}]c(A)\label{heuristic2}\\
&  =\frac{\hbar}{2\mathrm{i}}c(A)\frac{3}{4\pi}\sum_{v\in V(c)}N(x(v))\nu
_{v}^{-2/3}\sum_{I_{v}}\sum_{i}q_{I_{v}}^{i}\frac{\mathrm{e}^{\int_{\Sigma
}\left[  \cdots\right]  _{\delta}^{I_{v},i}}-1}{\delta}+O(\delta).\nonumber
\end{align}
%Here $\left[  \cdots\right]_{\delta}^{I_{v},i}$
%is obtained by replacing $N_{i}^{aI_{v}}$ by
%${\hat e}_{I_v}$ in $\left[  \cdots\right]
%_{N_{i}^{I_{v}}}$ and we have added the  subscript $v$ to specification of ${\hat e}_I, \nu, q_I$ in
%(\ref{evaluenI}) to remind the reader of the vertex in question.\\
We use (\ref{heuristic2}) as our starting point rather than (\ref{heuristic1})
for the following reasons: The quantum shift depends on the charge $q_{I}^{i}$
(see (\ref{niievalue})). In the SU$(2)$ case this would correspond to an
insertion of a Pauli matrix into the graph holonomy. Exponentiating such an
operation to obtain a linear combination of charge networks seems difficult to
us, so we leave $q_{I}^{i}$ as an overall factor. Considerations of
diffeomorphism covariance \cite{meinprep} lead us to leave the lapse (see
(\ref{niievalue})) as an overall factor as well.\newline

\noindent(ii) The vector ${\hat{e}}_{I_{v}}^{a}$ is tangent to the edge
$e_{I_{v}}$ at $v$. This suggests that the vertex $v$ is to be displaced along
the edge $e_{I_{v}}$ by $O(\delta)$. However (as the reader may verify
\emph{after} reading this section), this leads to a trivial transformation of
$c$. Therefore we will move the displaced vertex slightly off the edge (where,
by slightly, we mean within a distance of $O(\delta^{2})$). As will be
apparent towards the end of this paper, much of the finer details of this
choice will be washed away by the `diffeomorphism covariant' nature of the VSA
states.\newline

We now proceed to define the graph deformations suggested by (i) and (ii)
above. Let us restrict attention to the vicinity of a vertex $v$ (in what
follows we shall on occasion suppress the subscripts indicative of this
specific vertex). We interpret $\varphi(\vec{{\hat{e}}}_{I},\delta)$ to be a
`singular diffeomorphism' which drags the vertex $v$ (and the edges at $v$) a
distance of $O(\delta)$ `almost' (see (ii) above) along the edge $e_{I}$. We
would like this deformation to have support only at the vertex $v$ in the
continuum limit. The right hand side of Equation (\ref{heuristic2}), apart
from the `$-1$' term, is then essentially a sum over charge networks obtained
by multiplying three different graph holonomies. The first is the original
graph holonomy $c(A)\sim h_{c}(A)$; the second is a graph holonomy which sits
on the deformed graph and has charge flips of the type mentioned at the end of
Section \ref{heuristic} and the third is a graph holonomy which sits on the
original, undeformed graph but has (the inverse) charge flips. The
multiplication of the second and third graph holonomies result in non-trivial
charges only in the vicinity of the vertex $v$ and multiplication with the
first (original) graph holonomy results in a charge network state which lives
on the union of the undeformed graph and its deformation with appropriate sums
and difference of the charges coming from the three types of terms.

In Section \ref{placevertex} we detail the position of the displaced vertex
and in Section \ref{newedges} we detail the accompanying deformation of the
edges in the vicinity of $v$. In Section \ref{charges} we describe the charge
labels of the charge network alluded to above as arising from the product of
three graph holonomies and, finally, display the action of the Hamiltonian
constraint operator at finite triangulation on the charge network basis.

\subsubsection{Placement of the Translated Vertex}

\label{placevertex}Let $\dot{e}_{I_{v}}^{b}\equiv\dot{e}_{I}^{b}(v)\equiv
\dot{e}_{I}^{b}$ be the tangent vector of the $I^{\text{th}}$ edge at the
vertex $v$. Fix a Euclidean metric adapted to $\{x\}$ such that $\mathrm{d}%
s^{2}=\delta_{ab}\mathrm{d}x^{a}\mathrm{d}x^{b}$.
%Let $B_{\delta
%}(v)$ be a unit ball around $v$.
%Compute $\dot{e}_{I}^{a}(t_{\cap}),$ the
%tangent to the edge $e_{I}$ at $\partial B_{\delta}(v)\cap e_{I}%
%=:e_{I}(t_{\cap}).$
Choose some unit (normal) vector $\hat{n}_{I}^{a}$ such that%
\begin{equation}
\delta_{ab}\hat{n}_{I}^{a}\dot{e}_{I}^{b}=0 \label{nvec}%
\end{equation}
We have a circle's worth of these. Picking one as detailed in Appendix
\ref{deformApp}, we use it to single out the point%
\begin{equation}
v_{I}^{\prime a}=\delta\hat{e}_{I}^{a}+\delta^{p}\hat{n}_{I}^{a}
\label{posnvprimei}%
\end{equation}
which locates the displaced vertex. Here we choose $p>2$ and, as discussed in
Appendix \ref{deformApp}, $\hat{n}_{I}^{a}$ is chosen so that $v_{I}^{\prime}$
does not lie on the undeformed graph $\gamma(c)$. Also note that the straight
line from $v$ to $\delta\hat{e}_{I}^{a}$ can deviate from the edge $e_{I}$ to
$O(\delta^{2})$ so that $v_{I}^{\prime}$ certainly lies within a distance of
$O(\delta^{2})$ from $e_{I}.$ It is in this sense that $v_{I}^{\prime}$ lies
`almost' on $e_{I}$.
%
%%%%%BEG
Finally, for technical reasons (see Section C of the appendix) we choose $p\ll
k$ (recall that we use semianalytic, $C^{k}$ structures in this work).
%%%%END

\subsubsection{New Edges}

\label{newedges}
\begin{figure}[t]
\centering
\includegraphics[width=12cm]{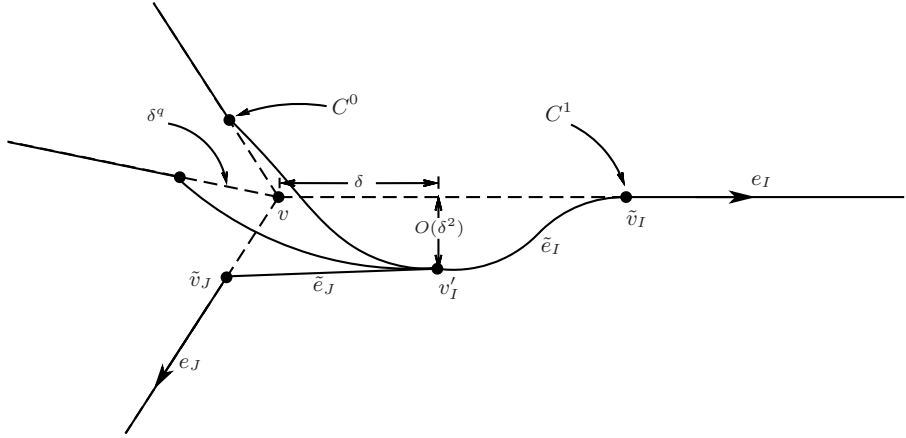}\caption{A sample deformation
produced by a single Hamiltonian constraint action at a non-degenerate vertex
$v$ along the edge $e_{I}$. The dashed edges emanating from the original
vertex $v$ are now only charged in two of the three U(1) factors, but
$v_{I}^{\prime}$ is expected to be, generically, non-degenerate. With respect to the coordinate system
fixed at $v$, $v_{I}^{\prime}$ is located a distance $\delta$ from $v$ along
${\dot e}_{I}$ and displaced off of $e_{I}$ a distance $O(\delta^{2})$. $\tilde
{e}_{I}$ and $e_{I}$ share a tangent at $\tilde{v}_{I}$, but $\tilde{e}_{J}$
and $e_{J}$ do not share a tangent. All of the $\tilde{e}_{J\neq I}$ have
tangents at $v_{I}^{\prime}$ which are `bunched' and lie within a cone of apex
angle $O(\delta^{q-1})$.}%
\label{1ham}%
\end{figure}

We imagine the deformed graph to be obtained by `pulling' the
original graph in the vicinity of the vertex $v$ \textquotedblleft
almost\textquotedblright\ along the direction of the edge $e_{I}$. Thus new
edges $\{\tilde{e}_{K}\}$ are obtained as the image of those parts of the old
edges $\{{e}_{K}\}$ which are in the vicinity of the vertex $v$. The new edges
connect the displaced vertex to the old edges as follows (see Figure
(\ref{1ham}))

%Now to connect the charge-flipped edges to $v_{I}^{\prime}$.
For $e_{I},$ we introduce a trivial vertex $\tilde{v}_{I}$ (on $e_I$), a coordinate
distance $2\delta$ from $v,$ and adjoin the new $C^{k}$-semianalytic edge
$\tilde{e}_{I}$ which connects $\tilde{v}_{I}$ and $v_{I}^{\prime}.$ Since we
want $\tilde{e}_{I}$ to \textquotedblleft almost\textquotedblright\ overlap
with (part of) the edge $e_{I}$, we demand that the transition from $\tilde
{e}_{I}$ to the original edge $e_{I}$ at $\tilde{v}_{I}$ be $C^{k}$-violating
in a strictly $C^{1}$ manner and that the tangent $\dot{\tilde{e}}{}_{I}^{a}$
at $v_{I}^{\prime}$ be proportional to $\dot{e}_{I}^{a}$ at $v$ (these vectors
are comparable in the coordinate system $\{x\}$), i.e.:
\begin{equation}
\hat{\dot{\tilde{e}}}{}_{I}^{a}|_{v_{I}^{\prime}}=\hat{\dot{e}}{}_{I}^{a}|_{v}
\label{doteI}%
\end{equation}
%though the transition is differentiable,
We will refer to such vertices $\tilde{v}_{I}$ as $C^{1}$-kink vertices or
simply as $C^{1}$-kinks.

Next, we introduce a coordinate ball $B_{\delta^{q}}(v)$ of radius $\delta
^{q}$ about $v$.

\bigskip

\textbf{Note:} We choose $q\geq2,q<p$. This choice is important for the
technicalities of Appendix \ref{deformApp}.

\bigskip

For the remaining edges $e_{J\neq I},$ we introduce trivial vertices
$\tilde{v}_{J}$ on all edges $e_{J}$ where they intersect $\partial
B_{\delta^{q}}(v)$; that is, $\tilde{v}_{J}=e_{J}\cap\partial B_{\delta^{q}%
}(v).$ At $\tilde{v}_{J}$ introduce new $C^{k}$-semianalytic edges $\tilde
{e}_{J}$ which split off from each $e_{J}$ and head off to meet $v_{I}%
^{\prime}.$ These edges also `almost' overlap (part of) the edge $e_{I}$,
reflecting our `singular pulling' along of the vicinity of $v$ along the
direction of the edge $e_{I}$. As a result we require that the $C^{k}$
violation at $\tilde{v}_{J}$ be strictly $C^{0}$. Such vertices $\tilde{v}%
_{J}$ will be referred to as $C^{0}$-kink vertices or simply as $C^{0}%
$-kinks.\footnote{We note that the $C^{1}$ or $C^{0}$ nature of the kink is a
diffeomorphism-invariant imprint of the graph deformation.}

Since we imagine $\tilde{e}_{J}$ to be almost along $e_{I}$, we require that
the tangents of new edges $\tilde{e}_{J}$ at $v_{I}^{\prime}$
%need to satisfy
%some properties for the subsequent calculation to go through. First, when they
%eventually meet $v_{I}^{\prime},$ we ask that the tangents
be `bunched' around the direction $-\dot{\tilde{e}}_{I}$ at $v_{I}^{\prime}$
within a cone with apex angle of $O(\delta^{q-1})$ ($q\geq2$) with respect to
$\{x\}$; i.e.
\begin{equation}
\hat{\dot{\tilde{e}}}^{a}_{J}|_{v_{I}^{\prime}}=-\hat{\dot{\tilde{e}}}^{a}%
_{I}|_{v_{I}^{\prime}}+O(\delta^{q-1}), \label{doteJ}%
\end{equation}

Further, since we think of the deformation as some sort of `singular'
diffeomorphism, we require that some subset of diffeomorphism-invariant
properties of the graph structure at $v$ be preserved by the new graph
structure at the displaced vertex $v_{I}^{\prime}$. In particular we require
that if the set of edge tangents $\{\dot{e}_{K}^{a}\}$ at $v$ are such that no
triple lie in a plane, the same should be true of the set $\{\dot{{\tilde{e}}%
}{}_{K}^{a}\}$ at $v_{I}^{\prime}$. Vertices which such properties arise in
the study of moduli in knot classes by Grot and Rovelli \cite{RG}. Grot and
Rovelli call this property `non-degeneracy'. Accordingly we term such vertices
%as GR non-degenerate vertices, or simply 
as GR vertices. Thus, we require that
the graph deformation preserve the GR nature of its vertices. Conversely, we
also require that non-GR vertices do not acquire the GR property under graph
deformation. We shall see that the GR property plays a key role in our
analysis of diffeomorphism covariance \cite{meinprep}.

Since we are thinking of the deformation as a (singular) diffeomorphism, we
also require that no new non-trivial vertices are formed other than
$v_{I}^{\prime}$; i.e., the new edges do not further intersect each other or
the original graph. This may be explicitly achieved as follows.
%Since all we have asked of the new edges $\tilde{e}_{J}$ is that they not
%share a tangent with the corresponding $e_{J}$ at their intersection, and that
%their tangents at $v_{I}^{\prime}$ make a particular angle with $\dot
%{\tilde{e}}_{I}(0),$ we have a great deal of freedom in choosing them. The
%plan of construction is the following: As a first pass,

Let the valence of $v$ be $M$.
%Let $\delta$ be small enough that the parts of
%the edges $e_{K}$ between $v$ and ${\tilde{v}_{K}}$ are analytic.
Consider the $M$ new edges in some order $\tilde{e}_{I},\tilde{e}_{J_{1}%
},\tilde{e}_{J_{2}},..,\tilde{e}_{J_{M-1}},J_{k}\neq I$. Let $\tilde{e}_{I}$
be a semianalytic curve which connects $\tilde{v}_{I}$ with $v_{I}^{\prime}$
in accordance with the requirements on its tangents at $\tilde{v}_{I}$%
,$v_{I}^{\prime}$.
%
%%BEG%%%
Let the coordinate plane (in the $\{x\}$ coordinates) which contains
${v_{I}^{\prime}}$ and which is normal to the direction $\hat{\dot{\tilde{e}}%
}_{I}^{a}|_{v_{I}^{\prime}}$ be $P$. We require that the curve $\tilde{e}_{I}$
intersects $P$ only at $v_{I}^{\prime}$ so that $\tilde{e}_{I}$ is always
\textquotedblleft above\textquotedblright\ $P$. As we show in Appendix C, for
small enough $\delta$ we can always find such an (almost straight in $\{x\}$)
curve.
%%END%%
Let $\tilde{e}_{J_{1}}$ be a straight line in $\{x\}$ which connects
$\tilde{v}_{J}$ with $v_{I}^{\prime}$. If no unwanted intersections are
produced, then we are done, and $\dot{\tilde{e}}_{J_{1}}$ at $v_{I}^{\prime}$
is approximately on the $O(\delta^{q-1})$ cone. If the so-constructed
$\tilde{e}_{J_{1}}$ happens to produce an intersection at some (isolated
because of the semianalyticity of the edges near $v$, see Appendix C) point
other than $\tilde{v}_{J_{1}}$ and $v_{I}^{\prime}$, we can modify it with a
bump function\footnote{By this we mean that we can always apply, \emph{only}
to $\tilde{e}_{J_{1}}$, a semi-analytic diffeomorphism which differs from the
identity in a small enough compact set containing the intersection point so as
to `lift' $\tilde{e}_{J_{1}}$ away from the intersection point. Such a
diffeomorphism can be generated by a vector field obtained by multiplying a
semianalytic, appropriately transverse vector field with a semianalytic
function of compact support.} so that the intersection is avoided. It is
always possible to tune the size of the bump so as to not produce any new
unwanted intersection, nor destroy its tangent-near-on-cone property. We
continue in this manner constructing each $\tilde{e}_{J_{K}}$ as a straight
line, modifying this line with bumps where necessary. Since the `bumping' is
achieved via semianalytic diffeomorphisms, the new edges remain $C^{k}%
$-semianalytic. It remains to show that the GR property (or lack thereof)
of $v$ is preserved. 
%%%OCT 19 2012%%
First consider the case when $v$ is GR. Then if $v_{I}^{\prime}$ is GR we are done.
If not, then as discussed further in Appendix C, we {\em assume} that the above prescription
can be modified in a small vicinity of $v_{I}^{\prime}$, without introducing any $C^0$ or $C^1$ kinks,
in such a way as to render $v_{I}^{\prime}$ GR while still retaining the 
properties described by equations (\ref{posnvprimei}), (\ref{doteI}), and
(\ref{doteJ}). Indeed just such a prescription is constructed in detail in Reference \cite{meinprep}
and we refer the interested reader to section 5.2 of that work.
On the other hand, if  $v$ is not GR, we show in Appendix C.3 
 that a minor
modification of  the prescription of the previous paragraph ensures that 
$v_{I}^{\prime}$ is also not GR.

Before we conclude this section, we note that the above prescriptions at
triangulation fineness $\delta=\delta_{1}$ and at $\delta=\delta_{2}$ with
$\delta_{2}<\delta_{1}$ are not necessarily related by a diffeomorphism. It
turns out that for future considerations, such as the construction of the
space of VSA states, as well as for our study of diffeomorphism covariance in
\cite{meinprep}, it is useful to construct prescriptions which \emph{are}
related by diffeomorphisms. In the appendix we show how this can be done in
such a way that Equations (\ref{posnvprimei}), (\ref{doteI}), and
(\ref{doteJ}) continue to hold.

%Let the `height' of the bump functions be
%$O(\delta^{2p})$ say, to be sure they shrink as $\delta\rightarrow0$ and don't
%produce any pathologies. At the end of the day this construction satisfies the
%properties needed for the calculation below.

\subsubsection{Charges}

\label{charges}Since $\left[  \cdots\right]  _{\delta}^{I_{v},i}$ contains the
difference between a deformed (and charge-flipped) charge network coordinate
and its undeformed relative (but still with flipped charges), $\mathrm{e}%
^{\int\left[  \cdots\right]  _{\delta}^{I_{v},i}}$ contains the product of the
deformed graph holonomy and the inverse of the undeformed relative, and so all
edges of the graph holonomy $\mathrm{e}^{\int\left[  \cdots\right]  _{\delta
}^{I_{v},i}}$ away from the deformation `erase' each other. That is, the
(colored) graph underlying $\mathrm{e}^{\int\left[  \cdots\right]  _{\delta
}^{I_{v},i}}$ itself can be described simply by a gauge-invariant `pyramid
skeleton' consisting of the thin `star' formed by $v$ and (for all original
edges except the $I^{\text{th}}$) coordinate length $\delta^{q}$ edge segments
from the original graph that connect $v$ and $\tilde{v}_{J}$ (for $e_{I},$ the
contribution to the star has coordinate length $2\delta$). The charges on the
star are minus the charge-flipped configuration charges; e.g., for the $i=1$
deformation, the star carries $\left(  -q^{1},q^{3},-q^{2}\right)  $ (with
respect to an original coloring $\left(  q^{1},q^{2},q^{3}\right)  $) on each
of its segments. The remaining edges (which meet $v^{\prime}$) carry the
flipped charges $\left(  q^{1},-q^{3},q^{2}\right)  .$ This pyramid charge
network is multiplied by the original charge network $c(A)$ and, in our
example of $i=1$, the star part of the resulting state carries $\left(
0,q^{2}+q^{3},q^{3}-q^{2}\right)  ,$ which means that $v$ is now a zero-volume
vertex (see Appendix \ref{invq}). A similar conspiration of the charges
results for the other values of $i.$ Our $\hat{q}^{-1/3}$ will now annihilate
this vertex (so the action of another Hamiltonian vanishes here).

%The action of the finite-regulated $\hat{C}_{\delta}%
%^{\prime}$ results in the difference of two charge networks (per each
%$v,i,I$) as we described in the previous paragraph.
We change notation slightly and drop from here on the prime on $\hat
{C}^{\prime}$. Equation (\ref{heuristic2}) then reads
\begin{equation}
\hat{C}_{\delta}[N]c(A)=\frac{\hbar}{2\mathrm{i}}c(A)\sum_{v\in V(c)}\frac
{3}{4\pi}N(x(v))\nu_{v}^{-2/3}\sum_{I_{v},i}q_{I_{v}}^{i}\frac{1}{\delta
}\left(  \exp\left(
%TCIMACRO{\tint }%
%BeginExpansion
{\textstyle\int}
%EndExpansion
\left[  \cdots\right]  _{\delta}^{I_{v},i}\right)  -1\right)  ,
\label{hamconstrdelta}%
\end{equation}
where we have made the regulating parameter $\delta$ explicit on the left hand
side and dropped the $O(\delta)$ term. $\left[  \cdots\right]  _{\delta
}^{I_{v},i}$ stands for the type of deformation described above (with charge
flips).
%, but it's magnitude is only
%determined by $\delta$.
The charge configurations on the edges that meet at $v_{I}^{\prime}$ for the
three quantum shifts $\vec{N}_{i}$ are%
\begin{align}
\vec{N}_{1}  &  :\left(  q^{1},-q^{3},q^{2}\right) \nonumber\\
\vec{N}_{2}  &  :\left(  q^{3},q^{2},-q^{1}\right) \\
\vec{N}_{3}  &  :\left(  -q^{2},q^{1},q^{3}\right) \nonumber
\end{align}
We can write this compactly as%
\begin{equation}
\left.  ^{(i)}\!q^{j}\right.  =\delta^{ij}q^{j}-%
%TCIMACRO{\tsum \nolimits_{k}}%
%BeginExpansion
{\textstyle\sum\nolimits_{k}}
%EndExpansion
\epsilon^{ijk}q^{k} \label{defchrgeflip}%
\end{equation}
where $(i)$ specifies which shift $\vec{N}_{(i)}$ acted.

In the next section we evaluate the action of a second Hamiltonian constraint
on the right hand side of Equation (\ref{hamconstrdelta}). In doing so it is
of advantage to further improve our notation as follows. Denote the charge
network corresponding to $\mathrm{e}^{\int_{\triangle_{\delta}(v)}\left[
\cdots\right]  _{\delta}^{I_{v},i}}c(A)$ by $c(i,v_{I_{v},\delta}^{\prime
a}=\delta\hat{e}_{I_{v}}^{a}+\delta^{p}\hat{n}_{I_{v}}^{a})$ so that Equation
(\ref{hamconstrdelta}) is written as:
\begin{equation}
\hat{C}_{\delta}[N]c(A)=\frac{\hbar}{2\mathrm{i}}\sum_{v\in V(c)}\frac{3}%
{4\pi}N(x(v))\nu_{v}^{-2/3}\sum_{I_{v},i}q_{I_{v}}^{i}\frac{1}{\delta}\left(
c(i,v_{I_{v},\delta}^{\prime})-c\right)  \label{cdeltanc}%
\end{equation}
The various quantifiers $\{I_{v},i,\delta\}$ in the argument of $c$ specify
%which vertex $\hat{C}_{\delta}[N]$ acted on,
the particular edge $e_{I_{v}}$ emanating from $v$ along which the deformation
(of magnitude $\sim\delta$) was performed, and the particular flipping of the
charges via $i.$ Finally note that $\sum_{I_{v}}q_{I_{v}}^{i}=0$ by gauge
invariance (all edges outgoing at $v$) so that:
\begin{equation}
\hat{C}_{\delta}[N]c(A)=\frac{\hbar}{2\mathrm{i}}\sum_{v\in V(c)}\frac{3}%
{4\pi}N(x(v))\nu_{v}^{-2/3}\sum_{I_{v},i}q_{I_{v}}^{i}\frac{1}{\delta
}c(i,v_{I_{v},\delta}^{\prime}) \label{cdeltancfinal}%
\end{equation}

\subsection{Second Hamiltonian}

\label{2ndh}We evaluate the action of a second regularized Hamiltonian
constraint, smeared with a lapse $M$ on the right hand side of
(\ref{cdeltancfinal}). Since we are interested in the continuum limit of (the
action of VSA dual states on) the commutator, we drop those terms in $\hat
{C}_{\delta^{\prime}}[M]\hat{C}_{\delta}[N]c(A)$ which vanish in the continuum
limit upon the antisymmetrization of $N$ and $M$ and `contraction' with a dual
state. The dropped terms are those in which $\hat{C}_{\delta^{\prime}}[M]$
acts at vertices \emph{not} moved by $\hat{C}_{\delta}[N]$; that is, the only
contributions to the commutator will be from terms where $\hat{C}%
_{\delta^{\prime}}[M]$ acts at a vertex newly created by $\hat{C}_{\delta}%
[N]$.\footnote{The reader may easily verify this fact after the perusal of the
next section.}

%Accordingly, we focus on one term in the sum $\hat
%{C}_{\delta}[N]c,$ for a fixed $v,I,i$.
%We denote this charge network, up to
%the overall multiplicative factors, by $c(i,v_{I_{v},\delta}^{\prime a}%
%=\delta\hat{e}_{I_{v}}^{a}+\delta^{p}\hat{n}_{I_{v}}^{a})$ so that
%\begin{equation}
%\hat{C}_{\delta}[N]c(A)=\frac{\hbar}{2\mathrm{i}}\sum_{v\in V(c)}\frac{3}%
%{4\pi}N(x(v))\nu_{v}^{-2/3}\sum_{I_{v},i}q_{I_{v}}^{i}\frac{1}{\delta}\left(
%c(i,v_{I_{v},\delta}^{\prime})-1\right)
%\end{equation}
%The various quantifiers $\{v,I_{v},i,\delta\}$ in the argument of $c$ specify
%which vertex $\hat{C}_{\delta}[N]$ acted on, the particular edge $e_{I_{v}}$
%emanating from $v$ along which the deformation (of magnitude $\sim\delta$) was
%performed, and the particular flipping of the charges via $i.$
%and  compute
Consider the term $\hat{C}_{\delta^{\prime}}[M]c(i,v_{I_{v},\delta}^{\prime}%
)$. Since $v$ now has vanishing inverse volume, the constraint acts at the
displaced vertex $v_{I_{v},\delta}^{\prime}$ as well as on all other vertices
of $c(i,v_{I_{v},\delta}^{\prime})$ which have non-vanishing inverse volume.
But these other vertices are precisely the non-degenerate vertices of $c$
other than $v$. As mentioned above, the contributions from these
non-degenerate vertices vanish in the continuum limit evaluation of the
commutator and so we do not display them here.

The deformations generated by the action of $\hat{C}_{\delta^{\prime}}[M]$ on
$c(i,v_{I_{v},\delta}^{\prime})$ at the vertex $v_{I_{v},\delta}^{\prime}$ are
defined in terms of the coordinate patch around $v_{I_{v},\delta}^{\prime}$
(see Section \ref{deformations}). We denote this coordinate system by
$\{x^{\prime a^{\prime}}\}_{v_{I_{v},\delta}^{\prime}}$ or simply by
$\{x^{\prime a^{\prime}}\}_{\delta}$ or just $\{x^{\prime}\}$ when the context
is clear.

\bigskip

%%BEG
\textbf{Note:} \emph{In this work we require that in their region of joint
validity $\{x_{\delta}^{{\prime}}\}$ and $\{x\}$ are related in a non-singular
fashion as $\delta\rightarrow0$ so that $\lim_{\delta\rightarrow0}\{x^{\prime
a^{\prime}}\}_{\delta}=:\{x^{\prime a^{\prime}}\}_{\delta=0}$ is a good
coordinate system. Specifically, we require that the Jacobian matrix $J^{\mu
}{}_{\nu^{\prime}}(x,x_{\delta}^{\prime}):=\partial x^{\mu}/\partial
x_{\delta}^{\nu^{\prime}}$ is continuous in $\delta$ with non-vanishing and
non-singular determinant. }

\bigskip

It follows from the Note above that
\begin{equation}
\lim_{\delta\rightarrow0}J^{\mu}{}_{\nu^{\prime}}(x,x_{\delta}^{\prime
})=J^{\mu}{}_{\nu^{\prime}}(x,x_{\delta=0}^{\prime}) \label{contj}%
\end{equation}
%%END

\bigskip

One possible way to construct such a set of coordinate patches is as follows:
Since $\Sigma$ is compact, it can be covered by finitely-many coordinate
charts. We pick one such set. Clearly (at least) one chart $\{x_{0}\}$ in this
set covers a neighborhood of $v$ with ${\vec{x}}_{0}(v)$ being the coordinates
of $v$. Rigidly translate $\{x_{0}\}$ by ${\vec{x}}_{0}(v)$ to obtain obtain
$\{x\}$. For small enough $\delta$, $\{x_{0}\}$ also covers small enough
neighborhoods of the new vertices $v_{I_{v},\delta}^{\prime}$ with ${\vec{x}%
}_{0}(v_{I_{v},\delta}^{\prime})$ being the coordinates of $v_{I_{v},\delta
}^{\prime}$. Rigidly translate $\{x_{0}\}$ by ${\vec{x}}_{0}(v_{I_{v},\delta
}^{\prime})$ to obtain $\{x^{\prime}\}_{\delta}$. Clearly, this ensures that
the Jacobian for the $\{x\}$ and $\{x^{\prime}\}_{\delta}$ charts is
unity.\footnote{This choice turns out to result in a conflict with
diffeomorphism covariance; we shall comment on this in the concluding section
and attempt to alleviate the problem in \cite{meinprep}.}

Recall that the edges $\{e_{J_{v}}\}$ at $v$ are deformed to the edges
$\{{\tilde{e}}_{J_{v}}\}$ at $v_{I_{v},\delta}^{\prime}$ so that the valence
of $v$ and $v_{I_{v},\delta}^{\prime}$ are equal and we may use the same index
$J_{v}$ to enumerate the edges at $v$ and their counterparts at $v_{I_{v}%
,\delta}^{\prime}$. In what follows the primed index $a^{\prime}$ denotes
components in the $\{x^{\prime}\}$ system and the primed `hat' superscript,
$^{\symbol{94}\prime}$, denotes unit norm as measured in the $\{x^{\prime}\}$
coordinate metric.

From Equation (\ref{hatnc=nc})
%(\ref{niievalue})
the quantum shift eigenvalues at $v_{I_{v},\delta}^{\prime}$ are defined
through:
\begin{equation}
\hat{M}_{i^{\prime}}^{a^{\prime}}(v_{I_{v},\delta}^{\prime})|c(i,v_{I_{v}%
,\delta}^{\prime})\rangle=\sum_{J_{v}}\left.  ^{(i)}\!M_{i^{\prime}%
}^{a^{\prime}J_{v}}\right.  (v_{I_{v},\delta}^{\prime})|c(i,v_{I_{v},\delta
}^{\prime})\rangle,
\end{equation}
and computed, via Equation (\ref{niievalue}) to be:
\begin{equation}
\left.  ^{(i)}\!M_{i^{\prime}}^{a^{\prime}J_{v}}\right.  (v_{I_{v},\delta
}^{\prime})=\frac{3}{4\pi}M(x_{\delta}^{\prime}(v_{I_{v},\delta}^{\prime}%
))\nu_{v_{I_{v},\delta}^{\prime}}^{-2/3}\left.  ^{(i)}\!q_{J_{v}}^{i^{\prime}%
}\right.  \hat{\tilde{e}}^{\prime}{}_{J_{v}}^{a^{\prime}} \label{mshiftevalue}%
\end{equation}
where $\hat{\tilde{e}}^{\prime}{}_{J_{v}}^{a^{\prime}}$ is the unit tangent to
the edge ${\tilde{e}}_{J_{v}}$ at $v_{I_{v},\delta}^{\prime},$ and where we
have used the fact that the inverse volume eigenvalue is independent of the
charge flips inherent in the $i$-dependence of $c(i,v_{I_{v},\delta}^{\prime
})$ (see Appendix \ref{invq}). The term that survives the antisymmetrization
and continuum limit is%
\begin{equation}
\hat{C}_{\delta^{\prime}}[M]c(i,v_{I_{v},\delta}^{\prime})=\frac{\hbar
}{2\mathrm{i}}\frac{3}{4\pi}M(x_{\delta}^{\prime}(v_{I_{v},\delta}^{\prime
}))\nu_{v_{I_{v},\delta}^{\prime}}^{-2/3}\sum_{J_{v},i^{\prime}}\left.
^{(i)}\!q_{J_{v}}^{i^{\prime}}\right.  \frac{1}{\delta^{\prime}}%
(c(i,i^{\prime},v_{(I_{v},\delta),(J_{v},\delta^{\prime})}^{\prime\prime
})-c(i,v_{I_{v},\delta}^{\prime})) \label{cmc}%
\end{equation}
where the arguments of $c$ denote the deformation and charge flips determined
by $\hat{C}_{\delta^{\prime}}[M].$ We detail their form below.

\begin{figure}[t]
\centering
\includegraphics[width=9cm]{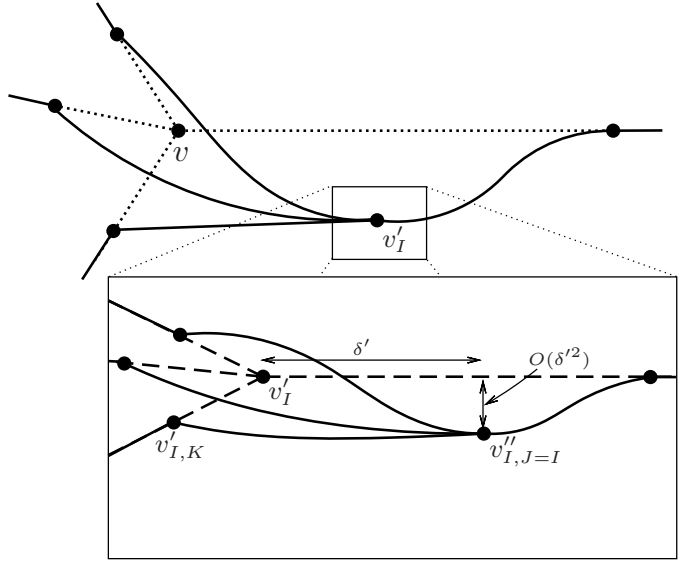}\caption{Detail of the deformation
generated by two successive Hamiltonian actions, in this case along the same
edge $J=I$. Here $\delta^{\prime}\ll\delta$.}%
\label{2ham}%
\end{figure}

We distinguish two types of charge network that appear in the sum:
$J_{v}=I_{v}$ and $J_{v}\neq I_{v}.$ Let $J_{v}=I_{v}$ (this situation is
depicted in Figure (\ref{2ham}))
%recall that we have required $\dot{\tilde e}^a_{I_{v}}|_{v^{\prime}_{I_v,\delta}}=
%\dot{e}_{I_{v}}|_v$ (as vectors computed in
%$\{x\}$)
and focus on the resulting charge network $c(i,i^{\prime},v_{(I_{v}%
,\delta),I_{v},\delta^{\prime})}^{\prime\prime}).$ Following the prescription
given above, $v_{I_{v},\delta}^{\prime}$ moves to (with respect to
$\{x^{\prime}\}$ with origin at $v_{I_{v},\delta}^{\prime}$)%
\begin{equation}
v_{(I_{v},\delta),(I_{v},\delta^{\prime})}^{\prime\prime a^{\prime}}%
=\delta^{\prime}\hat{\tilde{e}}^{\prime}{}_{I_{v}}^{a^{\prime}}+\delta^{\prime
p}\;\hat{n}_{I_{v}}^{\prime a^{\prime}} \label{vdoubledisplaceI}%
\end{equation}
for some ${\hat{n}}^{\prime}$ satisfying the conditions spelt out in Appendix
\ref{deformApp}.

For $J_{v}\neq I_{v},$ $v_{I_{v},\delta}^{\prime}$ gets displaced along one of
the `cone directions' ${\hat{\tilde{e}}}{}_{J_{v}\neq I_{v}}^{\prime
a^{\prime}}$:
\begin{equation}
v_{(I_{v},\delta),(J_{v},\delta^{\prime})}^{\prime\prime a^{\prime}}%
=\delta^{\prime}{\hat{\tilde{e}}}{}_{J_{v}}^{\prime a^{\prime}}+\delta^{\prime
p}{\;{\hat{n}}_{J_{v}}^{\prime a^{\prime}}} \label{vdoubledisplaceJ}%
\end{equation}
The structure of the deformations are as described for the first action, but
with $\delta$ replaced by $\delta^{\prime}.$ The particular charge
configurations at $v^{\prime\prime}$ resulting from each possible sequence of
charge flips is summarized in the following table:%
\begin{equation}%
\begin{tabular}
[c]{l|lll}
& $i=1$ & $i=2$ & $i=3$\\\hline
$i^{\prime}=1$ & $\left(  q^{1},-q^{2},-q^{3}\right)  $ & $\left(  q^{3}%
,q^{1},q^{2}\right)  $ & $\left(  -q^{2},-q^{3},q^{1}\right)  $\\
$i^{\prime}=2$ & $\left(  q^{2},-q^{3},-q^{1}\right)  $ & $\left(
-q^{1},q^{2},-q^{3}\right)  $ & $\left(  q^{3},q^{1},q^{2}\right)  $\\
$i^{\prime}=3$ & $\left(  q^{3},q^{1},q^{2}\right)  $ & $\left(  -q^{2}%
,q^{3},-q^{1}\right)  $ & $\left(  -q^{1},-q^{2},q^{3}\right)  $%
\end{tabular}
\end{equation}
Thus%
\begin{align}
&  \hat{C}_{\delta^{\prime}}[M]\hat{C}_{\delta}[N]c\nonumber\\
&  =\frac{\hbar}{2\mathrm{i}}\sum_{v\in V(c)}\frac{3}{4\pi}N(x(v))\nu
_{v}^{-2/3}\sum_{I_{v},i}q_{I_{v}}^{i}\frac{1}{\delta}\hat{C}_{\delta^{\prime
}}[M]c(i,v_{I_{v},\delta}^{\prime})\nonumber\\
&  =\left(  \frac{\hbar}{2\mathrm{i}}\frac{3}{4\pi}\right)  ^{2}\sum_{v\in
V(c)}\frac{1}{\delta\delta^{\prime}}N(x(v))\nu_{v}^{-2/3}\sum_{I_{v}}%
\nu_{v_{I_{v}}^{\prime}}^{-2/3}\sum_{i}q_{I_{v}}^{i}\label{cmcndelta}\\
&  \qquad\qquad\times\sum_{J_{v},i^{\prime}}\left(  \delta^{ii^{\prime}%
}q_{J_{v}}^{i^{\prime}}-%
%TCIMACRO{\tsum \nolimits_{j}}%
%BeginExpansion
{\textstyle\sum\nolimits_{j}}
%EndExpansion
\epsilon^{ii^{\prime}j}q_{J_{v}}^{j}\right)  M(x_{\delta}^{\prime}%
(v_{I_{v},\delta}^{\prime}))c(i,i^{\prime},v_{(I_{v},\delta),(J_{v}%
,\delta^{\prime})}^{\prime\prime}).\nonumber
\end{align}
%%
%BEG
%where we have set $\nu_{v_{I_{v},\delta}^{\prime}}^{-2/3}\equiv\nu_{v_{I_{v}%
%}^{\prime}}^{-2/3}$ and
Above, we have used gauge invariance to set $\sum_{J_{v}}\left.
^{(i)}\!q_{J_{v}}^{i^{\prime}}\right.  =0$. We have also set $\nu
_{v_{I_{v},\delta}^{\prime}}^{-2/3}\equiv\nu_{v_{I_{v}}^{\prime}}^{-2/3}$;
this follows from the diffeomorphism invariance of the inverse metric
eigenvalue (see Appendix A) together with the fact that the deformations at
different values of $\delta$ are related by diffeomorphisms (see Appendix
C.4).
%%END

\subsection{Continuum Limit}

\label{continuum}In this section we evaluate the continuum limit of the
commutator between a pair of finite triangulation Hamiltonian constraints
under certain assumptions with regard to the properties of the VSA states. In
Section \ref{vsasection} we shall construct a large class of VSA states which
satisfy these assumptions. As mentioned in Section \ref{logicsketchsteps}, the
VSA states are weighted sums over certain bra states. As we shall see in
Section \ref{vsasection}, the weights are obtained by the evaluation of a
smooth complex-valued function $f$ on the \emph{non-degenerate}
\footnote{%In the interests of pedagogy 
We assume that the deformed vertices 
created by the Hamiltonian constraint are nondegenerate (i.e. have
non- vanishing inverse volume) in the deformed chargenet if their undeformed counterparts are 
nondegenerate in the undeformed chargenet. While we expect this to be generically true, 
it is possible that this assumption is violated. However, the assumption is made only for
pedagogy. As the interested reader may verify ({\em after} a perusal of section 6), 
it suffices to replace this non- degeneracy
property by the property that the vertex is GR, has valence greater than 3
 and that there exists no $i$ such that the $i^{\rm th}$
charge vanishes on {\em all} edges emanating from it.
\label{fnotenondeg}
}
 vertices of
the bra it multiplies. More precisely, all bras in the sum have the same
number $n$ of non-degenerate vertices and the evaluation of $f:\Sigma
^{n}\rightarrow\mathbf{%
%TCIMACRO{\U{2102} }%
%BeginExpansion
\mathbb{C}
%EndExpansion
}$ on the $n$ points corresponding to the $n$ non-degenerate vertices of the
bra, provides the weight of that bra in the sum:
\begin{equation}
(\Psi_{B_{\mathrm{VSA}}}^{f}|~:=\sum_{\bar{c}\in B_{\mathrm{VSA}}}\kappa
_{\bar{c}}f(V(\bar{c}))\langle\bar{c}| \label{defvsastate}%
\end{equation}
For simplicity we restrict attention to those ${\bar c}$ such that there is no
symmetry of ${\bar c}$ which interchanges its nondegenerate vertices.
We will sometimes write $\Psi(c):=(\Psi|c\rangle.$ In (\ref{defvsastate}),
$\Psi_{B_{\mathrm{VSA}}}^{f}$ is a VSA state, $B_{\mathrm{VSA}}$ is the set of
bras being summed over, $V(\bar{c})$ denotes the set of non-degenerate
vertices of ${\bar{c},}$ and we have introduced the ${\bar{c}}$-dependent real
number $\kappa_{\bar{c}}$ into the expression. To avoid notational clutter we
have suppressed the $\kappa_{\bar{c}}$ dependence in $(\Psi_{B_{\mathrm{VSA}}%
}^{f}|$. The continuum limit of the commutator is
%(\ref{contlhsdef})%
\begin{equation}
\lim_{\delta\rightarrow0}\lim_{\delta^{\prime}\rightarrow0}\Psi^f
_{B_{\mathrm{VSA}}}(({\hat{C}}_{\delta^{\prime}}[M]{\hat{C}}_{\delta}%
[N]-{\hat{C}}_{\delta^{\prime}}[N]{\hat{C}}_{\delta}[M])c).
\end{equation}
Using Equation (\ref{cmc}), we first evaluate $\lim_{\delta^{\prime
}\rightarrow0}\Psi^f_{B_{\mathrm{VSA}}}({\hat{C}}_{\delta^{\prime}%
}[M]c(i,v_{I_{v},\delta}^{\prime}))$. We have that
\begin{equation}
\Psi_{B_{\mathrm{VSA}}}(\hat{C}_{\delta^{\prime}}[M]c(i,v_{I_{v},\delta
}^{\prime}))=\frac{\hbar}{2\mathrm{i}}\frac{3}{4\pi}M(x_{\delta}^{\prime
}(v_{I_{v},\delta}^{\prime}))\nu_{v_{I_{v}}}^{-2/3}\sum_{J_{v},i^{\prime}%
}\left.  ^{(i)}\!q_{J_{v}}^{i^{\prime}}\right.  \frac{1}{\delta^{\prime}}%
(\Psi_{B_{\mathrm{VSA}}}(c(i,i^{\prime},v_{(I_{v},\delta),(J_{v}%
,\delta^{\prime})}^{\prime\prime})) \label{psifcmcndelta1}%
\end{equation}
where we have set $\nu_{v_{I_{v},\delta}^{\prime}}=\nu_{v_{I_{v}}}$ and used
gauge invariance to drop the last term:
\begin{equation}
\sum_{J_{v}}q_{I_{v}}^{i}=0=\sum_{J_{v}}\left.  ^{(i)}\!q_{J_{v}}^{i^{\prime}%
}\right.  \label{flipchrgesum=0}%
\end{equation}

Next, we make the following assumptions which will be shown to hold in Section
\ref{vsasection}:
%
%%BEG

\begin{enumerate}
\item[(1)] For a point $v\in\Sigma$ and a charge network $c$, either there
exists $\delta_{0}(c)\equiv\delta_{0}$ such that $\forall\delta<\delta_{0}$
there exists $\delta_{0}^{\prime}(\delta)$ such that $\forall\delta^{\prime
}<\delta_{0}^{\prime}(\delta)$ we have that
\begin{equation}
\{\langle c(i,i^{\prime},v_{(I_{v},\delta),(J_{v},\delta^{\prime})}%
^{\prime\prime})~|~\forall~i,i^{\prime},I_{v},J_{v}\}\subset B_{\mathrm{VSA}},
\label{vsaassump1}%
\end{equation}
or $\forall\delta,\delta^{\prime}$ for which $c(i,i^{\prime},v_{(I_{v}%
,\delta),(J_{v},\delta^{\prime})}^{\prime\prime})$ is defined we have that:
\begin{equation}
\{\langle c(i,i^{\prime},v_{(I_{v},\delta),(J_{v},\delta^{\prime})}%
^{\prime\prime})~|~\forall~i,i^{\prime},I_{v},J_{v}\}\cap B_{\mathrm{VSA}%
}=\emptyset. \label{vsaassump2}%
\end{equation}

\item[(2)] If Equation (\ref{vsaassump1}) holds, then
\begin{equation}
\kappa_{c(i,i^{\prime},v_{(I_{v},\delta),(J_{v},\delta^{\prime})}%
^{\prime\prime})}=1\qquad\forall~i,i^{\prime},I_{v},J_{v} \label{kappa=1}%
\end{equation}

\end{enumerate}

%%END%%
%%

If (\ref{vsaassump2}) holds, the right hand side of (\ref{psifcmcndelta1})
vanishes. We shall see in Section \ref{vsasection} that in this case, the
corresponding `matrix element' for the RHS also vanishes. We continue the
calculation in the case that (\ref{vsaassump1}) holds. We have that:
\begin{equation}
\Psi_{B_{\mathrm{VSA}}}(\hat{C}_{\delta^{\prime}}[M]c(i,v_{I_{v},\delta
}^{\prime}))=\frac{\hbar}{2\mathrm{i}}\frac{3}{4\pi}M(x_{\delta}^{\prime
}(v_{I_{v},\delta}^{\prime}))\nu_{v_{I_{v}}}^{-2/3}\sum_{J_{v},i^{\prime}%
}\left.  ^{(i)}\!q_{J_{v}}^{i^{\prime}}\right.  \frac{1}{\delta^{\prime}%
}(f(v_{(I_{v},\delta),(J_{v},\delta^{\prime})}^{\prime\prime})-f(v_{I_{v}%
,\delta}^{\prime}))
\end{equation}
where, once again we have used gauge invariance to append the term
$f(v_{I_{v},\delta}^{\prime})$. In addition for notational convenience
only displayed the dependence of $f$ on the (doubly and singly) deformed images of $v$ and suppressed 
its dependence on the undeformed vertices. Using (\ref{vdoubledisplaceI}),
(\ref{vdoubledisplaceJ}) and the smoothness of $f$, we obtain
\begin{equation}
\lim_{\delta^{\prime}\rightarrow0}\Psi_{B_{\mathrm{VSA}}}(\hat{C}%
_{\delta^{\prime}}[M]c(i,v_{I_{v},\delta}^{\prime}))=\frac{\hbar}{2\mathrm{i}%
}\frac{3}{4\pi}M(x_{\delta}^{\prime}(v_{I_{v},\delta}^{\prime}))\nu_{v_{I_{v}%
}}^{-2/3}\sum_{J_{v},i^{\prime}}\left.  ^{(i)}\!q_{J_{v}}^{i^{\prime}}\right.
({\hat{\tilde{e}}}{}_{J_{v}}^{\prime})^{a}\partial_{a}f(v_{I_{v},\delta
}^{\prime})
\end{equation}
where $({\hat{\tilde{e}}}{}_{J_{v}}^{\prime})^{a}$ is the component of the
unit vector $\vec{{\hat{\tilde{e}}}}{}_{J_{v}}^{\prime}$ in the $\{x\}$
coordinate system. Here the vector $\vec{{\hat{\tilde{e}}}}{}_{J_{v}}^{\prime
}$ is obtained by normalizing the tangent vector to the edge ${\tilde{e}%
}_{J_{v}}$ at $v_{I_{v},\delta}^{\prime}$ in the $\{x^{\prime}\}$ system
(recall, from (\ref{vdoubledisplaceI}), (\ref{vdoubledisplaceJ}) that the
components of this vector in the $\{x^{\prime}\}$ system are given by
$({\hat{\tilde{e}}}{}_{J_{v}}^{\prime})^{a^{\prime}}$).

It follows from the above equation in conjunction with (\ref{cmcndelta}) that
\begin{align}
\lim_{\delta^{\prime}\rightarrow0}\Psi_{B_{\mathrm{VSA}}}(\hat{C}%
_{\delta^{\prime}}[M]\hat{C}_{\delta}[N]c)  &  =\left(  \frac{\hbar
}{2\mathrm{i}}\frac{3}{4\pi}\right)  ^{2}\frac{1}{\delta}\sum_{v}\nu
_{v}^{-2/3}N(x(v))\sum_{I_{v}}M(x_{\delta}^{\prime}(v_{I_{v},\delta}^{\prime
}))\nonumber\\
&  \qquad\times\sum_{i}q_{I_{v}}^{i}\nu_{v_{I_{v}}}^{-2/3}\sum_{J_{v}%
,i^{\prime}}\left.  ^{(i)}\!q_{J_{v}}^{i^{\prime}}\right.  ({\hat{\tilde{e}}%
}{}_{J_{v}}^{\prime})^{a}\partial_{a}f(v_{I_{v},\delta}^{\prime}).
\end{align}
Since $M$ is of density weight $-1/3$ we have:
\begin{equation}
M(x_{\delta}^{\prime}(v_{I_{v},\delta}^{\prime}))=M(x(v_{I_{v},\delta}%
^{\prime}))\left[  \det\left(  \frac{\partial x}{\partial x^{\prime}}\right)
_{v_{I_{v},\delta}^{\prime}}\right]  ^{-1/3}. \label{mdetxxprime}%
\end{equation}
Using this, we obtain
\begin{align}
\lim_{\delta^{\prime}\rightarrow0}\Psi_{B_{\mathrm{VSA}}}(\hat{C}%
_{\delta^{\prime}}[M]\hat{C}_{\delta}[N]c)  &  =\left(  \frac{\hbar
}{2\mathrm{i}}\frac{3}{4\pi}\right)  ^{2}\frac{1}{\delta}\sum_{v}\nu
_{v}^{-2/3}N(x(v))\label{cmcnbrkt}\\
&  \qquad\times\sum_{I_{v}}M(x(v_{I_{v},\delta}^{\prime}))\left[  \det\left(
\frac{\partial x}{\partial x^{\prime}}\right)  _{v_{I_{v},\delta}^{\prime}%
}\right]  ^{-1/3}\{\cdots\}_{I_{v},\delta}\nonumber
\end{align}
where
\begin{equation}
\{\cdots\}_{I_{v},\delta}:=\sum_{i}q_{I_{v}}^{i}\nu_{v_{I_{v}}}^{-2/3}%
\sum_{J_{v},i^{\prime}}\left.  ^{(i)}\!q_{J_{v}}^{i^{\prime}}\right.
({\hat{\tilde{e}}}{}_{J_{v}}^{\prime})^{a}\partial_{a}f(v_{I_{v},\delta
}^{\prime}) \label{defcurly}%
\end{equation}
Next, we use (\ref{posnvprimei}) to Taylor expand $M$ as:
\begin{equation}
M(x(v_{I,\delta}^{\prime}))=M(x(v))+(\delta\hat{e}_{I_{v}}^{a})\partial
_{a}M(x(v))+O(\delta^{2}). \label{taylorm}%
\end{equation}
Using the above Equation in (\ref{cmcnbrkt}) to evaluate the commutator, we
obtain in `bra-ket' notation:
\begin{align}
&  \lim_{\delta^{\prime}\rightarrow0}(\Psi_{B_{\mathrm{VSA}}}|(\hat{C}%
_{\delta^{\prime}}[M]\hat{C}_{\delta}[N]-\left(  N\leftrightarrow M\right)
)|c\rangle\nonumber\\
&  =\left(  \frac{\hbar}{2\mathrm{i}}\frac{3}{4\pi}\right)  ^{2}\sum_{v}%
\nu_{v}^{-2/3}\nonumber\\
&  \qquad\times\sum_{I_{v}}\{N(x(v))\hat{e}_{I_{v}}^{a}\partial_{a}%
M(x(v))-\left(  N\leftrightarrow M\right)  +O(\delta)\}\left[  \det\left(
\frac{\partial x}{\partial x^{\prime}}\right)  _{v_{I_{v},\delta}^{\prime}%
}\right]  ^{-1/3}\{\cdots\}_{I_{v},\delta} \label{cmcnbrkt2}%
\end{align}
We now compute the $\delta\rightarrow0$ limit of the above equation so as to
obtain the continuum limit of the commutator. By virtue of the smooth
dependence of $x$ on $x_{\delta}^{\prime}$ (see the note in Section
\ref{2ndh}) the determinant is a continuous function of $\delta$. It remains
to compute the $\delta\rightarrow0$ limit of $\{\cdots\}_{I_{v},\delta}$.

Since the $\{x\}$ and $\{x^{\prime}\}\equiv\{x^{\prime}\}_{\delta}$ systems
are not necessarily the same, we have that $({\hat{\tilde{e}}}{}_{J_{v}%
}^{\prime})^{a}$ is proportional to $({\hat{\tilde{e}}}_{J_{v}})^{a}%
|_{v_{I_{v},\delta}^{\prime}}$ where now the same tangent vector has been
normalized in the $\{x\}$ system.
%%%BEG
From the Note and equation (\ref{contj}) in Section \ref{2ndh},
%we have
%that $\{x\}$ and $\{x^{\prime}\}$ are smoothly related. This,
%%END
in conjunction with Equations (\ref{doteJ}) in Section \ref{newedges}, we have
that that at $v_{I_{v},\delta}^{\prime}$, for $J_{v}\neq I_{v}$
\begin{equation}
{\hat{\tilde{e}}}{}_{J_{v}}^{\prime a}=-{\hat{\tilde{e}}}{}_{I_{v}}^{\prime
a}+O(\delta^{q-1}),\qquad q\geq2. \label{tildebunching}%
\end{equation}
Using this in ({\ref{defcurly}) together with the smoothness of $\partial
_{a}f$, we obtain
\begin{equation}
\{\cdots\}_{I_{v},\delta}=\nu_{v_{I_{v}}}^{-2/3}\sum_{i}q_{I_{v}}^{i}%
\sum_{i^{\prime}}\left(  \left.  ^{(i)}\!q_{I_{v}}^{i^{\prime}}\right.
-\sum_{J_{v}\neq I_{v}}\left.  ^{(i)}\!q_{J_{v}}^{i^{\prime}}\right.  \right)
({\hat{\tilde{e}}}{}_{I_{v}}^{\prime})^{a}\partial_{a}f(v_{I_{v},\delta
}^{\prime})+O(\delta) \label{curly1}%
\end{equation}
Gauge invariance (\ref{flipchrgesum=0}) then implies that:
\begin{equation}
\{\cdots\}_{I_{v},\delta}:=2\nu_{v_{I_{v}}}^{-2/3}\sum_{i}q_{I_{v}}^{i}%
\sum_{i^{\prime}}\left.  ^{(i)}\!q_{I_{v}}^{i^{\prime}}\right.  ({\hat
{\tilde{e}}}{}_{I_{v}}^{\prime})^{a}\partial_{a}f(v_{I_{v},\delta}^{\prime
})+O(\delta) \label{curly2}%
\end{equation}
Finally, from (\ref{defchrgeflip}) it follows that
\begin{equation}
\lim_{\delta\rightarrow0}\{\cdots\}_{I_{v},\delta}:=2\nu_{v_{I_{v}}}%
^{-2/3}\sum_{i}(q_{I_{v}}^{i})^{2}({\hat{\tilde{e}}}{}_{I_{v}}^{\prime}%
)^{a}\partial_{a}f(v) \label{curly3}%
\end{equation}
Up to this point we have refrained from assuming any particular relation
between $\{x_{\delta=0}^{\prime}\}$ and $\{x\}$ in order to exhibit the
structure of the calculation as $\delta\rightarrow0$.
%%BEG%%
Section 4.1 together with equation (\ref{contj}) implies that
%Let us now assume that
%%END%%
the Jacobian between the two coordinate systems is the identity:
\begin{equation}
\frac{\partial x_{\delta=0}^{\prime\mu}}{\partial x^{\nu}}=\delta_{\nu}^{\mu}.
\label{xprimedelta=0}%
\end{equation}
Using this together with (\ref{curly3}) and (\ref{cmcnbrkt2}) we obtain the
continuum limit of the commutator under the assumption (\ref{vsaassump1}) to
be:
\begin{align}
&  (\Psi_{B_{\mathrm{VSA}}}^{f}|[\hat{C}[M],\hat{C}[N]]|c\rangle\nonumber\\
&  =\lim_{\delta\rightarrow0}\lim_{\delta^{\prime}\rightarrow0}(\Psi
_{B_{\mathrm{VSA}}}^{f}|(\hat{C}_{\delta^{\prime}}[M]\hat{C}_{\delta
}[N]-\left(  N\leftrightarrow M\right)  )|c\rangle\nonumber\\
&  =2\left(  \frac{\hbar}{2\mathrm{i}}\frac{3}{4\pi}\right)  ^{2}\sum_{v\in
V(c)}\sum_{I_{v},i}(q_{I_{v}}^{i})^{2}\nu_{v}^{-2/3}\nu_{v_{I_{v}}}^{-2/3}%
\hat{e}_{I_{v}}^{a}\hat{e}_{I_{v}}^{b}\left(  N\partial_{a}M-M\partial
_{a}N\right)  (x(v))\partial_{b}f(v) \label{lhsfinal}%
\end{align}
}

\section{RHS}

\label{rhs}In Section \ref{rhsid} we display a remarkable classical identity
which expresses the RHS as the Poisson bracket between a pair of
diffeomorphism constraints, each smeared with an electric shift. This implies,
that in the quantum theory, we may identify the RHS with commutator between
two such constraints. Accordingly, in Section \ref{datfinitet} we construct
the finite triangulation operator corresponding to single diffeomorphism
constraint smeared with an electric shift using arguments which parallel those
of Section \ref{hatfinitet}. We compute the finite-triangulation commutator
between two such operators in Section \ref{2ndd}. We compute the continuum
limit of this commutator in Section \ref{ddcontinuum} under certain
assumptions (whose validity is demonstrated in Section \ref{vsasection}) on
the VSA states.

\subsection{A Remarkable Identity}

\label{rhsid}It is straightforward to check that for
\begin{equation}
H[N]=\tfrac{1}{2}\int\mathrm{d}^{3}x~\frac{N}{q^{\alpha}}\epsilon^{ijk}%
E_{i}^{a}E_{j}^{b}F_{ab}^{k},
\end{equation}
we have
\begin{equation}
\{H[M],H[N]\}=\int\mathrm{d}^{3}x~\left(  N\partial_{c}M-M\partial
_{c}N\right)  \frac{E_{i}^{c}E_{i}^{b}}{q^{2\alpha}}F_{ba}^{j}E_{j}%
^{a}=:D[\vec{\omega}],
\end{equation}
where%
\[
\omega^{a}:=\left(  N\partial_{b}M-M\partial_{b}N\right)  q^{-2\alpha}%
E_{i}^{b}E_{i}^{a}.
\]

Let the diffeomorphism generator smeared with the \textquotedblleft electric
shift\textquotedblright\ (see Section \ref{heuristic}), $N_{i}^{a}%
:=q^{-\alpha}NE_{i}^{a}$, be denoted $D[\vec{N}_{i}]$:%
\begin{equation}
D[\vec{N}_{i}]=\int\mathrm{d}^{3}x~q^{-\alpha}NE_{i}^{a}F_{ab}^{j}E_{j}^{b},
\label{electricdiff}%
\end{equation}
We shall refer to $D[\vec{N}_{i}]$ as an electric diffeomorphism constraint.
The Poisson bracket between a pair of electric diffeomorphism constraints is
(summing over the internal index $i$):
\begin{align}
&  \{D[\vec{M}_{i}],D[\vec{N}_{i}]\}\nonumber\\
&  =\int\mathrm{d}^{3}x~\left(  \frac{\delta D[\vec{M}_{i}]}{\delta A_{a}%
^{j}(x)}\frac{\delta D[\vec{N}_{i}]}{\delta E_{j}^{a}(x)}-\left(
N\leftrightarrow M\right)  \right) \nonumber\\
&  =-\int\mathrm{d}^{3}x~2\delta_{\lbrack c}^{a}\partial_{b]}\left(
\frac{ME_{i}^{b}}{q^{\alpha}}E_{j}^{c}\right)  \left(  \frac{NE_{k}%
^{b^{\prime}}}{q^{\alpha}}\delta_{i}^{j}F_{ab^{\prime}}^{k}-\frac
{NE_{i}^{b^{\prime}}}{q^{\alpha}}F_{ab^{\prime}}^{j}+\int\mathrm{d}%
^{3}y~NE_{i}^{b^{\prime}}F_{b^{\prime}c^{\prime}}^{k}E_{k}^{c^{\prime}}%
\frac{\delta q^{-\alpha}(y)}{\delta E_{j}^{a}(x)}\right) \nonumber\\
&  \qquad\qquad\qquad\qquad-\left(  N\leftrightarrow M\right) \nonumber\\
&  =\int\mathrm{d}^{3}x~\left(  \frac{E_{j}^{a}E_{i}^{b}}{q^{\alpha}}%
\frac{E_{i}^{c}}{q^{\alpha}}F_{ac}^{j}N\partial_{b}M+\frac{2E_{[i}^{a}%
E_{j]}^{b}}{q^{\alpha}}\partial_{b}M\int\mathrm{d}^{3}y~NE_{i}^{b^{\prime}%
}F_{b^{\prime}c^{\prime}}^{k}E_{k}^{c^{\prime}}\frac{\delta q^{-\alpha}%
(y)}{\delta E_{j}^{a}(x)}-\left(  N\leftrightarrow M\right)  \right) \\
&  =\int\mathrm{d}^{3}x~\left(  \frac{E_{i}^{b}E_{i}^{c}}{q^{2\alpha}}%
F_{ca}^{j}E_{j}^{a}-2\alpha\frac{E_{i}^{b}E_{i}^{b^{\prime}}}{q^{2\alpha}%
}F_{b^{\prime}c^{\prime}}^{k}E_{k}^{c^{\prime}}\right)  \left(  M\partial
_{b}N-N\partial_{b}M\right) \nonumber\\
&  =\left(  1-2\alpha\right)  \int\mathrm{d}^{3}x~\left(  M\partial
_{b}N-N\partial_{b}M\right)  \frac{E_{i}^{b}E_{i}^{c}}{q^{2\alpha}}F_{ca}%
^{j}E_{j}^{a}\nonumber\\
&  =\left(  2\alpha-1\right)  D[\vec{\omega}],\nonumber
\end{align}
in which we have used%
\begin{equation}
\frac{\delta q^{\alpha}(y)}{\delta E_{i}^{a}(x)}=\alpha q^{\alpha}(E^{-1}%
)_{a}^{i}(y)\delta^{(3)}(x,y),
\end{equation}
where $(E^{-1})_{a}^{i}$ is the `inverse' of $E_{j}^{b}$ so that $E_{a}%
^{i}E_{i}^{b}=\delta_{a}^{b},E_{a}^{i}E_{j}^{a}=\delta_{j}^{i}$. Thus we may
write the RHS as
\begin{equation}
\{H[M],H[N]\}=\frac{1}{2\alpha-1}\sum_{i=1}^{3}\{D[\vec{M}_{i}],D[\vec{N}%
_{i}]\}. \label{u13rhsidentity}%
\end{equation}
In this work we are interested in $\alpha=\frac{1}{3}$ (see Equation
(\ref{ham4/3})). In Section \ref{ddcontinuum} we use this identity to express
the RHS operator as the commutator between two finite diffeomorphism
operators. As mentioned in Section \ref{logicsketchsteps} (see Step 3 of that
section), this facilitates the comparison of the LHS and RHS operators.

\emph{Note that this identity trivializes precisely for the case $\alpha
=\frac{1}{2}$; this is the case of Hamiltonian constraints of density weight
one considered hitherto in the literature}. We take this trivialization as
further support for the move away from the density one case. We also note
that, as shown in Appendix \ref{su2rhsid}, this identity holds for the SU$(2)$
case in $2+1$ and $3+1$ dimensions and in all cases trivializes for the
density weight one choice.

\subsection{The Electric Diffeomorphism Constraint Operator at Finite
Triangulation}

\label{datfinitet}We set $\alpha=\frac{1}{3}$ in (\ref{electricdiff}). Modulo
Gauss law terms we have that:
\begin{equation}
D[{\vec{N}}_{i}]=\int_{\Sigma}\mathrm{d}^{3}x(\pounds _{\vec{N}_{i}}A_{b}%
^{j})E_{j}^{b}%
\end{equation}
where ${\vec{N}}_{i}$ is the electric shift of Section \ref{hatfinitet}. This
motivates, analogous to (\ref{heurcn}), the following heuristic operator
action
\begin{equation}
{\hat{D}}[{\vec{N}}_{i}]c(A)=-\frac{\hbar}{\mathrm{i}}c(A)\int_{\Sigma
}\mathrm{d}^{3}x(\pounds _{\vec{N}_{i}}c_{i}^{a})A_{a}^{i}%
\end{equation}
Following an argumentation similar to that between Equations (\ref{heurcn}%
)-(\ref{liee}) leads us to the finite-triangulation electric diffeomorphism
constraint operator action:
\begin{align}
\hat{D}_{\delta}[\vec{N}_{i}]c  &  =\frac{\hbar}{\mathrm{i}}\frac{3}{4\pi}%
\sum_{v}N(x(v))\nu_{v}^{-2/3}\sum_{I_{v}}q_{I_{v}}^{i}\frac{1}{\delta
}(c(v_{I_{v},\delta}^{\prime})-c)\nonumber\\
&  =\frac{\hbar}{\mathrm{i}}\frac{3}{4\pi}\sum_{v}N(x(v))\nu_{v}^{-2/3}%
\sum_{I_{v}}q_{I_{v}}^{i}\frac{1}{\delta}c(v_{I_{v},\delta}^{\prime})
\label{dndelta}%
\end{align}
where we have used gauge invariance to drop the \textquotedblleft%
$-c$\textquotedblright\ term in the second line and where the charge network
coordinate underlying the state $c(v_{I_{v},\delta}^{\prime})$ is given by
\begin{equation}
(c_{v_{I_{v},\delta}^{\prime}})_{i}^{a}(x):=\varphi(\vec{{\hat{e}}}_{I}%
,\delta)^{\ast}c_{i}^{a}(x)
\end{equation}
where $\varphi(\vec{{\hat{e}}}_{I},\delta)$ deforms the graph underlying $c$
in the manner discussed in Section \ref{deformations}. More in detail, the
graph underlying $c(v_{I_{v},\delta}^{\prime})$ is obtained by removing the
segments of the graph underlying $c$ which connect $v$ to the points
${\tilde{v}}_{J}$ and adjoining new edges, ${\tilde{e}}_{J}$ which connect
${\tilde{v}}_{J}$ to the displaced vertex $v_{I_{v},\delta}^{\prime}$ as
explained in Section \ref{deformations}. The deformed graph is identical to
the one shown in Figure (\ref{1ham}), but with the dashed edges removed. Also
note that since $D[{\vec{N}}_{i}]$ is constructed by smearing the
\emph{diffeomorphism} constraint with an electric shift, the edges ${\tilde
{e}}_{J}$ carry the same charges as $e_{J}$ i.e. there are no
\textquotedblleft charge flips\textquotedblright.

\subsection{Second Electric Diffeomorphism}

\label{2ndd}We evaluate the action of a second electric diffeomorphism
constraint, smeared with the electric shift ${\vec{M}}_{i}$ on the right hand
side of (\ref{dndelta}). Since we are interested in the continuum limit of
(the action of VSA dual states on) the commutator between two electric
diffeomorphism constraints, we drop those terms in $\hat{D}_{\delta^{\prime}%
}[{\vec{M}}_{i}]\hat{D}_{\delta}[{\vec{N}}_{i}]c(A)$ which vanish in the
continuum limit upon the antisymmetrization of $N$ and $M$. The dropped terms
are those in which $\hat{D}_{\delta^{\prime}}[{\vec{M}}_{i}]$ acts at vertices
not moved by $\hat{D}_{\delta}[{\vec{N}}_{i}]$; that is, the only
contributions to the commutator will be from terms where $\hat{D}%
_{\delta^{\prime}}[{\vec{M}}_{i}]$ acts at a vertex which has been moved by
$\hat{D}_{\delta}[{\vec{N}}_{i}]$. Consider the term $\hat{D}_{\delta^{\prime
}}[{\vec{M}}_{i}]c(v_{I_{v},\delta}^{\prime})$. The constraint acts at the
displaced vertex $v_{I_{v},\delta}^{\prime}$ as well as on all other vertices
of $c(v_{I_{v},\delta}^{\prime})$ which have non-vanishing inverse volume. But
these other vertices are precisely the non-degenerate vertices of $c$ other
than $v$. As mentioned above, the contributions from these non-degenerate
vertices vanish in the continuum limit evaluation of the commutator and so we
do not display them here.

\begin{figure}[t]
\centering
\includegraphics[width=10.5cm]{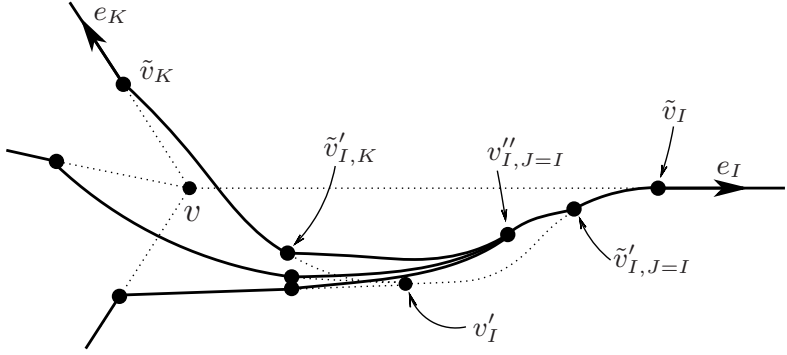}\caption{Sample deformation
produced by two successive singular diffeomorphisms along the edge $e_{I}$.
Here the dotted lines indicate the position of the graph before the
deformations; these are not part of the resulting graph. The structure of the
deformations is similar to that produced by the action of two successive
Hamiltonian-type deformations, except that now the original vertices $v$ and
$v^{\prime}_{I}$ are charged in no copies of U(1); the dotted edges are not
actually there. Note that the kink structure is the same as for the
Hamiltonian deformations: The edge $e_{I}$ is $C^{1}$ at $\tilde{v}_{I}$ and
$\tilde{v}^{\prime}_{I,J=I}$, but all other edges are $C^{0}$ at the various
$\tilde{v}_{K}$ and $\tilde{v}^{\prime}_{I,K}$.}%
\label{2diff}%
\end{figure}

The deformations generated by the action of $\hat{D}_{\delta^{\prime}}%
[{\vec{M}}_{i}]$ on $c(v_{I_{v},\delta}^{\prime})$ at the vertex
$v_{I_{v},\delta}^{\prime}$ are, as in the case of Hamiltonian constraint of
Section \ref{deformations}, defined in terms of the coordinate patch
$\{x^{\prime}\}$ around $v_{I_{v},\delta}^{\prime}$. From Equation
(\ref{hatnc=nc}), we have that
\begin{equation}
{\hat{M}}_{i}^{a^{\prime}}(v_{I_{v},\delta}^{\prime})|c(v_{I_{v},\delta
}^{\prime})\rangle=\sum_{J_{v}}M_{i}^{a^{\prime}J_{v}}(v_{I_{v},\delta
}^{\prime})|c(v_{I_{v},\delta}^{\prime})\rangle,
\end{equation}
with $M_{i}^{a^{\prime}J_{v}}(v_{I_{v},\delta}^{\prime})$ given by
\begin{equation}
M_{i}^{a^{\prime}J_{v}}(v_{I_{v},\delta}^{\prime})=\frac{3}{4\pi}M(x_{\delta
}^{\prime}(v_{I_{v},\delta}^{\prime}))\nu_{v_{I_{v},\delta}^{\prime}}%
^{-2/3}q_{J_{v}}^{i}{{\hat{\tilde{e}}}}{}^{\prime}{_{J_{v}}^{a^{\prime}}}%
\end{equation}
The term that survives the antisymmetrization and continuum limit is%
\begin{align}
\hat{D}_{\delta^{\prime}}[{\vec{M}}_{i}]c(v_{I_{v},\delta}^{\prime})  &
=\frac{\hbar}{\mathrm{i}}\frac{3}{4\pi}M(x_{\delta}^{\prime}(v_{I_{v},\delta
}^{\prime}))\nu_{v_{I_{v},\delta}^{\prime}}^{-2/3}\sum_{J_{v}}q_{J_{v}}%
^{i}\frac{1}{\delta^{\prime}}(c(v_{(I_{v},\delta),(J_{v},\delta^{\prime}%
)}^{\prime\prime})-c(v_{I_{v},\delta}^{\prime}))\nonumber\\
&  =\frac{\hbar}{\mathrm{i}}\frac{3}{4\pi}M(x_{\delta}^{\prime}(v_{I_{v}%
,\delta}^{\prime}))\nu_{v_{I_{v},\delta}^{\prime}}^{-2/3}\sum_{J_{v}}q_{J_{v}%
}^{i}\frac{1}{\delta^{\prime}}c(v_{(I_{v},\delta),(J_{v},\delta^{\prime}%
)}^{\prime\prime}) \label{dmc}%
\end{align}
where we have used gauge invariance to drop the last term in the second line.
Here $c(v_{(I_{v},\delta),(J_{v},\delta^{\prime})}^{\prime\prime})$ denotes
the charge network state obtained by deforming the state $c(v_{I_{v},\delta
}^{\prime})$ by the `singular' diffeomorphism generated by $\hat{D}%
_{\delta^{\prime}}[{\vec{M}}_{i}]$. The deformation moves the vertex
$v_{I_{v},\delta}^{\prime}$ of $c(v_{I_{v},\delta}^{\prime})$ to its new
position, $v_{(I_{v},\delta),(J_{v},\delta^{\prime})}^{\prime\prime}$ given by
Equation (\ref{vdoubledisplaceI}) when $J_{v}=I_{v}$ and by Equation
(\ref{vdoubledisplaceJ}) when $J_{v}\neq I_{v}$. The structure of the
deformations are as described for the first action in Section \ref{datfinitet}%
, but with $\delta\rightarrow\delta^{\prime}$ (see\ Figure (\ref{2diff})).

Restoring the sum over vertices
%and using gauge invariance to drop the last term in (\ref{dmc}),
we have, modulo terms which vanish upon antisymmetrization in the lapses and
the taking of the continuum limit:
\begin{align}
\hat{D}_{\delta^{\prime}}[\vec{M}_{i}]\hat{D}_{\delta}[\vec{N}_{i}]c  &  =
\left(  \frac{\hbar}{\mathrm{i}}\frac{3}{4\pi}\right)  \sum_{v}\frac{1}%
{\delta}N(x(v))\nu_{v}^{-2/3} \sum_{I_{v}}q_{I_{v}}^{i}\hat{D}_{\delta
^{\prime}}[\vec{M}_{i}]c(v_{I_{v},\delta}^{\prime})\nonumber\\
&  = \left(  \frac{\hbar}{\mathrm{i}}\frac{3}{4\pi}\right)  ^{2} \sum_{v}%
\frac{1}{\delta}N(x(v))\nu_{v}^{-2/3}\sum_{I_{v}}q_{I_{v}}^{i}\nu_{v_{I_{v}%
}^{\prime}}^{-2/3}\sum_{J_{v}} q_{J_{v}}^{i}\frac{1}{\delta^{\prime}%
}M(x_{\delta}^{\prime}(v_{I_{v},\delta}^{\prime}))c(v_{(I_{v},\delta
),(J_{v},\delta^{\prime})}^{\prime\prime}) \label{cmcndeltafinal}%
\end{align}

\subsection{Continuum Limit}

\label{ddcontinuum}In this section we evaluate the continuum limit of the
commutator between a pair of finite-triangulation electric diffeomorphism
constraints under certain assumptions with regard to the bra set
$B_{\mathrm{VSA}}$ which underlies the VSA states (see Section \ref{continuum}%
). These assumptions are in addition to Equations (\ref{vsaassump1}%
),(\ref{vsaassump2}) of Section \ref{continuum}.
%In section 6 we shall construct a large class of vsa states
%which satisfy all the  assumptions we have made.
The assumptions are as follows:\newline

\begin{enumerate}
\item[(1)] Given a point $v\in\Sigma$ and a charge network $c$, either, there
exists $\delta_{0}(c)\equiv\delta_{0}$ such that $\forall\delta<\delta_{0}$
there exists $\delta_{0}^{\prime}(\delta)$ such that $\forall\delta^{\prime
}<\delta_{0}^{\prime}(\delta)$ we have that
\begin{equation}
\{\langle c(v_{(I_{v},\delta),(J_{v},\delta^{\prime})}^{\prime\prime
})|~\forall~I_{v},J_{v}\}\subset B_{\mathrm{VSA}}, \label{vsaassump3}%
\end{equation}
or, $\forall\delta,\delta^{\prime}$ for which $c(v_{(I_{v},\delta
),(J_{v},\delta^{\prime})}^{\prime\prime})$ is defined, we have that
\begin{equation}
\{\langle c(v_{(I_{v},\delta),(J_{v},\delta^{\prime})}^{\prime\prime
})|~\forall~I_{v},J_{v}\}\cap B_{\mathrm{VSA}}=\emptyset. \label{vsaassump4}%
\end{equation}

\item[(2)] If Equation (\ref{vsaassump3}) holds, then
\begin{equation}
\kappa_{c(v_{(I_{v},\delta),(J_{v},\delta^{\prime})}^{\prime\prime})}%
=-\tfrac{1}{12},\qquad\forall~I_{v},J_{v}. \label{kappa=-1/12}%
\end{equation}

\item[(3)] Equation (\ref{vsaassump3}) holds if and only if Equation
(\ref{vsaassump1}) holds. Equation (\ref{vsaassump4}) holds if and only if
Equation (\ref{vsaassump2}) holds.
\end{enumerate}

\bigskip

If (\ref{vsaassump4}) holds, it is immediate to see that the continuum limit
of the commutator vanishes; from the assumption above, it follows that the LHS
also vanishes. We continue the calculation in the case that (\ref{vsaassump3})
holds (which also means that by assumption, (\ref{vsaassump1}) holds as well).

From Equation(\ref{dmc}), we have that
\begin{equation}
\Psi_{B_{\mathrm{VSA}}}^{f}(\hat{D}_{\delta^{\prime}}[{\vec{M}}_{i}%
]c(v_{I_{v},\delta}^{\prime}))=-\frac{1}{12}\frac{\hbar}{\mathrm{i}}\frac
{3}{4\pi}M(x_{\delta}^{\prime}(v_{I_{v},\delta}^{\prime}))\nu_{v_{I_{v}%
,\delta}^{\prime}}^{-2/3}\sum_{J_{v}}q_{J_{v}}^{i}\frac{1}{\delta^{\prime}%
}(f(v_{(I_{v},\delta),(J_{v},\delta^{\prime})}^{\prime\prime})-f(v_{I_{v}%
,\delta}^{\prime})) \label{vsadmc}%
\end{equation}
where, once again, we have used gauge invariance to append the last term. It
follows that
\begin{equation}
\lim_{\delta^{\prime}\rightarrow0}\Psi_{B_{\mathrm{VSA}}}^{f}(\hat{D}%
_{\delta^{\prime}}[{\vec{M}}_{i}]c(v_{I_{v},\delta}^{\prime}))=-\frac{1}%
{12}\frac{\hbar}{\mathrm{i}}\frac{3}{4\pi}M(x_{\delta}^{\prime}(v_{I_{v}%
,\delta}^{\prime}))\nu_{v_{I_{v},\delta}^{\prime}}^{-2/3}\sum_{J_{v}}q_{J_{v}%
}^{i}({\hat{\tilde{e}}}{}_{J_{v}}^{\prime})^{a}\partial_{a}f(v_{I_{v},\delta
}^{\prime}). \label{vsadmccont}%
\end{equation}
It follows from Equation (\ref{cmcndeltafinal}) that
\begin{align}
&  \lim_{\delta^{\prime}\rightarrow0}\Psi_{B_{\mathrm{VSA}}}(\hat{D}%
_{\delta^{\prime}}[{\vec{M}}_{i}]\hat{D}_{\delta}[{\vec{N}}_{i}]c)\nonumber\\
&  =-\frac{1}{12}\left(  \frac{\hbar}{\mathrm{i}}\frac{3}{4\pi}\right)
^{2}\frac{1}{\delta}\sum_{v}\nu_{v}^{-2/3}N(x(v))\sum_{I_{v}}M(x_{\delta
}^{\prime}(v_{I_{v},\delta}^{\prime}))q_{I_{v}}^{i}\nu_{v_{I_{v}}}^{-2/3}%
\sum_{J_{v}}q_{J_{v}}^{i}({\hat{\tilde{e}}}{}_{J_{v}}^{\prime})^{a}%
\partial_{a}f(v_{I_{v},\delta}^{\prime}).
\end{align}
Using (\ref{mdetxxprime}) in the above equation we have,
\begin{align}
&  \lim_{\delta^{\prime}\rightarrow0}\Psi_{B_{\mathrm{VSA}}}(\hat{D}%
_{\delta^{\prime}}[{\vec{M}}_{i}]\hat{D}_{\delta}[{\vec{N}}_{i}%
]c)\label{5.4.1}\\
&  =-\frac{1}{12}\left(  \frac{\hbar}{\mathrm{i}}\frac{3}{4\pi}\right)
^{2}\frac{1}{\delta}\sum_{v}\nu_{v}^{-2/3}N(x(v))\sum_{I_{v}}M(x(v_{I_{v}%
,\delta}^{\prime}))\left[  \det\left(  \frac{\partial x}{\partial x^{\prime}%
}\right)  _{v_{I_{v},\delta}^{\prime}}\right]  ^{-\frac{1}{3}}\{\cdots
\}_{i,I_{v},\delta}\nonumber
\end{align}
where
\begin{equation}
\{\cdots\}_{i,I_{v},\delta}:=q_{I_{v}}^{i}\nu_{v_{I_{v}}}^{-2/3}\sum_{J_{v}%
}q_{J_{v}}^{i}({\hat{\tilde{e}}}{}_{J_{v}}^{\prime})^{a}\partial_{a}%
f(v_{I_{v},\delta}^{\prime}). \label{defbrakiivdelta}%
\end{equation}
Using (\ref{taylorm}) in (\ref{5.4.1}) and antisymmetrizing in the lapses, one
obtains (in bra-ket notation):
\begin{align}
&  \lim_{\delta^{\prime}\rightarrow0}(\Psi_{B_{\mathrm{VSA}}}|(\hat{D}%
_{\delta^{\prime}}[{\vec{M}}_{i}]\hat{D}_{\delta}[{\vec{N}}_{i}]-\left(
N\leftrightarrow M\right)  )|c\rangle\nonumber\\
&  =-\frac{1}{12}\left(  \frac{\hbar}{\mathrm{i}}\frac{3}{4\pi}\right)
^{2}\sum_{v}\nu_{v}^{-2/3}\nonumber\\
&  \times\sum_{I_{v}}\{N(x(v))\hat{e}_{I_{v}}^{a}\partial_{a}%
M(x(v))-M(x(v))\hat{e}_{I_{v}}^{a}\partial_{a}N(x(v))+O(\delta)\}\left[
\det\left(  \frac{\partial x}{\partial x^{\prime}}\right)  _{v_{I_{v},\delta
}^{\prime}}\right]  ^{-\frac{1}{3}}\nu_{v_{I_{v}}}^{-2/3}\{\cdots
\}_{i,I_{v},\delta} \label{dmdnbrkt2}%
\end{align}
As in Section \ref{continuum}, the determinant is a continuous function of
$\delta$. It remains to evaluate the $\delta\rightarrow0$ limit of
$\{\cdots\}_{i,I_{v},\delta}$. Using Equation (\ref{tildebunching}) in
(\ref{defbrakiivdelta}) together with gauge invariance, one obtains:
\begin{equation}
\{\cdots\}_{i,I_{v},\delta}=2(q_{I_{v}}^{i})^{2}\nu_{v_{I_{v}}}^{-2/3}%
({\hat{\tilde{e}}}{}_{I_{v}}^{\prime})^{a}\partial_{a}f(v_{I_{v},\delta
}^{\prime})+O(\delta).
\end{equation}
Using this together with Equations (\ref{xprimedelta=0}), (\ref{dmdnbrkt2})
and (\ref{u13rhsidentity}), we obtain the continuum limit of the RHS, in the
case where (\ref{vsaassump3}) holds, to be:
\begin{align}
&  (\Psi_{B_{\mathrm{VSA}}}^{f}|{\hat{D}}[{\vec{\omega}}]|c\rangle\nonumber\\
&  =-3(\Psi_{B_{\mathrm{VSA}}}^{f}|\sum_{i=1}^{3}[\hat{D}[{\vec{M}}_{i}%
],\hat{D}[{\vec{N}}_{i}]]|c\rangle\nonumber\\
&  =-3\sum_{i=1}^{3}\lim_{\delta\rightarrow0}\lim_{\delta^{\prime}%
\rightarrow0}(\Psi_{B_{\mathrm{VSA}}}^{f}|(\hat{D}_{\delta^{\prime}}[{\vec{M}%
}_{i}]\hat{D}_{\delta}[{\vec{N}}_{i}]-\left(  N\leftrightarrow M\right)
)|c\rangle\nonumber\\
&  =2\left(  \frac{\hbar}{2\mathrm{i}}\frac{3}{4\pi}\right)  ^{2}\sum_{v\in
V(c)}\sum_{I_{v},i}(q_{I_{v}}^{i})^{2}\nu_{v}^{-2/3}\nu_{v_{I_{v}}}^{-2/3}%
\hat{e}_{I_{v}}^{a}\hat{e}_{I_{v}}^{b}\left(  N(x(v))\partial_{a}%
M(x(v))-M(x(v))\partial_{a}N(x(v))\right)  \partial_{b}f(v), \label{rhsfinal}%
\end{align}
which agrees with Equation (\ref{lhsfinal}).

\section{Existence of a Large Space of VSA States}

\label{vsasection}In this section we show the existence of VSA states which
satisfy the assumptions (1)-(2) of Section 4 and (1)-(3) of Section 5. As
mentioned in Sections 4 and 5, the VSA states are weighted sums over a set of
bras, the weights being vertex-smooth functions. In Section 6.1, we provide a
qualitative discussion of the issues which arise in the construction of an
appropriate set of VSA states. In Section 6.2 we construct sets of bras and
vertex-smooth functions which specify the VSA states of interest. In Section
6.3 we show that these states satisfy the assumptions of Sections 5 and 6.
While the states we construct span an infinite-dimensional vector space, they
are still of a restricted variety. Specifically, all elements of the sets of
bras under consideration have only one non-degenerate 
\footnote{See Footnote \ref{fnotenondeg} in section 4.6.}
vertex. While a
generalization to the
%We shall
%consider the
case of multiple non-degenerate vertices should not be too difficult, we shall
leave this for the future.

In what follows it is pertinent to recall that in this paper we consider
diffeomorphisms which are semianalytic and $C^{k},$ $k\gg1,$ $k\gg p$. Such
diffeomorphisms send a semianalytic edge into a semianalytic edge which is
$C^{k}$. This implies that the first $k$ derivatives along the edge are
continuous everywhere and at worst, in any semianalytic chart, there are a
finite number of points $p_{i}$ at which the $k_{i}^{\mathrm{th}}$ derivative
along the edge is discontinuous for some $k_{i}>k$.

\subsection{Discussion of Our Strategy}

While we do ignore issues of diffeomorphism covariance in this paper, we would
like to set things up in such a way that issues of diffeomorphism covariance
can be potentially addressed. As a result, we require that the set of bras,
$B_{\mathrm{VSA}}$, be closed under the action of diffeomorphisms. This,
together with a careful study of the assumptions of Sections 4 and 5 imply
that the set of bras should be such that whenever it contains any
doubly-deformed charge network obtained by two successive Hamiltonian
constraint-type deformations, on some charge network $|c\rangle$, it should
also contain (a) \emph{all} other doubly-deformed charge networks obtained by
the action of \emph{any} two successive Hamiltonian constraint-type
deformations on $|c\rangle,$ and (b) \emph{all} doubly-deformed charge
networks obtained by the action of \emph{any} two successive `singular'
diffeomorphism-type deformations which occur on the RHS.

Conversely, if the set contains any doubly-deformed charge network obtained by
two successive singular diffeomorphism-type deformations on some charge
network $|c\rangle$, it should also contain (a) \emph{all} other
doubly-deformed charge networks obtained by the action of \emph{any} two
successive `singular' diffeomorphism-type deformations, and (b) \emph{all}
doubly-deformed charge networks obtained by the action of \emph{any} two
successive Hamiltonian constraint-type deformations on $|c\rangle$.

In suggestive language we call $|c\rangle$ the parent, the single deformations
of $|c\rangle$ its children, and its double deformations its grandchildren.
Our problem then is to ensure that if any grandchild is present in the bra
set, \emph{all} grandchildren should be present. This in turn implies that one
should be able to infer all possible parent charge networks which could yield
a given grandchild. This sort of backward inference is direct for the case of
Hamiltonian constraint grandchildren because the parent charge network graph
is embedded in that of any grandchild, and the charge flips
(\ref{defchrgeflip}) are invertible. However, this embedding of parent into
grandchild is not available for singular diffeomorphism-type grandchildren,
and the bra set needs to be generated via double (Hamiltonian and) singular
diffeomorphism deformations of all possible parent charge networks which could
produce a specific grandchild. This is what we do.

In order to do this we start out with a set of parents from which the output
of grandchildren is well-controlled. Specifically, our starting point is a
parent which is an $n^{\mathrm{th}}$-generation child of a `primordial' charge
network (by `primordial' we mean the charge network is itself not generated by
the action of any Hamiltonian constraint/singular diffeomorphism-type of
deformations on some other charge network). This $n^{\mathrm{th}}$-generation
parent is chosen (for concreteness and simplicity) to be obtained from the
primordial charge network by $n$ Hamiltonian constraint-type deformations. Our
discussion indicates that the charge networks under consideration encode a
sort of `chronological heredity'. As a result, we introduce a suggestive
`causal' nomenclature for certain graph structures of interest in Section 6.2
which go into the construction of $B_{\mathrm{VSA}}$.

As mentioned above, in this paper, we restrict attention to primordial charge
networks with a singular non-degenerate GR vertex. While there seems to be no
barrier to the consideration of multi-vertex primordial charge networks, we
shall leave a generalization of our constructions to such charge networks for
future work.

\subsection{Construction of the VSA States}

Let $|c_{0}\rangle$ be a charge network with a single non-degenerate GR vertex
of valence $M,$
%and let its $M$ edges be semianalytic.
and let $|n,{\vec{\alpha}},c_{0}\rangle$ be the state obtained by $n$
successive finite-triangulation Hamiltonian constraint-type of deformations
applied to $|c_{0}\rangle$. Here, $\vec{\alpha}:=\{\alpha_{i}~|~i=1,\dots
,n\},$ and each $\alpha_{i}$ is a collection of labels corresponding to the
internal charge, vertex, edge, and deformation parameter which go into
specification of the Hamiltonian constraint-type deformations. For example,
for the state $c(i,i^{\prime},v_{(I_{v},\delta),(J_{v},\delta^{\prime}%
)}^{\prime\prime})$ in Equation (\ref{cmc}), we have that $n=2,$ $\alpha
_{1}=(i,v,I_{v},\delta),$ $\alpha_{2}=(i^{\prime},v_{I_{v},\delta}^{\prime
},J_{v},\delta^{\prime})$ and $c=c_{0}$. Let the set of all distinct
diffeomorphic images of $\langle n,{\vec{\alpha}},c_{0}|$ be $B_{[n,\vec
{\alpha},c_{0}]}$. For every element of this set, we generate a new family of
charge networks. In order to do so, for every $\langle c|\in B_{[n,\vec{\alpha
},c_{0}]}$ we now define some graph structures of interest.

Note that every $\langle c |\in B_{[n,\vec{\alpha},c_{0}]}$ has a unique `final'
non-degenerate vertex $v(c)$ of valence $M$ which is connected to one $C^{1}%
$-kink vertex and to $M-1$ $C^{0}$-kink vertices. Let the $I^{\mathrm{th}}$
edge from $v$, $e_{I},$ connect $v$ to the $C^{1}$-kink vertex. Let $e_{J\neq
I}$ connect $v$ to the $C^{0}$-kinks.

\bigskip

\noindent\textbf{Definition: }\emph{The 1-past of $\gamma(c)$}\footnote{Recall
that $\gamma(c)$ is the graph underlying $c$.}: The 1-past of $\gamma(c)$,
denoted by $\gamma_{\mathrm{1}\text{-}\mathrm{p}}(c)$, is the (closed) graph
obtained by removing the edges $e_{K},$ $K=1,..,M$ from $\gamma(c)$; i.e.
\begin{equation}
\gamma_{\mathrm{1}\text{-}\mathrm{p}}(c):=\overline{\gamma(c)-%
%TCIMACRO{\tbigcup \nolimits_{K=1}^{M}}%
%BeginExpansion
{\textstyle\bigcup\nolimits_{K=1}^{M}}
%EndExpansion
e_{K}}.
\end{equation}

\begin{figure}[ptbh]
\centering\begin{minipage}[t]{7.5cm}
\includegraphics[width=7cm]{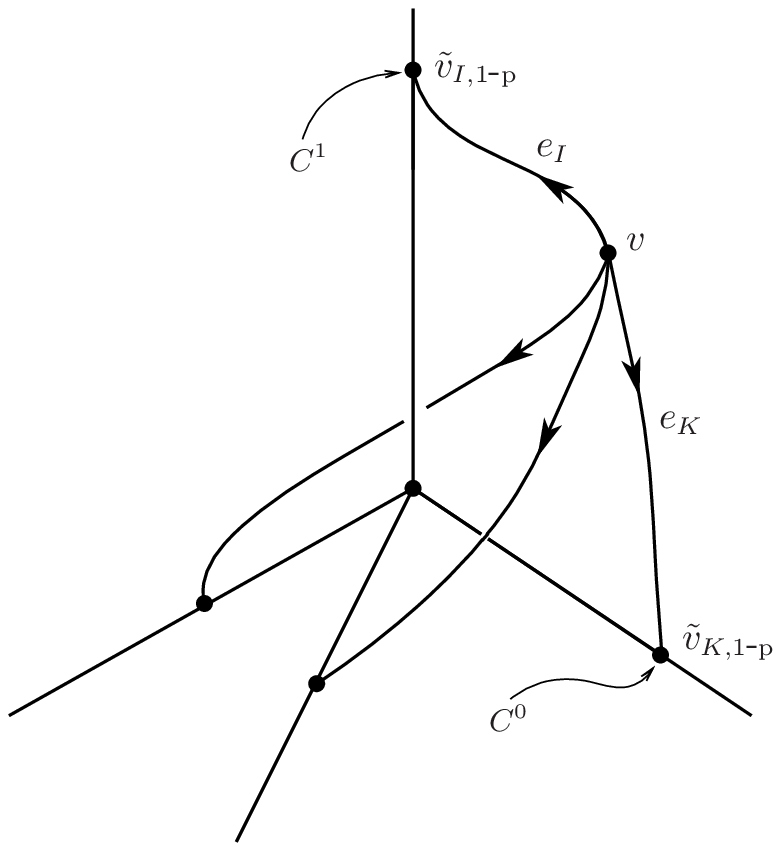}
\caption{The original graph $\gamma(c)$.}
\label{gammac}
\end{minipage}\begin{minipage}[t]{7.5cm}
\includegraphics[width=7cm]{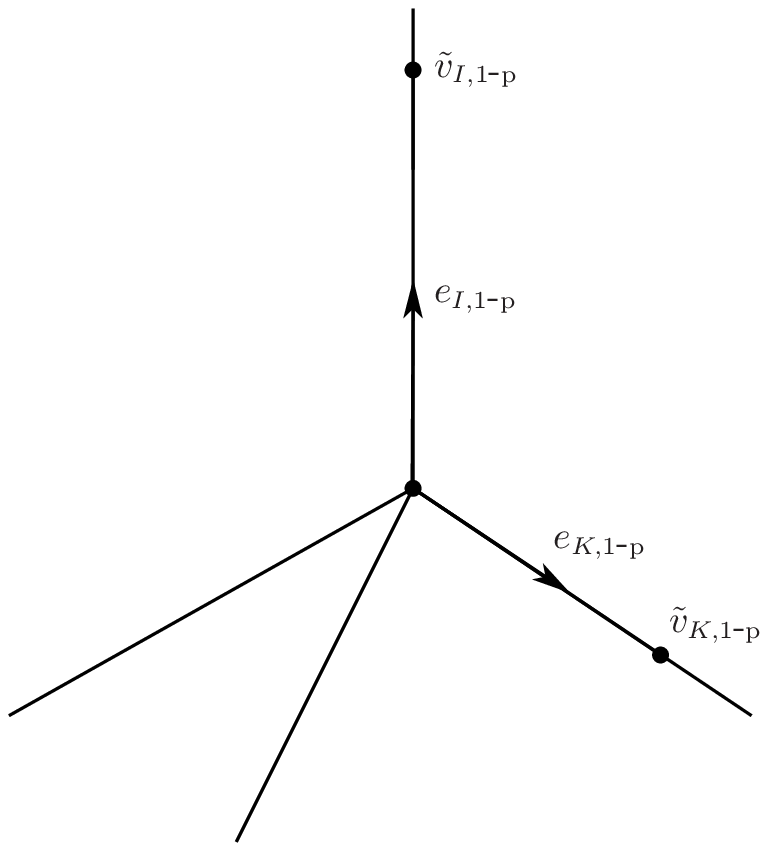}
\caption{The 1-past $\gamma_{1\textrm{-p}}(c)$.}
\label{1past}
\end{minipage}\end{figure}

\begin{figure}[ptbh]
\centering\begin{minipage}[t]{7.5cm}
\includegraphics[width=7cm]{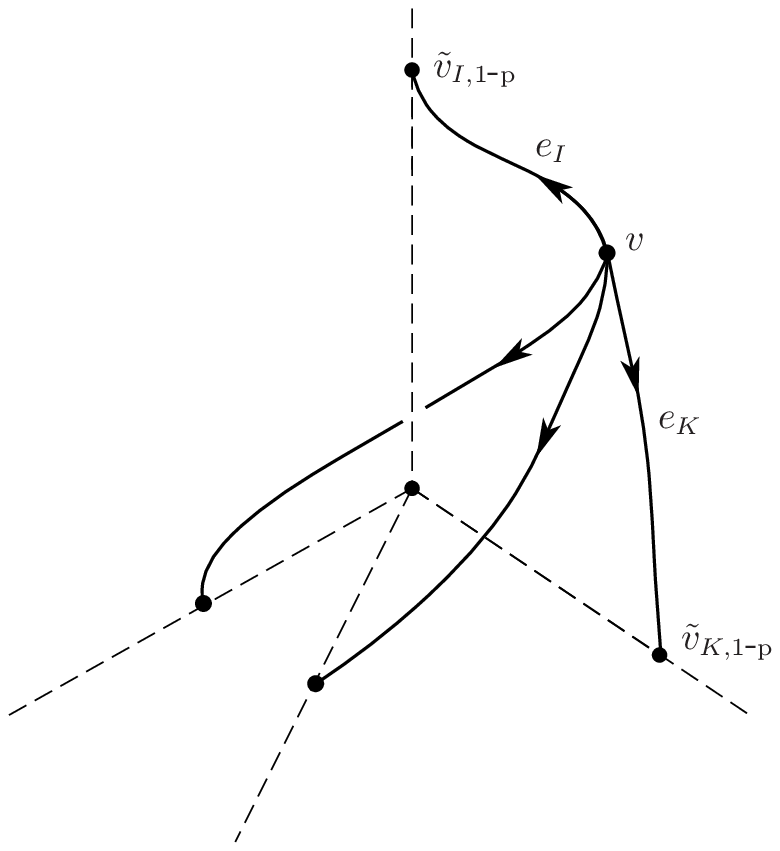} \caption{The future graph $\gamma^{f_0}_{\text{1-p}}$ of $\gamma_{\mathrm{1}%
\text{-}\mathrm{p}}(c)$ in $c$.}
\label{gammafutureinc}
\end{minipage}\begin{minipage}[t]{7.5cm}
\includegraphics[width=7cm]{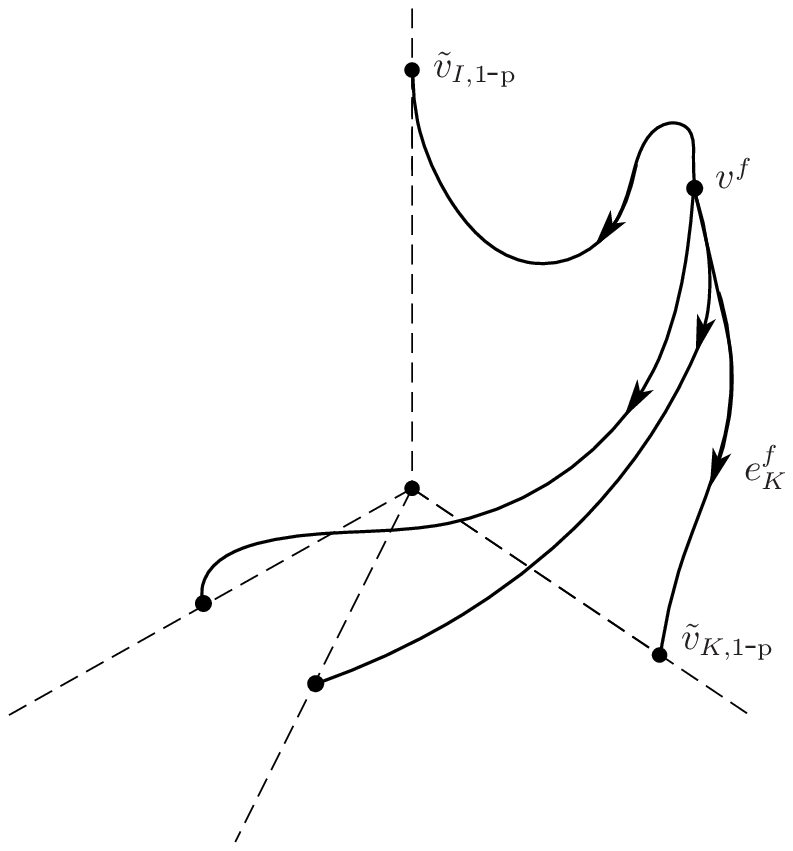} \caption{A future graph $\gamma^{f}_{\text{1-p},c}$ of $\gamma_{1\textrm{-p}}(c)$ with respect to $c$.}
\label{gammaotherfuture}
\end{minipage}\end{figure}

\begin{figure}[ptbh]
\centering
\includegraphics[width=7cm]{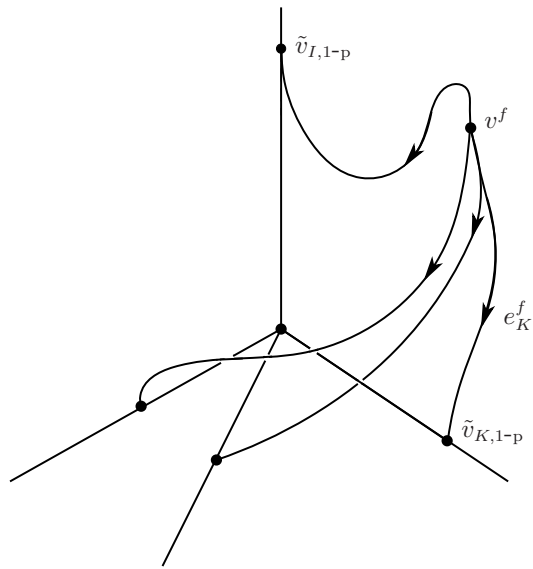} \caption{The graph underlying a
causal completion of the 1- past of $c$.}%
\label{gammafullfuture}%
\end{figure}

Let $e_{K}$ intersect $\gamma_{\mathrm{1}\text{-}\mathrm{p}}(c)$ at
${\tilde{v}}_{K,\mathrm{1}\text{-}\mathrm{p}}$ on the edge $e_{K,\mathrm{1}%
\text{-}\mathrm{p}}$ of $\gamma_{\mathrm{1}\text{-}\mathrm{p}}(c)$ so that
${\tilde{v}}_{I,\mathrm{1}\text{-}\mathrm{p}}$, ${\tilde{v}}_{J\neq
I,\mathrm{1}\text{-}\mathrm{p}}$ are the $C^{1},C^{0}$ kinks mentioned above.
Since $|c\rangle$ is diffeomorphic to $|n,{\vec{\alpha}},c_{0}\rangle$, it
follows that the edges $e_{K,\mathrm{1}\text{-}\mathrm{p}}$ intersect at a GR
vertex which we denote by $v_{\mathrm{1}\text{-}\mathrm{p}}(c)$. The following
definitions are illustrated in Figures (\ref{gammac})-(\ref{gammafullfuture}).

\bigskip

\noindent\textbf{Definition: }\emph{The future graph of $\gamma_{\mathrm{1}%
\text{-}\mathrm{p}}(c)$ in $c$}: The future graph of $\gamma_{\mathrm{1}%
\text{-}\mathrm{p}}(c)$ in $c$, denoted by $\gamma_{\mathrm{1}\text{-}%
\mathrm{p}}^{f_{0}}$, is defined by
\begin{equation}
\gamma_{\mathrm{1}\text{-}\mathrm{p}}^{f_{0}}:=\cup_{K=1}^{M}e_{K}%
={\overline{\gamma(c)-\gamma_{\mathrm{1}\text{-}\mathrm{p}}(c)}}.
\end{equation}
Thus, modulo the action of diffeomorphisms, $\gamma_{\mathrm{1}\text{-}%
\mathrm{p}}^{f_{0}}$ is just the nested graph structure produced by the action
of a particular Hamiltonian constraint-type deformation which acts on the
`parent' vertex $v_{\mathrm{1}\text{-}\mathrm{p}}(c)$ of the `parent' charge
network based on the graph $\gamma_{\mathrm{1}\text{-}\mathrm{p}}(c)$.

Next, we define a graph structure which is similar to $\gamma_{\mathrm{1}%
\text{-}\mathrm{p}}^{f_{0}}$ in terms of its \textquotedblleft
causal\textquotedblright\ properties.

\bigskip

\noindent\textbf{Definition: }\emph{A future graph of $\gamma_{\mathrm{1}%
\text{-}\mathrm{p}}(c)$ with respect to $c$}: A graph $\gamma_{\mathrm{1}%
\text{-}\mathrm{p},c}^{f}$ is a future graph of $\gamma_{\mathrm{1}%
\text{-}\mathrm{p}}(c)$ with respect to $c$ if and only if it has the
following properties:

\begin{enumerate}
\item[(i)] $\gamma_{\mathrm{1}\text{-}\mathrm{p},c}^{f}=\cup_{K=1}^{M}%
e_{K}^{f}$ where $e_{K}^{f}$ for each $K$ is a semianalytic $C^{k}$ edge such
that $e_{K}^{f}\cap\gamma_{\mathrm{1}\text{-}\mathrm{p}}(c)={\tilde{v}%
}_{K,\mathrm{1}\text{-}\mathrm{p}}$ and such that the edges $e_{K}^{f}$ do not
intersect each other except at the GR vertex $v^{f}\in\Sigma$ from which they
are outgoing.

\item[(ii)] If we color each $e_{K}^{f}$ with the same charge as $e_{K}$
carries in $c$ (with respect to the orientation in (i)), then $v^{f}$ is non-degenerate.

\item[(iii)] Define $\gamma_{c}^{f}$ as
\begin{equation}
\gamma_{c}^{f}:=\gamma_{\mathrm{1}\text{-}\mathrm{p}}(c)\cup\gamma
_{\mathrm{1}\text{-}\mathrm{p},c}^{f}. \label{defgammafc}%
\end{equation}
Then with respect to $\gamma_{c}^{f}$, the point ${\tilde{v}}_{I,\mathrm{1}%
\text{-}\mathrm{p}}$ is a trivalent $C^{1}$-kink vertex and the points
${\tilde{v}}_{J\neq I,\mathrm{1}\text{-}\mathrm{p}}$ are trivalent $C^{0}%
$-kink vertices.
\end{enumerate}

Note that the future graph of $\gamma_{\mathrm{1}\text{-}\mathrm{p}}(c)$ in
$c$ is a future graph of $\gamma_{\mathrm{1}\text{-}\mathrm{p}}(c)$ with
respect to $c$ but the converse is not necessarily true. In particular the set
of tangent vectors at the non-degenerate vertex $v_{f}$ (of a future graph of
$\gamma_{\mathrm{1}\text{-}\mathrm{p}}(c)$ with respect to $c$) need not be
obtained through the action of a diffeomorphism from the set of tangent
vectors at the non-degenerate vertex $v$ of $c$; i.e., the two sets may have
different moduli \cite{RG}.

Next, we define a charge network which is identical to $c$ in terms of its
causal properties and colorings.\newline

\noindent\textbf{Definition:}\emph{A causal completion of the 1-past of $c$:}
%A causal completion of the 1- past of $c$ is the charge net based on a graph
%which is the union of the 1- past of $\gamma (c)$ with a future graph of the 1- past of $\gamma (c)$.
%Specifically,
A causal completion of the 1-past of $c$, denoted by $c^{f}(c)$, is the charge
network based on the graph $\gamma_{c}^{f}$ (see Equation (\ref{defgammafc}))
%$\gamma_{1-p}(c)\cup  \gamma^f_{1-p, c}$
with charges on $\gamma_{\mathrm{1}\text{-}\mathrm{p}}(c)$ being the same as
those coming from $c$, and on $\{e_{K}^{f}\}$ being the same as those on
$\{e_{K}\}$ in $c$.
%We shall refer to $c^f(c)$ as an allowable completion of the 1 past of $c$.
\newline

Note that the definition of the 1-past in terms of the removal of immediate
edges from a final non-degenerate vertex to trivalent kink vertices extends
naturally to such causal completions and we shall assume that the definition
has been so extended.

We now use the above definitions to construct $B_{\mathrm{VSA}}$ as follows.
Consider all distinct causal completions, $\langle c^{f}(c)|$
%which satisfy (i)- (iii) above
for every $c\in B_{[n,{\vec{\alpha}},c_{0}]}$. Let the resulting set of bras
be $B_{\langle n,{\vec{\alpha}},c_{0}\rangle}$. Consider all possible single
Hamiltonian constraint-type deformations (i.e. for all values of `$\alpha$')
of elements of $B_{\langle n,{\vec{\alpha}},c_{0}\rangle}$ and take all
distinct diffeomorphic images of the resulting set of charge networks. Call
the resulting set $B_{[H\langle n,{\vec{\alpha}},c_{0}\rangle]}$. Repeat this
procedure again. That is, once again consider all Hamiltonian constraint-type
deformations of the elements of this set and then take distinct diffeomorphic
images of such deformed charge networks. Call this set $B_{[H[H\langle
n,{\vec{\alpha}},c_{0}\rangle]]}$.

%dmc $c(v_{(I_{v},\delta),(J_{v}%
%,\delta^{\prime})}^{\prime\prime})$

Next, we consider deformations of the type encountered in the RHS.
Accordingly, denote a double `singular' diffeomorphism-type of deformation of
any state $|c\rangle$ by $\hat{D}^{2}(\beta)|c\rangle$. Here $\beta$ is a
label which specifies the vertex at which the deformation takes place, the two
edge labels along which the deformations take place and the parameters
$\delta,\delta^{\prime}$ which quantify the amount of deformation. For
example, for the state $c(v_{(I_{v},\delta),(J_{v},\delta^{\prime})}%
^{\prime\prime})$ in Equation (\ref{dmc}), we have that
\begin{equation}
|c(v_{(I_{v},\delta),(J_{v},\delta^{\prime})}^{\prime\prime})\rangle=\hat
{D}^{2}(\beta)|c\rangle\;\;\;\mathrm{with}\;\;\;\beta=(v,I_{v},J_{v}%
,\delta,\delta^{\prime}) \label{defdsquarebeta}%
\end{equation}
Act by $\hat{D}^{2}(\beta)$ for all $\beta$ on elements of $B_{\langle
n,{\vec{\alpha}},c_{0}\rangle}$ and then take all distinct diffeomorphic
images thereof to form the set $B_{[D^{2}\langle n,{\vec{\alpha}},c_{0}%
\rangle]}$.

Finally define $B_{\mathrm{VSA}}$ as:
\begin{equation}
B_{\mathrm{VSA}}:=B_{[H[H\langle n,{\vec{\alpha}},c_{0}\rangle]]}\cup
B_{[D^{2}\langle n,{\vec{\alpha}},c_{0}\rangle]} \label{defbvsa}%
\end{equation}
Note that every element of $B_{\mathrm{VSA}}$ has a single `final'
non-degenerate GR vertex of valence $M$.

In terms of our discussion in Section 6.1, $|c_{0}\rangle$ is a primordial
charge network, $|n,{\vec{\alpha}},c_{0}\rangle$ is the parent in the
$n^{\mathrm{th}}$ generation, $B_{[n,{\vec{\alpha}},c_{0}]}$ is the set of all
diffeomorphic images of this parent. The role of $B_{\langle n,{\vec{\alpha}%
},c_{0}\rangle}$ is as follows. Recall from Section 6.1 that if a grandchild
is present in $B_{\mathrm{VSA}}$, we need to ensure that all possible related
grandchildren are present as well. This necessitates the identification of a
set of (grand)parents which give birth to all these grandchildren. Since the
specific (grand) parent which gives rise to a double singular diffeomorphism
grandchild is not embedded in the grandchild, it is difficult (and perhaps
impossible) to infer the identity of the specific (grand)parent which gave
birth to such a grandchild. The solution is then to accommodate \emph{all}
possible (grand)parents which could conceivably have given birth to the
grandchild in question. The set of all possible such (grand)parents is
$B_{\langle n,{\vec{\alpha}},c_{0}\rangle}$.

Before we proceed to the next section, we prove a Lemma which will be of use below.

\bigskip

\noindent\textbf{Lemma:} The set $B_{\langle n,{\vec{\alpha}},c_{0}\rangle}$
is closed under the action of diffeomorphisms; i.e., in the notation we have
used above, we have that $B_{\langle n,{\vec{\alpha}},c_{0}\rangle
}=B_{[\langle n,{\vec{\alpha}},c_{0}\rangle]}$.

\bigskip

\emph{Proof:} Let $\langle{\hat{c}}|~\in B_{\langle n,{\vec{\alpha}}%
,c_{0}\rangle}$. This means that $\hat{c}$ is the causal completion of the
1-past of some charge network $c$ such that $\langle{c}|~\in B_{[n,{\vec
{\alpha}},c_{0}]}$. Consider the charge network $\phi\circ c$ obtained by the
action of the diffeomorphism $\phi$ on $c$. It is then straightforward to
check that $\phi\circ{\hat{c}}$ is a causal completion of the 1-past of
$\phi\circ c$. This implies that $\langle\phi\circ\hat{c}|~\in B_{\langle
n,{\vec{\alpha}},c_{0}\rangle}$ which completes the proof.

\subsection{Demonstration of Assumed Properties of VSA States}

The VSA states are constructed as in Sections 4 and 5 by summing over all bras
in the set $B_{\mathrm{VSA}}$ defined by Equation (\ref{defbvsa}), with each
bra weighted by the evaluation of a vertex smooth function $f:\Sigma
\rightarrow%
%TCIMACRO{\U{2102} }%
%BeginExpansion
\mathbb{C}
%EndExpansion
$ on the single non-degenerate vertex of the bra it multiplies.

Let $|{\bar{c}}\rangle$ be a charge network state. Then the following cases
are of interest:

\begin{enumerate}
\item[(a)] ${\bar{c}}$ is such that some double Hamiltonian constraint
deformation of ${\bar{c}}$ is in $B_{\mathrm{VSA}}$; i.e., in the notation of
the previous section, $|{\bar{\alpha}}_{1},{\bar{\alpha}}_{2},{\bar{c}}%
\rangle\in B_{\mathrm{VSA}}$ for some ${\bar{\alpha}}_{1},{\bar{\alpha}}_{2}$
which specify the two successive Hamiltonian constraint-type deformations the
${\bar{\alpha}}_{2}$ deformation occurring after the ${\bar{\alpha}}_{1}$ deformation.

\item[(b)] ${\bar{c}}$ is such that some double singular diffeomorphism
deformation ${\bar{c}}$ is in $B_{\mathrm{VSA}}$; i.e., in the notation of the
previous section, $|{\bar{\beta}},{\bar{c}}\rangle\in B_{\mathrm{VSA}}$ for
some ${\bar{\beta}}$ which specifies the two successive singular
diffeomorphism-type deformations.

\item[(c)] ${\bar{c}}$ is such that some single Hamiltonian constraint
deformation of ${\bar{c}}$ is in $B_{\mathrm{VSA}}$; i.e., in the notation of
the previous section, $|{\bar{\alpha}},{\bar{c}}\rangle\in B_{\mathrm{VSA}}$
for some ${\bar{\alpha}}$ which specifies a Hamiltonian constraint-type of deformation.
\end{enumerate}

We now consider each of them in turn.

\bigskip

\noindent\textbf{Case (a)}: First note that $|{\bar{c}}\rangle$ can be
reconstructed from $|{\bar{\alpha}}_{1},{\bar{\alpha}}_{2},{\bar{c}}\rangle$
as follows. Let $\gamma({\bar{\alpha}}_{1},{\bar{\alpha}}_{2},{\bar{c}})$ be
the graph underlying $|{\bar{\alpha}}_{1},{\bar{\alpha}}_{2},{\bar{c}}\rangle
$. Clearly its 1-past is the graph $\gamma({\bar{\alpha}}_{1},{\bar{c}})$
which underlies the state $|{\bar{\alpha}}_{1},{\bar{c}}\rangle$. The colors
of $|{\bar{\alpha}}_{1},{\bar{c}}\rangle$ can be obtained as follows. Retain
the colors from $|{\bar{\alpha}}_{1},{\bar{\alpha}}_{2},{\bar{c}}\rangle$ on
those edges in its 1-past which do not emanate from the final vertex
$v_{\mathrm{1}\text{-}\mathrm{p}}({\bar{\alpha}}_{1},{\bar{\alpha}}_{2}%
,{\bar{c}})$ of this 1-past. Note that the edges $e_{K,\mathrm{1}%
\text{-}\mathrm{p}},$ $K=1,..,M$ emanating from the final vertex
$v_{\mathrm{1}\text{-}\mathrm{p}}({\bar{\alpha}}_{1},{\bar{\alpha}}_{2}%
,{\bar{c}})$ of this 1-past each acquire kink vertices,${\tilde{v}%
}_{K,\mathrm{1}\text{-}\mathrm{p}}$, in $|{\bar{\alpha}}_{1},{\bar{\alpha}%
}_{2},{\bar{c}}\rangle$. The part of $e_{K,\mathrm{1}\text{-}\mathrm{p}}$
which connects ${\tilde{v}}_{K,\mathrm{1}\text{-}\mathrm{p}}$ to
$v_{\mathrm{1}\text{-}\mathrm{p}}({\bar{\alpha}}_{1},{\bar{\alpha}}_{2}%
,{\bar{c}})$ suffers changes of its colors relative to its coloring in
$|{\bar{\alpha}}_{1},{\bar{c}}\rangle$, but the remaining part retains its
charges from $|{\bar{\alpha}}_{1},{\bar{c}}\rangle$. Hence we can read off the
coloring of each $e_{K,\mathrm{1}\text{-}\mathrm{p}}$ in $|{\bar{\alpha}}%
_{1},{\bar{c}}\rangle$ from this remaining part and hence reconstruct
$|{\bar{\alpha}}_{1},{\bar{c}}\rangle$. The same procedure can then be applied
to $|{\bar{\alpha}}_{1},{\bar{c}}\rangle$ to obtain $|{\bar{c}}\rangle$.

At this stage it is useful to introduce `deformation' operators as follows.
Let us indicate the action of a Hamiltonian constraint-type deformation
labelled by $\alpha$ on a state $|c\rangle$ (with a single non-degenerate
vertex) by ${\hat{C}}_{\alpha}|c\rangle$. So in this notation we have, for
example, that
\begin{equation}
|{\bar{\alpha}}_{1},{\bar{\alpha}}_{2},{\bar{c}}\rangle=:{\hat{C}}%
_{{\bar{\alpha}}_{2}}{\hat{C}}_{{\bar{\alpha}}_{1}}|{\bar{c}}\rangle
\label{hatcalphabar}%
\end{equation}

Next, note that the final vertex of $|{\bar{\alpha}}_{1},{\bar{\alpha}}%
_{2},{\bar{c}}\rangle$ is connected to
%the vertex $v_{1-p}({\bar \alpha}_1, {\bar \alpha}_2, {\bar c})$ of
its 1-past by edges which end on trivalent kinks. It is immediate to see that
the edges from the final vertex of any state in $B_{[D^{2}\langle
n,{\vec{\alpha}},c_{0}\rangle]}$ end in bivalent kinks. Hence, it must be the
case that $|{\bar{\alpha}}_{1},{\bar{\alpha}}_{2},{\bar{c}}\rangle\in
B_{[H[H\langle n,{\vec{\alpha}},c_{0}\rangle]]}$. In the `deformation
operator' notation we have this may be written as
\begin{equation}
|{\bar{\alpha}}_{1},{\bar{\alpha}}_{2},{\bar{c}}\rangle={\hat{U}}_{\phi_{2}%
}{\hat{C}}_{{\alpha}_{2}^{\prime}}{\hat{U}}_{\phi_{1}}{\hat{C}}_{{\alpha}%
_{1}^{\prime}}|c\rangle
\end{equation}
for some $\langle c|~\in B_{\langle n,{\vec{\alpha}},c_{0}\rangle}$,
appropriate deformation labels ${\alpha}_{1}^{\prime},{\alpha}_{2}^{\prime}$
and diffeomorphisms $\phi_{1},\phi_{2}$ with ${\hat{U}}_{\phi_{i}},i=1,2$
being the unitary operators which implement these diffeomorphisms.

Since the definition of the 1-past as well as the process of `unflipping
charges' are diffeomorphism invariant, it is straightforward to see that
follows that the above equation implies that
\begin{equation}
|{\bar{c}}\rangle={\hat{U}}_{\phi_{2}}{\hat{U}}_{\phi_{1}}|c\rangle
\end{equation}
From the Lemma at the end of Section 6.2, it follows that $\langle{\bar{c}%
}|~\in B_{\langle n,{\vec{\alpha}},c_{0}\rangle}$. Hence \emph{all} its double
Hamiltonian constraint-type deformations and \emph{all} its double singular
diffeomorphism-type deformations are in $B_{\mathrm{VSA}}$. This immediately
implies that the assumptions of Section 4, 5 are satisfied in this case.

\bigskip

\textbf{Case (b)}:
%Recall, from equation (\ref{defdsquarebeta}) that,
%similar to the ${\hat C}(\alpha )$ operators which we introduced above,
In terms of the double singular diffeomorphism operators of Equation
(\ref{defdsquarebeta}) we have that
\begin{equation}
|{\bar{\beta}},{\bar{c}}\rangle=\hat{D}^{2}({\bar{\beta}})|{\bar{c}}\rangle.
\end{equation}
%Note that since $\delta , \delta^{\prime}$ can be decreased by semianalytic diffeomorphisms,
%it is immediate that the above equation holds for all (arbitrarily small) $\delta, \delta^{\prime}$.
Since $|{\bar{\beta}},{\bar{c}}\rangle$ is in $B_{\mathrm{VSA}}$, it has only
one non-degenerate vertex of valence $M$ which we denote by $v^{\prime\prime
}({\bar{c}})$, and this vertex is GR. Therefore ${\bar{c}}$ also has a single
non-degenerate $M$-valent vertex, which we denote by $v({\bar{c}})$ and, from
Section 4.4.2, this vertex must also be GR. In what follows we denote the
graphs underlying $|{\bar{\beta}},{\bar{c}}\rangle$, $|{\bar{c}}\rangle$ by
$\gamma({\bar{\beta}},{\bar{c}}),\gamma({\bar{c}})$.

The last part of Section 4.4.2 implies that the graph structure of
$\gamma({\bar{\beta}},{\bar{c}})$ in the vicinity of $v^{\prime\prime}%
({\bar{c}})$ is as follows. Each of the $M$ semianalytic $C^{k}$ edges
emanating from $v^{\prime\prime}({\bar{c}})$ ends in a bivalent $C^{r}$-kink
vertex where $r=0$ or $1$. The remaining semianalytic $C^{k}$ edge from each
such kink when followed `into the past' also ends in a bivalent $C^{r}$-kink
vertex with $r=0$ or $1$. The remaining semianalytic $C^{k}$ edge at
\emph{this} kink is part of the graph $\gamma({\bar{c}})$ and each of these
remaining edges when followed to `the past' connect to the rest of
$\gamma({\bar{c}})$. We denote the part of $\gamma({\bar{c}})$ which connects
to the past endpoints of these edges by $\gamma_{\mathrm{rest}}({\bar{c}})$.

\begin{figure}[t]
\centering
\includegraphics[width=10.5cm]{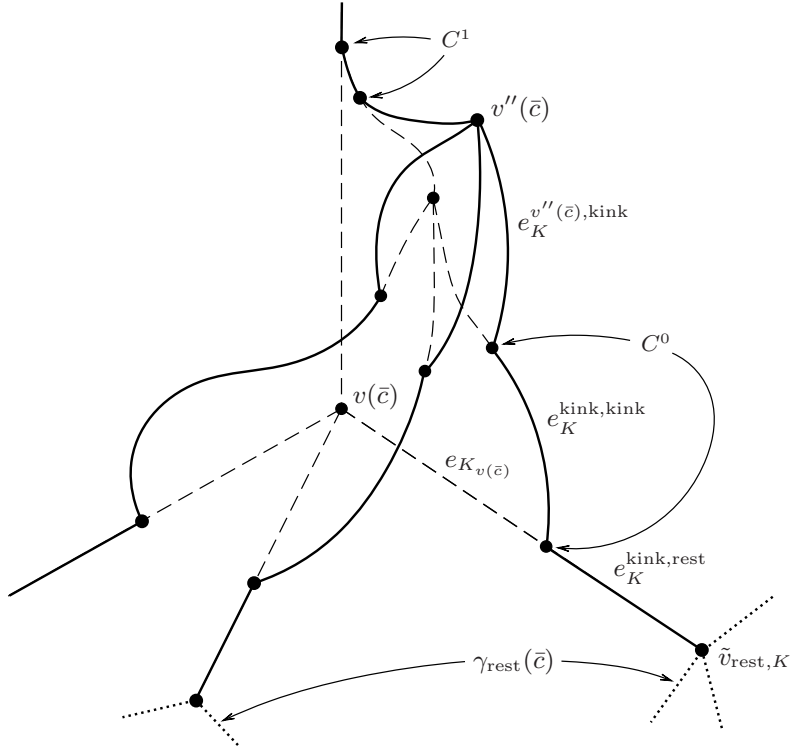} \caption{The result of a
double singular diffeomorphism action on $\bar{c}$ labeled corresponding to
definitions used in this section.}%
\label{Dsquared}%
\end{figure}

To summarize: We have that (see Figure (\ref{Dsquared}))

\begin{equation}
\gamma({\bar{\beta}},{\bar{c}})=\gamma_{\mathrm{rest}}({\bar{c}})\cup
\gamma_{\mathrm{rest}}^{D^{2}({\bar{\beta}})}({\bar{c}})
\end{equation}
where
\begin{equation}
\gamma_{\mathrm{rest}}^{D^{2}({\bar{\beta}})}({\bar{c}})=\cup_{K}%
e_{K}^{v^{\prime\prime}(\bar{c}),\mathrm{kink}}\circ e_{K}^{\mathrm{kink}%
,\mathrm{kink}}\circ e_{K}^{\mathrm{kink},\mathrm{rest}}%
\label{kinkrest}
\end{equation}
where $e_{K}^{v^{\prime\prime}(\bar{c}),\mathrm{kink}}$ connects
$v^{\prime\prime}({\bar{c}})$ to the first $C^{r}$ ($r=0$ or 1) kink to its
past, $e_{K}^{\mathrm{kink},\mathrm{kink}}$ connects this kink to the second
one and $e_{K}^{\mathrm{kink},\mathrm{rest}}\in\gamma({\bar{c}})$ connects
this second kink to $\gamma_{\mathrm{rest}}({\bar{c}})$.

%%%%%%%%REMOVED THIS SEP 28%%%%%%%%%%%%%%%%%55
%It also follows from the last part of Section 4.4.2 that had we analytically
%extended each of the edges $e_{K}^{\mathrm{kink},\mathrm{rest}}$ to the
%future, they would have first intersected with each other in $v({\bar{c}})$
%without intersecting anywhere with $\gamma({\bar{c}})$ except at their past
%endpoints. More precisely, if we denote by $e_{K_{v({\bar{c}})}}$ the edge
%which is obtained by analytically extending $e_{K}^{\mathrm{kink}%
%,\mathrm{rest}}$ to the point $v({\bar{c}})$, then the last part of Section
%4.4.2 implies that
%\begin{equation}
%\gamma({\bar{c}})=\gamma_{\mathrm{rest}}({\bar{c}})\cup\gamma_{\mathrm{rest}%
%}^{f_0}({\bar{c}}) \label{6.3.3}%
%\end{equation}
%where
%\begin{equation}
%\gamma_{\mathrm{rest}}^{f_0}({\bar{c}}):=\cup_{K=1}^{M}e_{K_{v({\bar{c}})}},
%\label{6.3.4}%
%\end{equation}
%and
%\begin{equation}
%\gamma_{\mathrm{rest}}({\bar{c}})\cap e_{K_{v({\bar{c}})}}={\tilde{v}%
%}_{\mathrm{rest},K} \label{6.3.5}%
%\end{equation}
%with ${\tilde{v}}_{\mathrm{rest},K}$ being the past end point of
%$e_{K_{v({\bar{c}})}}$ (which is the same as the past end point of
%$e_{K}^{\mathrm{kink},\mathrm{rest}}$).

Next, note that by virtue of the connection of its non-degenerate vertex two
successive bivalent kinks, it must be the case that $|{\bar{\beta}},{\bar{c}%
}\rangle\in B_{[D^{2}\langle n,{\vec{\alpha}},c_{0}\rangle]}$ so that
\begin{equation}
|{\bar{\beta}},{\bar{c}}\rangle={\hat{U}}(\phi)\hat{D}^{2}(\beta
)|c\rangle=:{\hat{U}}(\phi)|\beta,c\rangle\label{6.3.6}%
\end{equation}
for some appropriate diffeomorphism $\phi$, deformation label $\beta$ and
state $\langle c|~\in B_{\langle n,{\vec{\alpha}},c_{0}\rangle}$.

Next, note that it is possible to reconstruct the 1-past of $|c\rangle$ from
$|{\beta},{c}\rangle$ by following exactly the same procedure which resulted
in obtaining $\gamma_{\mathrm{rest}}({\bar{c}})$ from $\gamma({\bar{\beta}%
},{\bar{c}})$. Thus any edge emanating from the final (non-degenerate, GR,
$M$-valent) vertex of $|{\beta},{c}\rangle$ followed \textquotedblleft back in
time\textquotedblright\ connects to a bivalent $C^{1}$- or $C^{0}$-kink which,
in turn, connects to another bivalent $C^{1}$- or $C^{0}$-kink, which is then
connected to $\gamma_{\mathrm{1}\text{\textrm{-}}\mathrm{p}}(c)$ by an edge
which lies in $\gamma(c)$. Removing the $M$ sets of such triplets of
successive edges which connect the final vertex of $|{\beta},{c}\rangle$ to
$\gamma_{\mathrm{1}\text{\textrm{-}}\mathrm{p}}(c)$ yields $\gamma
_{\mathrm{1}\text{\textrm{-}}\mathrm{p}}(c)$. Since this procedure (of
removing the triplets of successive $C^{k}$ semianalytic edges which emanate
from the final non-degenerate vertex) is diffeomorphism-invariant, the same
procedure applied to ${\hat{U}}(\phi)D^{2}(\beta)|c\rangle$ yields the 1-past
of ${\hat{U}}(\phi)|c\rangle$. But, using Equation (\ref{6.3.6}), this very
same procedure resulted in the graph $\gamma_{\mathrm{rest}}({\bar{c}})$.
Hence we have that
\begin{equation}
\gamma_{\mathrm{rest}}({\bar{c}})=\gamma_{\mathrm{1}\text{\textrm{-}%
}\mathrm{p}}(c_{\phi}).
\label{rest=1p}
\end{equation}
where $|c_{\phi}\rangle:={\hat{U}}(\phi)|c\rangle$.
Moreover, from equations (\ref{6.3.6}) and (\ref{kinkrest})
and the nature of double singular diffeomorphisms, it follows that the 
edges $e_{K}^{\mathrm{kink},\mathrm{rest}}, K=1,..,M$ of equation (\ref{kinkrest}) are
a part of $\gamma ({\bar c})$ as well as $\gamma (c_{\phi})$. This, together with 
(\ref{rest=1p}), (\ref{6.3.6}) and the last definition of Section 6.2, implies that 
${\bar c}$ is the causal completion of the 1- past of $c_{\phi}$.
%
%From the above equation in conjunction with
%%%@@ OCT 2,OCT 21 2012 removed these lines:%%
%Equations (\ref{6.3.3}%
%)-(\ref{6.3.6}) and
%the last definition of Section 6.2, it follows that
%${\bar{c}}$ is \emph{a causal completion of the 1-past of $c_{\phi}$}. 
Since
$\langle c_{\phi}|~\in B_{\langle n,{\vec{\alpha}},c_{0}\rangle}$ by virtue of
the Lemma of Section 6.2, this means that $\langle{\bar{c}}|$ is in
$B_{\langle n,{\vec{\alpha}},c_{0}\rangle}$. Hence, once again all double
Hamiltonian constraint as well singular diffeomorphism-type deformations of
$\langle{\bar{c}}|$ are in $B_{\mathrm{VSA}}$ in accord with the assumptions
of Section 4 and 5.

\bigskip

\noindent\textbf{Case (c):} Since $|{\bar{\alpha}},{\bar{c}}\rangle$ is
obtained by the action of a single Hamiltonian constraint, each of the $M$
($C^{k}$, semianalytic) edges emanating from its final vertex is connected to
a trivalent kink. This, together with $\langle{\bar{\alpha}},{\bar{c}}|~\in
B_{\mathrm{VSA}}$ implies that $\langle{\bar{\alpha}},{\bar{c}}|~\in
B_{[H[H\langle n,{\vec{\alpha}},c_{0}\rangle]]}$ which means that for some
$\langle c|~\in B_{[H\langle n,{\vec{\alpha}},c_{0}\rangle]}$, some
Hamiltonian constraint deformation $\alpha_{1}$ and some diffeomorphism $\phi$
we have that
\begin{equation}
|{\bar{\alpha}},{\bar{c}}\rangle={\hat{U}}(\phi)|\alpha_{1},c\rangle
\label{casec}%
\end{equation}
Using argumentation similar to that for Case (a), it follows that
$\gamma({\bar{c}})=\gamma_{\mathrm{1}\text{\textrm{-}}\mathrm{p}}(\bar{\alpha
},{\bar{c}})$, that ${\bar{c}}$ can be reconstructed by appropriately coloring
$\gamma({\bar{c}})$ through the procedure of retaining the colors of
$|{\bar{\alpha}},{\bar{c}}\rangle$ away from the vicinity of its final
degenerate vertex and coloring those edges which emanate from this vertex with
the colors of their continuations past the immediate kinks they connect to,
and that all this, together with the diffeomorphism invariance of the
reconstruction procedure and Equation (\ref{casec}), implies that
\begin{equation}
|{\bar{c}}\rangle={\hat{U}}(\phi)|c\rangle.
\end{equation}
Since $B_{[H\langle n,{\vec{\alpha}},c_{0}\rangle]}$ is closed under the
action of semianalytic $C^{k}$ diffeomorphisms, it follows that $\langle
{\bar{c}}|~\in B_{[H\langle n,{\vec{\alpha}},c_{0}\rangle]}$ and, hence, that
$B_{[H[H\langle n,{\vec{\alpha}},c_{0}\rangle]]}$ contains all single
Hamiltonian constraint deformations of $\langle{\bar{c}}|$. It is then easy to
see that the considerations of Sections 4.1 and 4.6 imply that the continuum
limit of the `matrix element' of a single finite triangulation Hamiltonian
constraint operator is well defined and non-trivial i.e. $\lim_{\delta
\rightarrow0}\Psi_{B_{\mathrm{VSA}}}^{f}({\hat{C}}_{\delta}(N)|{\bar{c}%
}\rangle)$ is well defined and non-vanishing for suitable $f,N$ (by suitable
we mean that $N$ and the first derivative of $f$ do not vanish at the final
non-degenerate GR vertex of ${\bar{c}}$).

Note that Equation (\ref{casec}) implies that ${\bar{c}}$ has $n+1$ degenerate
GR vertices and that if either of Cases (a) or (b) hold, ${\bar{c}}$ must have
$n$ degenerate GR vertices which means that the matrix element for the single
Hamiltonian constraint action vanishes for Cases (a) and (b).

\bigskip

Cases (a)-(c) exhaust all possibilities of interest and imply that for any VSA
state and any charge network state:

\begin{enumerate}
\item[(i)] The continuum limits of the finite-triangulation operators
corresponding to the single Hamiltonian constraint, the commutator between two
Hamiltonian constraints (i.e. the LHS) and the operator corresponding to the
RHS, are all well defined.

\item[(ii)] For appropriate choices of lapses, vertex smooth functions and
charge networks, these limits are non-trivial.

\item[(iii)] These limits agree for the LHS and RHS operators.

\item[(iv)] Whenever they are non-trivial for the LHS and RHS they trivialize
for the single Hamiltonian constraint.
\end{enumerate}

It is straightforward to see that (i)-(iii) above imply that (i)-(iii) of
Section 3.2 hold. In particular point (iii) shows that, as stated towards the
end of Section 1, \emph{our considerations yield a non-trivial anomaly free
representation of the Poisson bracket between a pair Hamiltonian constraints.}

\section{Discussion}

\label{discussion}In any gauge theory, anomalies in the algebra of quantum
constraints typically point to a reduction of the number of true degrees of
freedom in the quantum theory. The quantization is then unphysical and,
depending on the severity of the anomalies, inconsistent. Hence, typically,
the viability of a quantum gauge theory is dependent on its support of an
anomaly-free representation of the classical constraint algebra. If the gauge
arises from general covariance, the constraint algebra has an additional role
to play \cite{hkt}: It encodes spacetime covariance in the Hamiltonian
formulation. We elaborate on this additional role below.

Any Hamiltonian formulation splits spacetime into space and time. As a result,
spacetime symmetries which are manifest in the Lagrangian description are not
explicit in the Hamiltonian formulation. For theories in flat spacetime, the
availability of preferred inertial times allows the straightforward recovery
of spacetime fields from spatial ones. However, in theories \emph{of}
spacetime, such as general relativity (or even in generally-covariant
reformulations of field theories on a fixed spacetime, such as PFT), the
absence of a preferred time, with respect to which the Hamiltonian theory is
to be defined, makes this loss of manifest spacetime covariance more acute.
One may then ask the following question: Which structure in the Hamiltonian
description of a generally covariant theory encodes spacetime covariance? The
answer to this question is provided by the seminal work of Hojman,
Kucha{\v{r},} and Teitelboim (HKT) \cite{hkt}. In the Hamiltonian description
of a generally-covariant theory of spacetime, initial data is prescribed on a
spatial slice embedded in spacetime, the spacetime itself emerging out of the
dynamics of the theory. HKT note that this dynamics pushes the spatial slice
`forward' in spacetime to the next one. In order that the spatial slices so
generated, stack up in a suitably consistent manner so as to yield a
spacetime, HKT show that the Poisson bracket algebra of the generators of
dynamics must be isomorphic to the commutator algebra of \emph{deformations}
of the spatial slice within the (emergent) spacetime. These deformations may
be separated into those which are tangential and those which are normal to the
slice. Their algebra has the characteristic structure that the commutator
between two tangential deformations is a tangential one, that between a
tangent and normal deformation is normal and, most non-trivially, the
commutator of two normal deformations is a tangential deformation which
depends on the spatial metric on the slice. This is, of course, exactly the
structure of the constraint algebra generated by the diffeomorphism and
Hamiltonian constraints of general relativity.\footnote{Recall that in any
generally-covariant theory, dynamics is generated by the constraints.} In
particular, the Hamiltonian constraint generates normal deformations and the
Poisson bracket between a pair of Hamiltonian constraints is proportional to a
diffeomorphism constraint, the proportionality involving a spatial
metric-dependent structure function. The generality and robustness of the
arguments of HKT lead one to believe that in the quantum theory, any notion of
spacetime covariance is predicated on the commutator algebra of the quantum
constraints exactly mirroring the classical Poisson bracket algebra, thus
providing a deep physical reason for the requirement of anomaly freedom.

In this work we studied a generally-covariant model with the same constraint
algebra as gravity. We concentrated on the most non-trivial aspect of this
algebra, namely the Poisson bracket between two Hamiltonian constraints, and
attempted to define the Hamiltonian constraint operator in an LQG-like
quantization in such a way that this Poisson bracket was represented in an
anomaly-free manner. Note that at a mathematical level, it would be enough to
provide a quantization of the RHS such that it agrees with the LHS. However,
the simple geometrical picture of spacetime deformations provided by HKT,
suggests that, in addition, \emph{the RHS operator should generate a
deformation akin to a spatial diffeomorphism}. The presence of `quantum
geometry'-dependent operator correspondents of the structure functions on the
RHS, together with the fact that the quantum geometry is excited along sets of
zero measure, unlike the classical ones, suggests that the deformation should
be some sort of `singular, quantum' version of a smooth diffeomorphism rather
than a smooth diffeomorphism. As seen in Sections 4 and 5, the choices we have
made in the construction of the Hamiltonian constraint and the RHS incorporate
this suggestion.

The physical viability of these choices can only be determined once a complete
quantization of the system is available. Specifically the work here needs to
be completed so as to provide:

\begin{enumerate}
\item[(i)] A large enough (by which we mean large enough to proceed to a
non-trivial implementation of (ii) below) space of solutions to the constraints.

\item[(ii)] A complete set of Dirac observables which preserve the space in
(i) and an inner product on (i) which implements the adjointness properties of
the Dirac observables.
\end{enumerate}

First consider issue (i). The VSA states of Section 6 provide off-shell
closure of the commutator between a pair of Hamiltonian constraints. Since
$B_{\mathrm{VSA}}$ contains entire diffeomorphism classes, it is
straightforward to check \cite{habitat1,mvaldiffeo} that the commutator
between two diffeomorphism constraints closes without anomalies as well. It is
also straightforward to check that the continuum limit actions of the
Hamiltonian and diffeomorphism constraints on a VSA state yield derivatives of
its vertex-smooth function so that off-shell VSA states obtained from a
specific choice of $B_{\mathrm{VSA}}$ can be `moved' on shell by setting the
vertex smooth functions to be a constant. Since we have infinitely-many
inequivalent choices of the parameters $c_{0},{\vec{\alpha}},n$ which go into
the construction of $B_{\mathrm{VSA}}$, this procedure yields a large class of
solutions to the constraints.\footnote{As mentioned in Section 6, while these
states are built from single-vertex primordial states, we expect our
considerations to easily generalize to a very large family of multi-vertex
primordial states.} These solutions may, of course, prove to be unphysical
once we attempt the incorporation of issue (ii). However, it seems plausible
that the chances of their physical relevance would be enhanced if it could be
shown that their off-shell deformations support the closure of the commutator
between the Hamiltonian and the diffeomorphism constraint, this being the only
remaining part of the constraint algebra. Clearly, showing this is equivalent
to the condition that the Hamiltonian constraint is diffeomorphism covariant;
i.e., that $\hat{U}(\phi)\hat{H}[N]\hat{U}^{\dag}(\phi)=\hat{H}[\phi_{\ast}N]$
for all (semianalytic $C^{k}$) diffeomorphisms $\phi$ and all (density
$-\frac{1}{3}$) lapses $N$.

As mentioned in Section 1, we have ignored precisely this issue of
diffeomorphism covariance in our constructions. While the issue will be
studied in a future publication \cite{meinprep}, we briefly comment on the
problems inherent in generalizing our constructions here to incorporate
diffeomorphism covariance. The primary non-covariant structure we use is the
regulating coordinate patches. These patches are chosen once and for all in
some arbitrary manner. It turns out (as is eminently plausible) that
diffeomorphism covariance requires that coordinate patches associated with
diffeomorphic vertex structures (by which we mean the graph structure of a
charge network in the vicinity of its (GR, non-degenerate
\footnote{See Footnote \ref{fnotenondeg}, section 4.6.}
) vertex) should be
related by diffeomorphisms. The ensuing problems are two fold:

\begin{enumerate}
\item[(a)] There are infinitely-many diffeomorphisms which map one vertex
structure to another.

\item[(b)] The vertex structure at a `daughter' vertex created by the
Hamiltonian constraint ${\hat{C}}_{\delta_{1}}[N]$ at triangulation
$T_{\delta_{1}}$ is mapped to the corresponding structure created by ${\hat
{C}}_{\delta_{2}}[N]$ at $T_{\delta_{2}}$, with $\delta_{2}<\delta_{1}$, by a
diffeomorphism which `scrunches' the edges at the vertex together along the
axis of the cone as described in Section 3 and Appendix \ref{scrunchAppendix}.
This fact, together with the necessity of relating the corresponding
coordinate patches through diffeomorphisms, implies that in the calculation of
commutators the coordinate patch $\{x^{\prime a^{\prime}}\}_{\delta}$ (see the
second paragraph of Section 4.5) goes bad as $\delta\rightarrow0$. This in
turn implies that the continuum limit of the commutator between two
Hamiltonian constraints blows up due to the $x^{\prime}$-dependence of the
calculation (for example, the Jacobian in Equation (\ref{cmcnbrkt2}) blows up).
\end{enumerate}

A solution to both these problems can be found \cite{meinprep}. It turns out
that progress on problem (a) is related to the GR property of the
non-degenerate vertices of the VSA states and that a possible way out of
problem (b) is to enlarge the dependence of the vertex smooth functions to
certain additional vertices of the graph and require some additional
regularity properties of the ensuing functional dependence \cite{meinprep}.
This concludes our comments on the problem of diffeomorphism covariance and
its relation to issue (i).

Another key open problem with regard to issue (i) has to do with the very
definition of the continuum limit we use (see Section 3.2).
%sequences (see section 3.2). Recall that in addition to the existence of the continuum limit of a single Hamiltonian constraint operator, we have only shown that the commutator of a pair of finite triangulation Hamiltonian
%constraints admits a continuum limit in the VSA topology; we have {\em not} shown that the product of
%pair of Hamiltonian constraints admits a limit and, indeed, it is straightforward to check that this limit
%An even more pressing problem (which encompasses the `product' problem)
This definition, while in the spirit of Thiemann's considerations involving
the URS topology, is far from conventional \cite{nic}. Notwithstanding the
fact that it \emph{is} extremely non-trivial to obtain an anomaly-free
representation in the context of this definition of the continuum limit, we
believe that a proper resolution of the problem of an anomaly-free off-shell
closure of the constraint algebra requires a representation of the latter on
some suitable vector space, which, as mentioned towards the end of Section
3.2, we call a `habitat'. In the case of the Husain-Kucha{\v{r}} model
\cite{habitat1,mvaldiffeo} as well as PFT \cite{ppftham}, the habitat is
spanned by vertex-smooth algebraic states of the type considered here. It is
our hope that these states can be suitably generalized (say, to accommodate
not only a dependence of the vertex smooth functions on vertices but, perhaps,
on other properties of the state at the vertex such as its edge tangents and
their charges) so that our calculations are supported on a genuine habitat. An
important aspect of such a generalization would be to ensure that not only the
commutator, but also the product of two Hamiltonian constraints has a
well-defined action.\footnote{Recall that we have only shown that the
commutator of a pair of finite-triangulation Hamiltonian constraints admits a
continuum limit in the VSA topology; the reader may readily verify that the
product of a pair of Hamiltonian constraints does not admit a continuum limit
in this topology.} Preliminary calculations suggest that ensuring this (not
only in the context of a habitat but also in the VSA topology considerations
of this work) requires a slight modification in the definition of the
Hamiltonian constraint operator at finite triangulation from the `$\delta-1$
form of Equation (\ref{heuristic1}) to a `$2\delta-\delta$' form.
%(we shall
%discuss this in \cite{meinprep}).

Next, we turn to issue (ii). The first step towards the construction of Dirac
observables is a detailed analysis of the equations of motion of the classical
theory.\footnote{While a few Dirac observables are available through Smolin's
work \cite{leeg=0}, infinitely-many are needed since the system has
infinitely-many true degrees of freedom.} Such an analysis has been initiated
by Barbero and Villase\~{n}or \cite{fereduardo} and we hope that their work
will stimulate further progress on issue (ii). As a side remark, we note that
a detailed understanding of the classical dynamics of the model would also
stimulate progress on Smolin's original idea \cite{leeg=0} of approaching
Euclidean gravity via an expansion in powers of Newton's constant.

Besides the open issues (i) and (ii), our work can also be improved upon in
the following aspects. We have required that the `singular' diffeomorphism
type deformations of Sections 4 and 5 preserve the GR (or non-GR) nature of
the non-degenerate vertex. This is a rather coarse requirement and it would be
good to further restrict the deformation so that it preserves a larger subset
of diffeomorphism-invariant properties. This would also lead to a tighter and
better-motivated prescription for connecting the original graph to the
displaced vertex.
%For example the knotting and linking properties of the new
%edges with each other has been left completely open in this work.
A tighter prescription would presumably lead to a smaller bra set
$B_{\mathrm{VSA}}$. One may even envisage that the current $B_{\mathrm{VSA}}$
can be split into `minimal' subsets.

We now turn to a discussion of various novel features of our constructions and
considerations. Our exposition will consist of a series of scattered remarks.
First, independent of any ramifications for quantum theory, it would be good
to understand if there is a deeper reason behind the existence of the
remarkable classical identity of Section 5 and Appendix \ref{su2rhsid}. Next,
as discussed in Section 6, we note a beautiful feature of repeated actions of
our Hamiltonian constraint on an `initial state'; namely that the resulting
`final' state encodes its own `chronological history' dating back to the
initial state. Finally, we note that while there does seem to be a significant
freedom in the details of the choices we have made, the class of choices
suggested by our considerations of Section 4.1 are qualitatively different
from those considered in the standard treatments of the Hamiltonian constraint
\cite{qsd,habitat1,aajurekreview}. Our considerations here rest on a number of
new ideas suggested by earlier studies of toy models \cite{ppftham,mvaldiffeo}%
. A few of them are: The consideration of higher density weight constraints, a
continuum limit defined by VSA states, deformations of charge networks which
depend on their charge labels, and a Hamiltonian constraint action which is
such that a second such action acts on deformations produced by the first.

In summary, while there are many open problems and obstructions to be
overcome, we believe that there is room for cautious optimism that the
considerations of this work and of the recent work \cite{hendersonal,alokrecent} present
the first necessary steps to define the correct quantum dynamics of this
model, and, perhaps, offer hope that the lessons learnt from this and
subsequent studies of the model will provide inputs for the much harder
context of gravity.

\bigskip

\section*{Acknowledgements}

CT is deeply indebted to Alok Laddha for bringing this model to his attention,
for numerous extremely useful discussions at every stage of this work, and for
general moral support and mentorship. CT is grateful to Miguel
Campiglia-Curcho, Martin Bojowald, Przemys\l aw Mal\l kiewicz, and Keith
Thornton for useful discussions. CT is supported by NSF grant PHY-0748336 and
a Mebus Fellowship, and acknowledges the generous hospitality and friendly
working environment provided by the Raman Research Institute. MV thanks Alok
Laddha for numerous discussions, for going through several of the arguments in
a draft version of this work and for his extreme generosity with regard to his
time and mentorship. MV thanks Abhay Ashtekar, Fernando Barbero, and Eduardo
Villase\~{n}or for their constant encouragement and FB and EV for going
through a preliminary version of this manuscript. MV thanks Jurek Lewandowski,
Christian Fleischhack and, especially, Hanno Sahlmann for help with the
semianalytic category.

%\newpage

\section*{Appendices}

\appendix

\section{A $q^{-1/3}$ Operator}

\label{invq} In this Appendix we derive some Thiemann-like classical
identities for negative powers of the metric determinant that we then quantize
on $\mathcal{H}_{\mathrm{kin}}$. These identities involve a volume operator,
which we take to be the Ashtekar-Lewandowski volume operator $\hat{V}$, with
SU(2) replaced by U(1)$^{3}.$ The construction of $\hat{V}$ in the case of
U(1)$^{3}$ proceeds just as for SU(2), so we direct the reader to
\cite{aajurekvol}\ for details. Here we merely cite the result in the
U(1)$^{3}$ case. Given a region $R\subset\Sigma,$ the volume operator $\hat
{V}(R)$ associated to that region, acting on the charge network state
$|c\rangle$ is given by%
\begin{equation}
\hat{V}(R)|c\rangle=\varepsilon_{(\mu)}%
%TCIMACRO{\tsum \nolimits_{v\in c\cap R}}%
%BeginExpansion
{\textstyle\sum\nolimits_{v\in c\cap R}}
%EndExpansion
\sqrt{\left\vert \hat{q}_{\mathrm{AL}}(v)\right\vert }|c\rangle.
\end{equation}
Here, $\varepsilon_{(\mu)}$ is a constant which depends on the choice of an
integration measure $\mu$ on a finite-dimensional `background
structure-averaging' space (if one subscribes to a consistency check in the
sense of \cite{volconsistency}, then this factor can be fixed to be equal to
one); the sum extends over all vertices $v$ of $c$ contained in the region
$R$. $\hat{q}_{\mathrm{AL}}\left(  v\right)  $ is diagonal in the charge
network basis and acts at vertices $v$ of $|c\rangle$ by
\begin{equation}
\hat{q}_{\mathrm{AL}}(v)|c\rangle=\tfrac{1}{48}(\hbar\gamma\kappa)^{3}%
%TCIMACRO{\tsum \nolimits_{IJK}}%
%BeginExpansion
{\textstyle\sum\nolimits_{IJK}}
%EndExpansion
\epsilon^{IJK}\epsilon_{ijk}q_{I}^{i}q_{J}^{j}q_{K}^{k}|c\rangle, \label{qal}%
\end{equation}
where each of the three sums (over $I,J,K$) extends over the valence of $v,$
with $I,J,K$ labeling (outgoing) edges $e_{I},e_{J},e_{K}$ emanating from $v.$
$\epsilon^{IJK}=0,+1,-1$ depending on whether the tangents of $e_{I}%
,e_{J},e_{K}$ are linearly dependent, define a right-handed frame (with
respect to the orientation of the underlying manifold), or define a
left-handed frame, respectively. As in the main text, $q_{I}^{i}$ is the
U(1)$_{i}$ charge on the edge $e_{I}.$ Before moving to inverse metric
operators, we note two properties of $\hat{V}$ that are shared with the SU(2) theory:

\begin{enumerate}
\item[(i)] Trivalent gauge-invariant vertices are annihilated by $\hat{V}.$
This follows immediately by using the gauge-invariance property $\sum_{I\neq
J}q_{I}^{i}=-q_{J}^{i}$ in (\ref{qal}).

\item[(ii)] `Planar' vertices (those for which the set of edge tangents spans
at most a plane) are annihilated by $\hat{V},$ since each orientation factor
$\epsilon^{IJK}$ in this case vanishes.
\end{enumerate}

%%@@INSERTED SEP 6,2012.
We now turn to the construction of negative powers of the spatial metric
determinant at any point in $\Sigma$. Let $U\subset\Sigma$ be an open set with
coordinate system $\{x\}$. Let any $p\in U$ have coordinates ${\vec{x}%
}(p)=\{x^{1},x^{2},x^{3}\}$. Since the analysis below is expressed in the
$\{x\}$ coordinates, we use the notation $p\equiv\vec{x}(p)\equiv x$. The
first step is to express negative powers of the classical volume in terms of
Poisson bracket identities involving quantities which have unambiguous quantum
analogs. Classically, the volume $V\left(  R\right)  $ of a region
$R\subset\Sigma$ is given by
\begin{equation}
V(R)=\int_{R}\sqrt{q}\equiv\int_{R}\mathrm{d}^{3}x\sqrt{|\det E|}:=\int
_{R}\mathrm{d}^{3}x\sqrt{\left\vert \tfrac{1}{3!}\eta_{abc}\epsilon^{ijk}%
E_{i}^{a}E_{j}^{b}E_{k}^{c}\right\vert }%
\end{equation}
Let $B_{\epsilon}(x)\subset\Sigma$ be a coordinate ball of radius $\epsilon$,
centered at $x$. Its volume $V_{\epsilon}(x)$ is then:
\begin{equation}
V_{\epsilon}(x):=\int_{B_{\epsilon}(x)}\mathrm{d}^{3}y~\sqrt{q(y)}.
\end{equation}
It follows that for smooth $q\left(  y\right)  $ and some $\alpha\in%
%TCIMACRO{\U{211d} }%
%BeginExpansion
\mathbb{R}
%EndExpansion
,$%
\begin{equation}
\frac{V_{\epsilon}(x)^{2\alpha}}{(\frac{4}{3}\pi\epsilon^{3})^{2\alpha}%
}=q(x)^{\alpha}+O(\epsilon). \label{qapprox}%
\end{equation}
Now it is straightforward to verify that
\begin{equation}
\eta^{abc}\epsilon_{ijk}\{A_{a}^{i}(x),V_{\epsilon}(x)^{\alpha}\}\{A_{b}%
^{j}(x),V_{\epsilon}(x)^{\alpha}\}\{A_{c}^{k}(x),V_{\epsilon}(x)^{\alpha
}\}=\tfrac{3}{4}\sigma\alpha^{3}V_{\epsilon}(x)^{3(\alpha-1)}\sqrt{q(x)},
\end{equation}
where we have defined $\sigma:=\mathrm{sgn}(\det E),$ and neglected terms such
as $\frac{\delta\sigma}{\delta E_{i}^{a}}.$ Using (\ref{qapprox}) we may then
write%
\begin{align}
q(x)^{-p}  &  =\epsilon^{3(2p+1)}\frac{(\tfrac{4}{3}\pi)^{2p+1}\eta
^{abc}\epsilon_{ijk}}{\tfrac{3}{4}(\tfrac{2}{3}\left(  1-p\right)  )^{3}%
}\sigma\nonumber\\
&  \qquad\times\{A_{a}^{i}(x),V_{\epsilon}(x)^{\frac{2}{3}\left(  1-p\right)
}\}\{A_{b}^{j}(x),V_{\epsilon}(x)^{\frac{2}{3}\left(  1-p\right)  }%
\}\{A_{c}^{k}(x),V_{\epsilon}(x)^{\frac{2}{3}\left(  1-p\right)  }\}+O\left(
\epsilon^{3}\right)
\end{align}
where the first term is $O(1)$. With an eye on quantizing this expression as
an operator on $\mathcal{H}_{\mathrm{kin}}$, we replace $A_{a}^{i}(x)$ with
holonomy approximants as follows: Let $e_{I},I=1,2,3$ be a triplet of edges,
each of coordinate length $B_{I}\epsilon$, emanating from the point $x$ (here
$B_{I}$ are a triple of dimensionless $\epsilon$-independent numbers). Let
their unit tangents, normalized with respect to the coordinate metric be
${\hat{e}}_{I}^{a}$ and let $e_{I}$ be such that the triple of their edge
tangents at $x$ is linearly independent. It is easy to check the following
identity:
\begin{equation}
\eta^{abc}=\frac{\epsilon^{IJK}{\hat{e}}_{I}^{a}{\hat{e}}_{J}^{b}{\hat{e}}%
_{K}^{c}}{\lambda(\vec{e})} \label{etaabc}%
\end{equation}
where $\lambda({\vec{e}})$ is given by
\begin{equation}
\lambda({\vec{e}})=\tfrac{1}{6}\eta_{fgh}\epsilon^{LMN}{\hat{e}}_{L}^{f}%
{\hat{e}}_{M}^{g}{\hat{e}}_{N}^{h} \label{Deflambda}%
\end{equation}
Here $\epsilon^{IJK}$ is antisymmetric with respect to interchange of its
indices with $\epsilon^{123}=1$ and the argument $\vec{e}:=\{e_{1},e_{2}%
,e_{3}\}$ signifies the dependence of $\lambda$ on the triplet of edges. Using
equation (\ref{etaabc}) and approximating $A_{a}^{i}{\hat{e}}_{I}^{a}$ in
terms of the edge holonomies $h_{I}^{i}$ along $e_{I}$, we obtain:
\begin{align}
q(x)^{-p}  &  =\epsilon^{3(2p+1)}\frac{9\epsilon^{IJK}\epsilon_{ijk}(\tfrac
{4}{3}\pi)^{2p+1}}{2(1-p)^{3}\lambda({\vec{e}})}\sigma\frac{h_{I}^{(i)}%
}{-\mathrm{i}\kappa\gamma B_{I}\epsilon}\{(h_{I}^{i})^{-1},V_{\epsilon
}(x)^{\frac{2}{3}\left(  1-p\right)  }\}\nonumber\\
&  \qquad\times\frac{h_{J}^{(j)}}{-\mathrm{i}\kappa\gamma B_{J}\epsilon
}\{(h_{J}^{j})^{-1},V_{\epsilon}(x)^{\frac{2}{3}\left(  1-p\right)  }%
\}\frac{h_{K}^{(k)}}{-\mathrm{i}\kappa\gamma B_{K}\epsilon}\{(h_{K}^{k}%
)^{-1},V_{\epsilon}(x)^{\frac{2}{3}\left(  1-p\right)  }\}+O\left(
\epsilon\right)
\end{align}
%Here $h_{I}^{i}$ a coordinate length $B_{(a)}\epsilon$ holonomy
%($B_{(a)}$ is a dimensionless number (independent of $\epsilon$) that we
%postpone fixing for now) along a curve in the $a$-direction beginning at $x.$
Setting $p=\frac{1}{3}$ we arrive at%
\begin{align}
q(x)^{-1/3}  &  =\epsilon^{2}\frac{9(\frac{4}{3}\pi)^{5/3}}{2(-\mathrm{i}%
\frac{2}{3}\kappa\gamma)^{3}\lambda({\vec{e}})B_{I}B_{J}B_{K}}\sigma
\epsilon^{IJK}\epsilon_{ijk}\label{dingdong}\\
&  \qquad\times h_{I}^{(i)}\{(h_{I}^{i})^{-1},V^{4/9}\}h_{J}^{(j)}\{(h_{J}%
^{j})^{-1},V^{4/9}\}h_{K}^{(k)}\{(h_{K}^{k})^{-1},V^{4/9}\}+O(\epsilon
)\nonumber
\end{align}
Now we define an $\epsilon$-regularized operator on $\mathcal{H}%
_{\mathrm{kin}}$ by taking all quantities to their operator correspondents,
$\{\cdot,\cdot\}\rightarrow(\mathrm{i}\hbar)^{-1}[\cdot,\cdot]$, and dropping
the classical $O(\epsilon)$ contribution:%
\begin{equation}
\hat{q}^{\prime}(x)_{\epsilon}^{-1/3}:=\epsilon^{2}\frac{9(\frac{4}{3}%
\pi)^{5/3}}{2(\frac{2}{3}\hbar\kappa\gamma)^{3}\lambda(\vec{e})B_{IJK}%
}\epsilon^{IJK}\epsilon_{ijk}\hat{\sigma}h_{I}^{(i)}[(h_{I}^{i})^{-1},\hat
{V}^{4/9}]h_{J}^{(j)}[(h_{J}^{j})^{-1},\hat{V}^{4/9}]h_{K}^{(k)}[(h_{K}%
^{k})^{-1},\hat{V}^{4/9}] \label{invqop}%
\end{equation}
with $B_{IJK}:=B_{I}B_{J}B_{K}$ (the prime in $\hat{q}^{\prime}$ appears
because this operator is not the final one we will employ in the main body).
As it stands, this operator is tied to the coordinate system $\{x\}$, which
should come as no surprise, since the classical quantity is a scalar density
with density weight not equal to 1. In keeping with the general philosophy of
this work, in which operators on $\mathcal{H}_{\mathrm{kin}}$ are tailored to
the underlying charge networks that they act on, we will choose the holonomy
segments of $\hat{q}^{\prime-1/3}$ to partially overlap edges of charge
networks (when this is possible).

Let us first consider charge network vertices $v\in c$ whose edge tangents
span at most a plane (we deem these planar (or linear) vertices); this
includes interior points of edges. Since there are not three
linearly-independent directions defined by the edge tangents of $c$ at $v,$ we
should have to choose the extra segment(s) needed for $\hat{q}^{\prime
}(x)_{\epsilon}^{-1/3}$ by hand, but this choice is arbitrary, since for the
ordering shown in (\ref{invqop}), there will be some factor $[(h_{I}^{i}%
)^{-1},\hat{V}^{4/9}]$ acting on $|c\rangle,$ where $(h_{I}^{i})^{-1}$
overlaps an existing edge of $c,$ and since $\hat{V}^{4/9}$ acts trivially at
planar (and linear) vertices, $[(h_{I}^{i})^{-1},\hat{V}^{4/9}]$ annihilates
$|c\rangle$ (perhaps even more simply, since planar vertices have zero volume,
$\hat{\sigma}$ is the zero operator). We henceforth restrict the discussion to
charge network vertices with at least one linearly-independent triple of edge tangents.

We write equation (\ref{invqop}) as:
\begin{equation}
\hat{q}^{\prime}(v)_{\epsilon}^{-1/3}|c\rangle=B^{\prime}\frac{\epsilon^{2}%
}{(\hbar\kappa\gamma)^{3}}\sum_{I,J,K=1}^{3}\epsilon^{IJK}\epsilon_{ijk}%
\hat{\sigma}h_{(I)}^{(i)}[(h_{I}^{i})^{-1},\hat{V}^{4/9}]h_{(J)}^{(j)}%
[(h_{J}^{j})^{-1},\hat{V}^{4/9}]h_{(K)}^{(k)}[(h_{K}^{k})^{-1},\hat{V}%
^{4/9}]|c\rangle, \label{q13'}%
\end{equation}
where $B_{123}\lambda({\vec{e}})$ has absorbed some dimensionless constants
and become $B^{\prime}$. Note that $\lambda({\vec{e}})$ depends on the charge
network $c$ through its dependence on the edge triplet ${\vec{e}}$. It also
depends on the choice of regulating coordinate patch $\{x\}$ through its
dependence on the \emph{unit} edge tangents which are normalized with respect
to the coordinate metric defined by $\{x\}$.

Next, define $Q$ to be the dimensionless rescaled eigenvalue of $\hat
{q}_{\mathrm{AL}}(v)$
\begin{equation}
\hat{q}_{\mathrm{AL}}(v)|c\rangle=\tfrac{1}{48}(\hbar\gamma\kappa)^{3}%
%TCIMACRO{\tsum \nolimits_{IJK}}%
%BeginExpansion
{\textstyle\sum\nolimits_{IJK}}
%EndExpansion
\epsilon^{IJK}\epsilon_{ijk}q_{I}^{i}q_{J}^{j}q_{K}^{k}|c\rangle=:\tfrac
{1}{48}(\hbar\gamma\kappa)^{3}Q|c\rangle, \label{defQ}%
\end{equation}
so that
\begin{equation}
\hat{V}(v)|c\rangle=\varepsilon_{(\mu)}\sqrt{\left\vert \tfrac{1}{48}%
(\hbar\gamma\kappa)^{3}Q\right\vert }|c\rangle, \label{Vevalue}%
\end{equation}
and let $Q_{I}^{i}$ be the rescaled eigenvalue of $\hat{q}_{\mathrm{AL}}(v)$
when the regulating holonomy $h_{I}^{i}$ is\ first laid on the edge $e_{I}$
(and $Q_{I}^{-i}$ when $(h_{I}^{i})^{-1}$ is laid). Let $\sigma$ be the
eigenvalue of $\hat{\sigma}$ (which is also the sign operator of $\det E$).
Then (\ref{q13'}) acts on $|c\rangle$ by
\begin{align}
\hat{q}^{\prime}(v)_{\epsilon}^{-1/3}|c\rangle &  =B^{\prime}\frac
{\epsilon^{2}}{(\hbar\kappa\gamma)^{3}}\left(  \varepsilon_{(\mu)}%
^{4/9}(\tfrac{1}{48}(\hbar\gamma\kappa)^{3})^{2/9}\right)  ^{3}\nonumber\\
&  \qquad\qquad\times\epsilon^{IJK}\epsilon_{ijk}\sigma(|Q|^{2/9}-|Q_{I}%
^{-i}|^{2/9})(|Q|^{2/9}-|Q_{J}^{-j}|^{2/9})(|Q|^{2/9}-|Q_{K}^{-k}%
|^{2/9})|c\rangle\nonumber\\
&  =B\frac{\epsilon^{2}}{\hbar\gamma\kappa}\epsilon^{IJK}\epsilon_{ijk}%
\sigma(|Q|^{2/9}-|Q_{I}^{-i}|^{2/9})(|Q|^{2/9}-|Q_{J}^{-j}|^{2/9}%
)(|Q|^{2/9}-|Q_{K}^{-k}|^{2/9})|c\rangle
\end{align}
where we have absorbed some numerical factors into $B^{\prime}$ to obtain
$B$.
%is a positive constant which depends only on the diffeomorphism class $[c]$ of $c$, or, equivalently,
%on the reference chargenet for $[c]$ (modulo the restrictions
%coming from equation (\ref{resref}).

We could stop here, but it turns out that this particular form of $\hat
{q}^{\prime-1/3}$ is not quite what we want, as it is not invariant under the
charge flips produced by the Hamiltonian constraint, a property that we
require in the main body of the paper. However, we can modify the preceding
construction slightly to obtain another $q^{-1/3}$ operator which is
insensitive to the charge flips. Consider the classical expression
(\ref{dingdong}). Instead of using inverse holonomies inside the Poisson
brackets, suppose we average over combinations in other representations;
specifically $q_{I}^{i}=\pm1$ for each $i,I$. Making this change and following
the remaining steps to arrive at the operator action, we find%
\begin{equation}
\hat{q}(v)_{\epsilon}^{-1/3}|c\rangle=-B\frac{\epsilon^{2}}{8\hbar\gamma
\kappa}\epsilon^{IJK}\epsilon_{ijk}\sigma\left(  O_{I}^{i}O_{J}^{j}O_{K}%
^{k}-3O_{I}^{-i}O_{J}^{j}O_{K}^{k}+3O_{I}^{-i}O_{J}^{-j}O_{K}^{k}-O_{I}%
^{-i}O_{J}^{-j}O_{K}^{-k}\right)  |c\rangle\label{gooberish}%
\end{equation}
where%
\begin{equation}
O_{I}^{\pm i}:=|Q|^{2/9}-|Q_{I}^{\pm i}|^{2/9}.
\end{equation}
The overall factor of $\frac{1}{8}$ comes from averaging over the eight
different combinations of $O_{I}^{\pm i}$, and the relative signs arise from
the classical Poisson bracket identity, depending on whether we choose to put
a fundamental representation holonomy, or its inverse, inside the bracket (an
odd number of minus superscripts yields a minus sign). We will see in the next
section that these eigenvalues are invariant under charge flips. If there is a
choice of edge triplets of $c$ at $v$ such that $\hat{q}(v)_{\epsilon}%
^{-1/3}|c\rangle\neq0$ , we term the vertex $v$ as \emph{non-degenerate}.
Henceforth, we restrict attention to charge networks with a single
non-degenerate vertex. For the purposes of this paper, this restriction
suffices because the continuum limit action of the quantum Hamiltonian
constraint and the quantum electric diffeomorphism vanish on all other charge
networks. which in turn stems from the fact that $B_{\mathrm{VSA}}$ has states
with (at most) only a single non-degenerate vertex. We leave a generalization of our
considerations to the multi-vertex case for future work.

Note that the inverse metric eigenvalue $\nu^{-\frac{2}{3}}$ in Section 4.2 is
defined through the equation
\begin{equation}
\hat{q}(v)_{\epsilon}^{-1/3}|c\rangle=\frac{\epsilon^{2}}{\hbar\gamma\kappa
}\nu^{-\frac{2}{3}}|c\rangle\label{DEFNU1}%
\end{equation}
We now show how to choose the triplet of edge holonomies in (\ref{invqop}) in
such way that this inverse metric eigenvalue is (a) diffeomorphism invariant,
and (b) the same for the (single non-degenerate vertex) charge networks
$c(i,v_{I_{v},\epsilon}^{\prime}),c(v_{I_{v},\epsilon}^{\prime})$ of Sections
4 and 5.

In each diffeomorphism class of charge networks $[{\bar{c}}]$ we pick a
reference charge network $c_{0}$ and a set of diffeomorphisms $\mathcal{D}%
_{[{\bar{c}}]}$ such that for any element $c\neq c_{0},c\in\lbrack{\bar{c}}]$
there is a unique diffeomorphism in $\mathcal{D}_{[{\bar{c}}]}$ which maps
$c_{0}$ to $c$. Our choice of reference charge networks is further restricted
as follows. Let $[{\bar{c}}_{i}],i=1,2,3,$ $[{\hat{\bar{c}}}],$ be such that
there exist $c_{i}\in\lbrack{\bar{c}}_{i}],{\hat{c}}\in\lbrack{\hat{\bar{c}}%
}],$ and a charge network $c$ with non-degenerate vertex $v$ such that for
some $I_{v},\epsilon$ we have that
\begin{equation}
c_{i}=c(i,v_{I_{v},\epsilon}^{\prime}),\;\;\;\;\;\;{\hat{c}}=c(v_{I_{v}%
,\epsilon}^{\prime}), \label{res}%
\end{equation}
where $c(i,v_{I_{v},\epsilon}^{\prime}),c(v_{I_{v},\epsilon}^{\prime})$ are
the deformations of $c$ as defined in Sections 4 and 5. If equation
(\ref{res}) holds, we require that the reference charge networks $c_{i0}%
,{\hat{c}}_{0}$ for $[{\bar{c}}_{i}],[{\hat{\bar{c}}}]$ be chosen such that
there exists a charge network $\mathbf{c}$ with a single non-degenerate vertex
$v_{0}$ and some parameter value $\delta$ for which it holds that:
\begin{equation}
c_{i0}=\mathbf{c}(i,v_{I_{v_{0}},\delta}^{\prime}),\;\;\;\;\;\;{\hat{c}}%
_{0}=\mathbf{c}(v_{I_{v_{0}},\delta}^{\prime}). \label{resref}%
\end{equation}

Next, we choose a triplet of edges for each reference charge network and
define the triplet of edges for any $c\in\lbrack c_{0}]$ as the image of these
edges by that diffeomorphism in $\mathcal{D}_{[{c}_{0}]}$ which maps $c_{0}$
to $c$. We restrict our choice of edge triplets as follows. Consider the
diffeomorphism classes $[{\bar{c}}_{i}],i=1,2,3$, $[{\hat{\bar{c}}}]$ and the
charge networks $c_{i0},i=1,2,3$, ${\hat{c}}_{0}$, $\mathbf{c}$, subject to
equations (\ref{res}), (\ref{resref}). The structure of the deformations
sketched in Sections 4,5 (and further elaborated upon in Appendix C) permits
the identification of the $J_{v_{0}}^{\text{th}}$ edge emanating from
$v_{I_{v},\delta}^{\prime}$ in $\mathbf{c}(1,v_{I_{v},\delta}^{\prime})$ with
the $J_{v_{0}}^{\text{th}}$ edges emanating from $v_{I_{v_{0}},\delta}%
^{\prime}$ in $\mathbf{c}(2,v_{I_{v_{0}},\delta}^{\prime}),\mathbf{c}%
(3,v_{I_{v_{0}},\delta}^{\prime})$ and $\mathbf{c}(v_{I_{v_{0}},\delta
}^{\prime})$; this edge is uniquely identified, in the notation of Sections
4,5 as the deformed counterpart of the $J_{v_{0}}^{\text{th}}$ edge emanating
from the vertex $v_{0}$ of $\mathbf{c}$. We choose a triplet of edge labels
$J_{v_{0}}^{K},K=1,2,3$ and choose the triplet of edge holonomies for $c_{10}$
to be along the $J_{v_{0}}^{K}{}^{\text{th}}$ edges emanating from
$v_{I_{v_{0}},\delta}^{\prime}$. Our choice for the triplet of edge holonomies
for the reference charge networks $c_{20},c_{30},{\hat{c}}_{0}$ is then
restricted to also be along the $J_{v_{0}}^{K}{}^{\text{th}}$ edges emanating
from $v_{I_{v_{0}},\delta}^{\prime}$ in $c_{20},c_{30},{\hat{c}}_{0}$. We do
not, however, restrict the choice of the sets of the reference diffeomorphisms
in any way.

Once we have made choices subject to the above restrictions, let us, for
convenience, once again number our edges in such a way that the triplet of
(positively oriented) edges for any charge network $c$ is $\{e_{1},e_{2}%
,e_{3}\}$ so that the action of the inverse metric operator is as denoted in
equation (\ref{gooberish}). Recall that the parameter $B$ in that equation is,
apart from an overall numerical factor, equal to $B_{123}\lambda({\vec{e}})$.
Recall, from equation (\ref{Deflambda}) that $\lambda({\vec{e}})$ depends on
the triplet of unit edge tangents normalized in the coordinate metric
associated with the coordinate patch around the vertex $v$ of the charge
network $c$ being acted upon. Hence $\lambda({\vec{e}})$ varies as the charge
network varies over its diffeomorphism class. \emph{We choose $B_{123}$ so
that $B_{123}\lambda({\vec{e}})$ is constant over each diffeomorphism class.}
Thus, depending on the charge network $c\in\lbrack c]$, we obtain some
$\lambda({\vec{e}})$ and `compensate' for this $\lambda({\vec{e}})$ by
appropriately varying the edge length parameters $B_{1},B_{2},B_{3}$ so that
$B_{123}\lambda({\vec{e}})=B_{1}B_{2}B_{3}\lambda({\vec{e}})$ is constant over
$[c]$. Hence the parameter $B$ in equation (\ref{gooberish}) also depends only
on $[c]$, or, equivalently, on the reference charge network $c_{0}\in\lbrack
c]$. Finally, require that choice of $B_{123}$ be identical for the reference
charge networks related by (\ref{resref}). As we shall see now, these choices
ensure that the inverse volume eigenvalue has the properties referred to above.

From equation (\ref{DEFNU1}) we have that
\begin{equation}
\nu^{-\frac{2}{3}}=-8B\epsilon^{IJK}\epsilon_{ijk}\sigma\left(  O_{I}^{i}%
O_{J}^{j}O_{K}^{k}-3O_{I}^{-i}O_{J}^{j}O_{K}^{k}+3O_{I}^{-i}O_{J}^{-j}%
O_{K}^{k}-O_{I}^{-i}O_{J}^{-j}O_{K}^{-k}\right)  \label{Defnu}%
\end{equation}
%%BEG
The factor $\sigma$ is equal to the sign of the eigenvalue $Q$ in Equation
(\ref{defQ}). From Equation (\ref{defQ}), it is easy to check that $Q$ is
diffeomorphism-invariant. Moreover, it is straightforward to check that $Q$ is
also invariant under the `charge flips' of equation (\ref{defchrgeflip}). This
shows that $\sigma$ is invariant under diffeomorphisms and charge flips.
%%END%%
As we showed above, the factor $B$ is invariant under diffeomorphisms. 
The rest of the expression
consists of various combinations of charge labels of $c$, and as a result of
our choice of regulating edge holonomies, is equal to its evaluation on the
reference charge network $c_{0}\in\lbrack c]$ \emph{irrespective of the choice
of the set of reference diffeomorphisms}, $\mathcal{D}_{[c]}$. Thus
$\nu^{-\frac{2}{3}}$ is diffeomorphism invariant. 
In
addition, by construction, $B$ is the same for the quadruple of charge
networks $c(i,v_{I_{v},\delta}^{\prime}),c(v_{I_{v},\delta}^{\prime})$ which
arise from the action of the Hamiltonian constraint and the action of the
electric diffeomorphisms on any charge network $c$. 
It follows that, 
since the charges in
equation (\ref{Defnu}) for the charge networks of equation (\ref{res}),
(\ref{resref}) are related by `charge flips', the next section also
establishes that, as assumed in Sections 4 and 5, $\nu^{-\frac{2}{3}}$ is also
the same for the diffeomorphism classes of the charge networks of equations
(\ref{res}), (\ref{resref}).

\subsection{Symmetries}

We are interested in the eigenvalues of $\hat{q}^{-1/3}$ for a vertex deformed
by the Hamiltonian. There is one important property we are looking for: For
the LHS and RHS to match in the main calculation, a charge-flipped vertex
produced by the Hamiltonian must have the same $\hat{q}^{-1/3}$ eigenvalue as
the unflipped configuration. Recall the structure of the charge flips:
Depending on the value of $i$ appearing in the quantum shift, edges charged in
$\left(  q^{1},q^{2},q^{3}\right)  $ go to%
\begin{align}
i  &  =1:\left(  q^{1},-q^{3},q^{2}\right) \nonumber\\
i  &  =2:\left(  q^{3},q^{2},-q^{1}\right) \\
i  &  =3:\left(  -q^{2},q^{1},q^{3}\right) \nonumber
\end{align}
First note that the sign eigenvalue $\sigma$ of $\hat{\sigma}$ is unchanged;
each flipped configuration differs in sign in one entry, and there is a
transposition of two charges. Also note that $\left\vert Q\right\vert $ itself
is unchanged by similar arguments. Let us now consider $Q_{I}^{i}$ for some
fixed $I=\bar{I}$ and $i=\bar{\imath}$:%
\begin{equation}
Q_{\bar{I}}^{\pm\bar{\imath}}=Q\pm3\epsilon^{\bar{I}JK}\epsilon_{\bar{\imath
}jk}q_{J}^{j}q_{K}^{k}%
\end{equation}
Here the unbarred indices are summed only over unbarred values. What happens
to this value under charge flips? We have argued that $Q$ is unchanged under
flips, so focusing on the remainder under
\begin{equation}
q_{J}^{j}\rightarrow\left.  ^{(\tilde{\imath})}\!q_{J}^{j}\right.
=\delta^{\tilde{\imath}j}q_{J}^{(j)}-\epsilon^{\tilde{\imath}jk^{\prime}}%
q_{J}^{k^{\prime}},
\end{equation}
we find%
\begin{equation}
3\epsilon^{\bar{I}JK}\epsilon_{\bar{\imath}jk}\left.  ^{(\tilde{\imath}%
)}\!q_{J}^{j}\right.  \left.  ^{(\tilde{\imath})}\!q_{K}^{k}\right.
=6\epsilon^{\bar{I}JK}q_{J}^{\bar{\imath}}q_{K}^{\tilde{\imath}}+3\delta
_{\bar{\imath}}^{\tilde{\imath}}\epsilon^{\bar{I}JK}\epsilon_{\bar{\imath}%
jk}q_{J}^{j}q_{K}^{k},
\end{equation}
hence%
\begin{equation}
\left.  ^{(\tilde{\imath})}\!Q_{\bar{I}}^{\pm\bar{\imath}}\right.
=Q\pm\left(  3\delta_{\bar{\imath}}^{\tilde{\imath}}\epsilon^{\bar{I}%
JK}\epsilon_{\bar{\imath}jk}q_{J}^{j}q_{K}^{k}+6\epsilon^{\bar{I}JK}%
q_{J}^{\bar{\imath}}q_{K}^{\tilde{\imath}}\right)  .
\end{equation}
Notice that for $\tilde{\imath}=\bar{\imath},$ $\left.  ^{(\tilde{\imath}%
)}\!Q_{\bar{I}}^{\pm\bar{\imath}}\right.  =Q_{\bar{I}}^{\pm\bar{\imath}},$ so
at least one factor in each term in (\ref{gooberish}) is invariant. $\left.
^{(\tilde{\imath})}\!Q_{\bar{I}}^{\pm\bar{\imath}\neq\tilde{\imath}}\right.  $
changes, but it transforms into one of the other $Q_{\bar{I}}^{\pm\bar{\imath
}}$ such that the eigenvalue of $\hat{q}^{-1/3}$ is invariant. In particular,
it is immediate to check that%
\begin{align*}
\left.  ^{(i)}\!Q_{I}^{\pm j}\right.   &  =Q_{I}^{\mp k},\qquad\text{for
}\epsilon^{ijk}=+1,\\
\left.  ^{(i)}\!Q_{I}^{\pm j}\right.   &  =Q_{I}^{\pm k},\qquad\text{for
}\epsilon^{ijk}=-1.
\end{align*}
The $O_{I}^{\pm i}$ also obey these flip rules, so armed with these
properties, it is straightforward to expand%
\begin{align}
&  \left.  ^{(\tilde{\imath})}\!\hat{q}^{-1/3}\right.  |c\rangle\nonumber\\
&  =-B\frac{\epsilon^{2}}{8\kappa\hbar}\epsilon^{IJK}\epsilon_{ijk}\sigma\\
&  \times\left(  \left.  ^{(\tilde{\imath})}\!O_{I}^{i}\right.  \left.
^{(\tilde{\imath})}\!O_{J}^{j}\right.  \left.  ^{(\tilde{\imath})}\!O_{K}%
^{k}\right.  -3\left.  ^{(\tilde{\imath})}\!O_{I}^{-i}\right.  \left.
^{(\tilde{\imath})}\!O_{J}^{j}\right.  \left.  ^{(\tilde{\imath})}\!O_{K}%
^{k}\right.  +3\left.  ^{(\tilde{\imath})}\!O_{I}^{-i}\right.  \left.
^{(\tilde{\imath})}\!O_{J}^{-j}\right.  \left.  ^{(\tilde{\imath})}\!O_{K}%
^{k}\right.  -\left.  ^{(\tilde{\imath})}\!O_{I}^{-i}\right.  \left.
^{(\tilde{\imath})}\!O_{J}^{-j}\right.  \left.  ^{(\tilde{\imath})}%
\!O_{K}^{-k}\right.  \right)  |c\rangle\nonumber
\end{align}
and verify that it is in fact equal to $\hat{q}^{-1/3}|c\rangle,$ and we
conclude that $\hat{q}^{-1/3}$ has the symmetry property we need.

We close this subsection by noting that the eigenvalues of (the symmetrized)
$\hat{q}^{-1/3}$ at zero volume vertices vanish. Indeed, in the zero volume
case $Q=0$, we have that the $Q_{I}^{\pm i}$ and $O_{I}^{\pm i}$ eigenvalues
defined above evaluate to
\begin{equation}
Q_{I}^{\pm i}=\pm3\epsilon^{IJK}\epsilon_{ijk}q_{J}^{j}q_{K}^{k}%
,\qquad\Rightarrow\qquad O_{I}^{\pm i}=-|Q_{I}^{\pm i}|^{2/9}=-|Q_{I}%
^{i}|^{2/9}=-|3\epsilon^{IJK}\epsilon_{ijk}q_{J}^{j}q_{K}^{k}|^{2/9}.
\end{equation}
In particular, $O_{I}^{+i}=O_{I}^{-i},$ and since $q^{-1/3}$ goes as
\begin{equation}
q^{-1/3}\sim\epsilon^{IJK}\epsilon_{ijk}\sigma\left(  O_{I}^{i}O_{J}^{j}%
O_{K}^{k}-3O_{I}^{-i}O_{J}^{j}O_{K}^{k}+3O_{I}^{-i}O_{J}^{-j}O_{K}^{k}%
-O_{I}^{-i}O_{J}^{-j}O_{K}^{-k}\right)  ,
\end{equation}
we see that the insensitivity of $O_{I}^{\pm i}$ to the sign of the
representation of the regulating holonomy leads to the vanishing of this quantity.

\subsubsection{Non-Triviality}

The eigenvalues $q^{-1/3}$ are rather complicated functions of the charges,
and it is not clear a priori whether the symmetrization procedure followed
above perhaps leads to an operator action which is trivially zero through some
cancellations. Here we attempt to quell this apprehension somewhat by
exhibiting a class of states\footnote{We thank Alok Laddha for this example.}
with large non-zero volume, and small but non-zero $q^{-1/3}.$

Let $v$ be a vertex of $c$ from which emanate $N+3$ edges, three of which,
$e_{1},e_{2},e_{3}$, define the (positively-oriented) coordinate axes of the
system we evaluate $\hat{q}^{-1/3}$ with respect to, and let these edges have
charges $q_{1}^{1}=q_{2}^{2}=q_{3}^{3}=N\gg1.$ Let the other charges on these
edges be zero and let the remaining $N$ edges be charged as $\vec{q}=\left(
-1,-1,-1\right)  $ (so that the state is gauge-invariant). Then we can
compute
\begin{align}
Q  &  =\epsilon^{IJK}\epsilon_{ijk}q_{I}^{i}q_{J}^{j}q_{K}^{k}\nonumber\\
&  =6\epsilon_{ijk}\left(  \epsilon^{123}q_{1}^{i}q_{2}^{j}q_{3}^{k}+%
%TCIMACRO{\tsum \nolimits_{K^{\prime}\neq1,2,3}}%
%BeginExpansion
{\textstyle\sum\nolimits_{K^{\prime}\neq1,2,3}}
%EndExpansion
\left(  \epsilon^{12K^{\prime}}q_{1}^{i}q_{2}^{j}q_{K^{\prime}}^{k}%
+\epsilon^{23K^{\prime}}q_{2}^{i}q_{3}^{j}q_{K^{\prime}}^{k}+\epsilon
^{31K^{\prime}}q_{3}^{i}q_{1}^{j}q_{K^{\prime}}^{k}\right)  \right)
\nonumber\\
&  =6\left(  N^{3}-N^{2}%
%TCIMACRO{\tsum \nolimits_{K^{\prime}\neq1,2,3}}%
%BeginExpansion
{\textstyle\sum\nolimits_{K^{\prime}\neq1,2,3}}
%EndExpansion
\left(  \epsilon^{12K^{\prime}}+\epsilon^{23K^{\prime}}+\epsilon^{31K^{\prime
}}\right)  \right)  \label{glue}%
\end{align}
where the terms quadratic and cubic in the remaining edge charges have
vanished as they all have identical charges. We notice that as long as the sum
over orientation factors is not negative and $O(N),$ then indeed $Q\sim
N^{3}.$ One way to ensure this is to demand that the remaining edges be
distributed roughly evenly throughout the octants defined by
%%%%%%CHANGE MADE SEP 28%%%%%
the tangents to $e_{1},e_{2},e_{3}$ at $v$.
%(and their analytic extensions past $v$).
In this case the sum over $K^{\prime}$ of each orientation factor is $O(1)$
(or perhaps vanishing).

For the sake of calculation, let us suppose that $N$ is in fact divisible by
8, and consider the case in which $N/8$ of the small-charge edges lie in each
octant. Then the sum over orientation factors in (\ref{glue}) in fact
vanishes, and we have $Q=6N^{3}.$ We now wish to compute $q^{-1/3}$ for this
configuration. We have, for example%
\begin{align}
Q_{1}^{\pm i}-Q  &  =\pm3\epsilon^{1JK}\epsilon_{ijk}q_{J}^{j}q_{K}%
^{k}\nonumber\\
&  =\pm6\left(  \epsilon_{i23}N^{2}+N%
%TCIMACRO{\tsum \nolimits_{K^{\prime}\neq1,2,3}}%
%BeginExpansion
{\textstyle\sum\nolimits_{K^{\prime}\neq1,2,3}}
%EndExpansion%
%TCIMACRO{\tsum \nolimits_{j}}%
%BeginExpansion
{\textstyle\sum\nolimits_{j}}
%EndExpansion
\left(  \epsilon^{12K^{\prime}}\epsilon_{ij2}+\epsilon^{13K^{\prime}}%
\epsilon_{ij3}\right)  \right)
\end{align}
so that%
\begin{equation}
Q_{1}^{\pm i=1}-Q=\pm6\left(  N^{2}+N%
%TCIMACRO{\tsum \nolimits_{K^{\prime}\neq1,2,3}}%
%BeginExpansion
{\textstyle\sum\nolimits_{K^{\prime}\neq1,2,3}}
%EndExpansion
\left(  \epsilon^{13K^{\prime}}-\epsilon^{12K^{\prime}}\right)  \right)
=\pm6N^{2},\qquad Q_{1}^{\pm i\neq1}-Q=0,
\end{equation}
with analogous results for $I=2,3.$ Then%
\begin{align}
\left\vert Q\right\vert ^{2/9}-\left\vert Q_{I}^{\pm i}\right\vert ^{2/9}  &
=\left\vert 6N^{3}\right\vert ^{2/9}-\left\vert 6N^{3}\pm6N^{2}\right\vert
^{2/9}=(6N^{3})^{2/9}\left(  1-\left(  1\pm\frac{1}{N}\right)  ^{2/9}\right)
\nonumber\\
&  =(6N^{3})^{2/9}\left(  \mp\frac{2}{9N}+O(N^{-2})\right)
\end{align}
for $I=i,$ and zero otherwise. Thus%
\[
O_{I}^{\pm i}O_{J}^{\pm j}O_{K}^{\pm k}=\frac{2^{3}6^{\frac{2}{3}}}{3^{6}}%
(\mp)_{ijk}\frac{1}{N}+O(N^{-2}),
\]
where $(\mp)_{ijk}$ denotes the product of the (negative of the) signs in the
$O$ superscripts, whence%
\begin{equation}
q^{-1/3}=B\frac{\epsilon^{2}}{\hbar\gamma\kappa}\left(  \frac{2^{4}6^{\frac
{2}{3}}}{3^{5}}\frac{1}{N}+O(N^{-2})\right)  ,
\end{equation}
and we conclude that $\hat{q}^{-1/3}$ constructed above is not trivially vanishing.

In fact, if one allows (an $N$-independent) tuning of the parameter $B,$ this
class of states may be considered as satisfying a crude notion of
semiclassicality (to leading order in $N$), in the sense that
\begin{equation}
q^{-1/3}\simeq(\tfrac{4}{3}\pi\epsilon^{3})^{2/3}V^{-2/3}=\frac{48^{1/3}%
}{\varepsilon_{(\mu)}^{2/3}}(\tfrac{4}{3}\pi)^{2/3}\frac{\epsilon^{2}}%
{\hbar\gamma\kappa}\left\vert Q\right\vert ^{-1/3}%
\end{equation}
if one chooses%
\begin{equation}
B=\left(  \frac{3^{11}}{2^{7}}\right)  ^{1/3}\left(  \frac{\pi}{\varepsilon
_{(\mu)}}\right)  ^{2/3}%
\end{equation}

\section{RHS Identity: SU(2)}

\label{su2rhsid}Consider the diffeomorphism generator (modulo Gauss
constraint) of the SU(2) theory smeared with the electric shift $N_{i}%
^{a}:=q^{-\alpha}NE_{i}^{a}$, where $N$ has density weight $\left(
2\alpha-1\right)  $:%
\begin{equation}
D[\vec{N}_{i}]:=\int\mathrm{d}^{3}x~q^{-\alpha}NE_{i}^{a}F_{ab}^{j}E_{j}^{b}%
\end{equation}
Here $F_{ab}^{i}:=2\partial_{\lbrack a}A_{b]}^{i}+G_{\mathrm{N}}\epsilon
^{ijk}A_{a}^{j}A_{b}^{k}$ and the connection again has units of
$[\mathrm{length}\times G_{\mathrm{N}}]^{-1}.$ It is straightforward to
compute the Poisson bracket of two such objects, summing over the SU(2)
index:
\begin{align}
&  \{D[\vec{N}_{i}],D[\vec{M}_{i}]\}\nonumber\\
&  =\int\mathrm{d}^{3}x~\left(  \frac{\delta D[\vec{N}_{i}]}{\delta A_{a}%
^{j}(x)}\frac{\delta D[\vec{M}_{i}]}{\delta E_{j}^{a}(x)}-\left(
N\leftrightarrow M\right)  \right) \nonumber\\
&  =2\int\mathrm{d}^{3}x~\left[  \left(  \partial_{d}\left(  \frac{NE_{i}%
^{[a}E_{j}^{d]}}{q^{\alpha}}\right)  +G_{\mathrm{N}}\epsilon^{jml}A_{d}%
^{m}\frac{NE_{i}^{[a}E_{l}^{d]}}{q^{\alpha}}\right)  \right. \nonumber\\
&  \qquad\qquad\qquad\qquad\times\left.  \frac{M}{q^{\alpha}}\left(
\delta_{i}^{j}F_{ab}^{k}E_{k}^{b}+F_{ba}^{j}E_{i}^{b}-\alpha E_{i}^{c}%
E_{a}^{j}F_{cb}^{k}E_{k}^{b}\right)  -\left(  N\leftrightarrow M\right)
\right] \nonumber\\
&  =2\int\mathrm{d}^{3}x~\left(  \frac{1}{q^{2\alpha}}\left(  E_{i}^{[a}%
E_{j}^{d]}\delta_{i}^{j}F_{ab}^{k}E_{k}^{b}+E_{i}^{[a}E_{j}^{d]}F_{ba}%
^{j}E_{i}^{b}-\alpha E_{i}^{[a}E_{j}^{d]}E_{i}^{c}E_{a}^{j}F_{cb}^{k}E_{k}%
^{b}\right)  M\partial_{d}N-\left(  N\leftrightarrow M\right)  \right)
\nonumber\\
&  =\left(  2\alpha-1\right)  \int\mathrm{d}^{3}x~q^{-2\alpha}E_{i}^{a}%
E_{i}^{c}F_{cb}^{j}E_{j}^{b}\left(  M\partial_{a}N-N\partial_{a}M\right)
\nonumber\\
&  =\left(  2\alpha-1\right)  \{H[N],H[M]\}
\end{align}
where we have used $\delta q/\delta E_{i}^{a}=q(E^{-1})_{a}^{i},$ with
$(E^{-1})_{a}^{i}$ the matrix inverse of $E_{i}^{a}.$ The U(1)$^{3}$ case
results by taking $G_{\mathrm{N}}\rightarrow0.$ In 2+1 dimensions, this
identity also holds in SU(2) and U(1)$^{3}$:%
\begin{align}
&  \{D[\vec{N}_{i}],D[\vec{M}_{i}]\}\nonumber\\
&  =\int\mathrm{d}^{2}x~\left(  \frac{\delta D[\vec{N}_{i}]}{\delta A_{a}%
^{j}(x)}\frac{\delta D[\vec{M}_{i}]}{\delta E_{j}^{a}(x)}-\left(
N\leftrightarrow M\right)  \right) \nonumber\\
&  =2\int\mathrm{d}^{3}x~\left(  q^{-\alpha}E_{i}^{[a}E_{j}^{d]}\left(
q^{-\alpha}\delta_{i}^{j}F_{ab}^{k}E_{k}^{b}+q^{-\alpha}F_{ba}^{j}E_{i}%
^{b}+2\alpha q^{-\alpha-1}\eta_{ab^{\prime}}\epsilon^{jj^{\prime}k^{\prime}%
}E^{j^{\prime}}E_{k^{\prime}}^{b^{\prime}}E_{i}^{c}F_{cb}^{k}E_{k}^{b}\right)
M\partial_{d}N-\left(  N\leftrightarrow M\right)  \right) \nonumber\\
&  =\left(  2\alpha-1\right)  \int\mathrm{d}^{3}x~\left(  M\partial
_{c}N-N\partial_{c}M\right)  q^{-2\alpha}E_{i}^{c}E_{i}^{b}F_{ba}^{j}E_{j}^{a}%
\end{align}
where we have used $q=E^{i}E^{i}$, $E^{i}:=\frac{1}{2}\eta_{ab}\epsilon
^{ijk}E_{j}^{a}E_{k}^{b}$ and $E^{i}\eta^{ab}=\epsilon^{ijk}E_{j}^{a}E_{k}%
^{b}$ (see \cite{qsd4}).

\section{Deformations: Further Technical Details}

\label{deformApp}

\subsection{Preliminary Remarks}

We use the notation of Section 4. Let $B_{4\delta}(v)$ be the ball of
coordinate radius $4\delta$, with respect to the metric $\delta_{ab}$
associated with the coordinates $\{x\}$, centered at $v$. Our considerations
are confined to the interior of this ball for sufficiently small $\delta$. We
shall choose
%%@@sep2012%%
$\delta$ to be small enough that the boundary of $B_{4\delta}(v)$ intersects
the interior of every edge emanating from $v$ once and only once.

Let the edge $e_{I}$ be parameterized by the parameter $t_{I}$ such that
$e_{I}(t_{I}=0)= v$. Let the interior of the edge be $e^{\mathrm{int}}_{I}$.
Let the coordinates of the point $e_{I}(t_{I})$ be denoted by $x^{\mu}
(t_{I})$ in the coordinate system $\{x\}$. Then for small enough $\delta$ it
follows from the semianalyticity of the edges that the parameterization
$t_{I}$ can be chosen in such a way that $x^{\mu}(t_{I}) \forall I$ are
analytic functions on $e^{\mathrm{int}}_{I}\cap B_{4\delta}(v)$. Accordingly
we choose $\delta$ small enough that the edges within $B_{4\delta}(v)$ are
analytic in the coordinate system $\{x\}$ except perhaps at $v$.
%%%%%%%%%

We assume for simplicity that $v$ resides at the origin of the coordinate
patch $\{x\}$ . We shall often denote the coordinates $\{x\}$ of a point by
the vector $\vec{x}$ from the origin to that point. Since the coordinates
range in some open subset of $%
%TCIMACRO{\U{211d} }%
%BeginExpansion
\mathbb{R}
%EndExpansion
^{3},$ we freely use the ensuing $%
%TCIMACRO{\U{211d} }%
%BeginExpansion
\mathbb{R}
%EndExpansion
^{3}$ structures, such as constant vectors, vectors connecting a pair of
points, straight lines, planes, etc. Recall that ${\dot{e}}_{I}^{a}%
(v)=:\vec{\dot{e}}_{I}(v)$ is the tangent vector of the $I^{\mathrm{th}}$ edge
at $v$. If $\vec{a}$ is a vector we denote its component perpendicular to
$\vec{\dot{e}}_{I}(v)$ by $\vec{a}_{\perp}$. The vector connecting a point
$P_{1}$ to the point $P_{2}$ is denoted as $\vec{l}_{P_{1}P_{2}}$.

\subsection{GR-Preserving Deformation}

\noindent\emph{1. The GR condition:} The set of tangent vectors $\vec{\dot{e}%
}_{K}$ at $v$ is GR if and only if no triplet lies in a plane. It is easy to
verify that this condition implies the pair of conditions:

\begin{enumerate}
\item[\emph{1.1}] $\vec{\dot{e}}_{J\perp}\neq0,\ J\neq I$.

\item[\emph{1.2}] No pair $(\vec{\dot{e}}_{J_{1}\perp},\vec{\dot{e}}%
_{J_{2}\perp}),\;J_{1}\neq J_{2}\neq I$ exists such that $\vec{\dot{e}}%
_{J_{1}\perp},\vec{\dot{e}}_{J_{2}\perp}$ are linearly-dependent.
\end{enumerate}

\noindent\emph{2. Choice of $\vec{\hat{n}}_{I}$ in equation (\ref{posnvprimei}%
):} We choose $\vec{\hat{n}}_{I}$ in a direction such that $v_{I}^{\prime}$ is
not on ${\gamma(c)}$. Clearly, this is possible because there are a finite
number of edges at $v$ and for small enough $\delta$ these edges are `almost'
straight lines. In Section C.4 we shall need to specify $\vec{\hat{n}}_{I}$
more precisely; for this section, it is enough that $v_{I}^{\prime}$ is not on
$\gamma(c)$.

%%BEG%%
%\noindent\emph{3. GR property of $v_{I}^{\prime}$}: 
\noindent\emph{3. Connecting $v_{I}^{\prime}$ to $\gamma (c)$}: 
Let $v_{I}^{\prime}$ be
connected to ${\tilde{v}}_{J},J=1,\dots,M$ in accordance with the prescription
of Section 4.4.2. In more detail, we have, from Section 4.4.2, that for $J\neq
I$, $\{{\tilde{v}}_{J}\}=B_{\delta^{q}}(v)\cap e_{J}$ and that $v_{I}^{\prime
}$ is connected to ${\tilde{v}}_{J}$ by the straight lines ${\vec{l}}%
_{v\tilde{v}_{J}}$. The $C^{k},k\gg1$ nature of $e_{J\neq I}$ near $v$ implies
that
\begin{equation}
\delta^{q}\vec{\hat{\dot{e}}}_{J}=\vec{l}_{v\tilde{v}_{J}}+O(\delta^{2q})
\label{A4.1}%
\end{equation}
where the hat $\hat{}$, as usual, denotes the unit vector in the direction of
$\vec{{\dot{e}}}_{J}.$ Equation (\ref{posnvprimei}) implies that for $J\neq
I$,
\begin{align}
\vec{l}_{v_{I}^{\prime}\tilde{v}_{J}\perp}  &  =-\delta^{p}\vec{\hat{n}}%
_{I}+\vec{l}_{v{\tilde{v}}_{J}\perp}\nonumber\\
&  =-\delta^{p}\vec{\hat{n}}_{I}+\delta^{q}(\vec{\hat{\dot{e}}}_{J})_{\perp
}+O(\delta^{2q})\nonumber\\
&  =\delta^{q}(\vec{\hat{\dot{e}}}_{J})_{\perp}+O(\delta^{2q})+O(\delta^{p}).
\label{deltaqeperp}%
\end{align}
Here $(\vec{\hat{\dot{e}}}_{J})_{\perp}$ is the perpendicular component of the
unit vector $\vec{\hat{\dot{e}}}_{J}$ and we have used (\ref{A4.1}) in the
second line. Note that $p>q$ in the last line so that the first term is the
leading order term.

As asserted in Section 4.4.2, the lines $\vec{l}_{v_{I}^{\prime}\tilde{v}_{J}%
},J\neq I$, intersect the graph $\gamma$ underlying the (undeformed) charge
network $c$ at most only at a finite number of points. This can be seen from
the following argument. If this was not the case, the analyticity of the edges
$\{e_{K}\}$ (see C.1) and the analyticity of the lines $\{\vec{l}%
_{v_{I}^{\prime}\tilde{v}_{J}}\}$ in the chart $\{x\}$ implies that a segment
of some line $\vec{l}_{v_{I}^{\prime}\tilde{v}_{J}}$ must overlap with a
segment of some edge $e_{K}$ in $B_{4\delta}(v)$. Equation (\ref{deltaqeperp})
together with the GR property of $v$ implies that if this overlap happens it
must be for $K=J\neq I$. But, whereas $||\vec{\dot{e}}_{J\perp}||/||\vec
{\dot{e}}_{J}||$ is of $O(1)$, equation (\ref{deltaqeperp}) implies that
$||\vec{l}_{\tilde{v}_{J}v_{I}^{\prime}\perp}||/||\vec{l}_{\tilde{v}_{J}%
v_{I}^{\prime}}||,$ is of $O(\delta^{q-1})$ (here $||\vec{a}||$ refers to the
norm of the vector $\vec{a}$).

We also note that the lines $\vec{l}_{v_{I}^{\prime}\tilde{v}_{J}},J\neq I$
cannot intersect each other (except at $v_{I}^{\prime}$) since equation
(\ref{deltaqeperp}) implies that they have different slopes. Moreover, since
$||\vec{l}_{\tilde{v}_{J}v_{I}^{\prime}\perp}||/||\vec{l}_{\tilde{v}_{J}%
v_{I}^{\prime}}||,$ is of $O(\delta^{q-1})$ it follows that these lines (and
any bumps thereof can be chosen so that they) are always below the plane $P$
(see Section 4.4.2). Hence these lines cannot intersect the curve ${\tilde{e}%
}_{I}$ of Section 4.4.2. Finally, it easy to see that ${\tilde{e}}_{I}$ can
indeed be constructed in accordance with the requirements of Section 4.4.2. To
do so, we join ${\tilde{v}}_{I}$ to $v_{I}^{\prime}$ by a straight line and
apply appropriate semianalytic diffeomorphisms of compact support in the
vicinity of ${\tilde{v}}_{I}$, $v_{I}^{\prime}$ \emph{only} to this line so as
to bring its tangents at these points in line with ${\vec{\hat{e}}}_{I}(v)$ as
required by equation (\ref{doteI}) and the requirement that ${\tilde{v}}_{I}$
be a $C^{1}$ kink. It is straightforward to see that this can be achieved in
such a way that ${\tilde{e}}_{I}$ remains above $P$.\\
%%END%%

\noindent{\em 4. GR property of $v_{I}^{\prime}$:} It remains to show that 
$v_{I}^{\prime}$ is GR.
%above ensures that $v_{I}^{\prime}$ is GR. 
Since we are unable to ascertain
if $v_{I}^{\prime}$ is GR when connected to $\gamma (c)$ as in $3.$ above, 
we seek a suitable modification of $3.$ which ensures that $v_{I}^{\prime}$
is GR while preserving the key equations 
(\ref{posnvprimei}), (\ref{doteI}), and
(\ref{doteJ}). Since the GR property is generic (as opposed to its negation
which requires the condition of coplanarity of some triplet to be enforced)
we expect that there should be several ways to do this. However, we do not analyse the
issue here and point the reader to Reference \cite{meinprep} wherein we 
present a detailed resolution of the issue, the particular choice of which is motivated
by our considerations in that work. Here, we only note that Reference \cite{meinprep}
applies  semianalytic diffeomorphisms supported away from identity in a small vicinity
of $v_{I}^{\prime}$ (only) to each edge in turn which renders the edge tangent configuration
`conical' and hence GR \cite{meinprep}. Each such diffeomorphism is of the type encountered in 
section C.3 below.

%Next, note that from equation (\ref{deltaqeperp}), the GR property of $v$ and
%conditions 1.1 and 1.2, it follows that the set of vectors $\{\vec{\dot{e}%
%}_{I},\vec{l}_{v_{I}^{\prime}\tilde{v}_{J\neq I}}\}$ is GR. Finally, note that
%the GR property of this set, together with the fact that $\vec{l}%
%_{v_{I}^{\prime}\tilde{v}_{J}}$ is proportional to $\vec{\dot{\tilde{e}}}%
%_{J}(v_{I}^{\prime}),$\footnote{Recall that the \textquotedblleft
%bumping\textquotedblright\ at the end of Section 4.4.2 does not change the
%edge tangents at $v_{I}^{\prime}$ and that these tangents were obtained by
%joining $v_{I}^{\prime}$ to $\tilde{v}_{J}$ by straight lines.} and that
%$\vec{\dot{\tilde{e}}}_{I}(v_{I}^{\prime})$ and $\vec{\dot{e}}_{I}(v)$ point
%in the same direction, implies that the displaced vertex $v_{I}^{\prime}$ is GR.

\subsection{Non-GR Case}

As in the previous section we choose $\vec{\hat{n}}_{I}$ in a direction such
that $v_{I}^{\prime}$ is not on ${\gamma(c)}$ and follow the prescription of
Section 4.4.2 to join $v_{I}^{\prime}$ to $\tilde{v}_{J},J\neq I$ by straight
lines. Note that, as asserted in Section 4.4.2, any such line $\vec{l}%
_{v_{I}^{\prime}\tilde{v}_{J}},J\neq I$ can intersect any edge $e_{K}$ at most
in a finite number of points. To see this assume the contrary. Analyticity of
the lines and edges (see C.1) in the $\{x\}$ coordinates implies that the line
$\vec{l}_{v_{I}^{\prime}\tilde{v}_{J}}$ overlaps with the edge $e_{K}$. If
${\vec{\dot{e}}}_{K}|_{v}$ is proportional to ${\vec{\dot{e}}}_{I}|_{v}$,
analyticity of $e_{K},\vec{l}_{v_{I}^{\prime}\tilde{v}_{J}}$ implies that
$\vec{l}_{v_{I}^{\prime}\tilde{v}_{J}}$ is contained in the line which joins
$v$ to $v_{I}^{\prime}$ along the direction ${\vec{\dot{e}}}_{I}(v)$. From
(\ref{posnvprimei}), no such line exists. If ${\vec{\dot{e}}}_{K\perp}%
(v)\neq0$ then $||\vec{\dot{e}}_{K\perp}||/||\vec{\dot{e}}_{K}||$ is of
$O(1)$, while equation (\ref{deltaqeperp}) implies that $||\vec{l}_{\tilde
{v}_{J}v_{I}^{\prime}\perp}||/||\vec{l}_{\tilde{v}_{J}v_{I}^{\prime}}||,$ is
of $O(\delta^{q-1})$, which, once again, rules out overlap.

Next, any possible overlap between the lines $\{\vec{l}_{v_{I}^{\prime}%
\tilde{v}_{J}},J\neq I\}$ can be removed by slightly altering the positions of
their vertices $\tilde{v}_{J}$ as follows. Suppose that $\vec{l}%
_{v_{I}^{\prime}\tilde{v}_{J_{1}}},\vec{l}_{v_{I}^{\prime}\tilde{v}_{J_{2}}}$
overlap. Their analyticity and the existence of a common end point
$v_{I}^{\prime}$ imply that one must be contained in the other. Accordingly,
assume that $\vec{l}_{v_{I}^{\prime}\tilde{v}_{J_{1}}}$ is contained in
$\vec{l}_{v_{I}^{\prime}\tilde{v}_{J_{2}}}$ so that $\vec{l}_{v_{I}^{\prime
}\tilde{v}_{J_{2}}}$ passes through $\tilde{v}_{J_{1}}$. Since $\tilde
{v}_{J\neq I}\in\partial B_{\delta^{q}}(v)$, it follows that this pair of
lines cannot overlap with any other line. If we now move $\tilde{v}_{J_{1}}$
slightly along $e_{J_{1}}$, this overlap is necessarily removed. For, if it
were not, then $\vec{l}_{v_{I}^{\prime}\tilde{v}_{J_{2}}}$ would overlap with
$e_{J_{1}}$ which is ruled out by the arguments of the previous paragraph.
Thus, with this modification, the lines $\{\vec{l}_{v_{I}^{\prime}\tilde
{v}_{J}},J\neq I\}$ intersect each other as well as $\gamma(c)$ at most at a
finite number of points and these intersections can be removed by appropriate
\textquotedblleft bumping\textquotedblright\ such that the bumps are all below
the plane $P$ of Section 4.4.2.

Next, we show that ${\tilde{e}}_{I}$ may be chosen so as to satisfy the
requirements of Section 4.4.2 on its tangents at its end points while
intersecting $\gamma(c)$ at most at a finite number of points and while being
positioned above the plane $P$ of Section 4.4.2. Connect ${\tilde{v}}_{I}$ to
$v_{I}^{\prime}$ by the straight line $\vec{l}_{v_{I}^{\prime}\tilde{v}_{I}}$.
Analyticity implies either a finite number of intersections with $\gamma(c)$
or overlap. Let $\vec{l}_{v_{I}^{\prime}\tilde{v}_{I}}$ overlap some edge
$e_{K}$. As above, if ${\vec{\dot{e}}}_{K}|_{v}$ is proportional to
${\vec{\dot{e}}}_{I}|_{v}$, analyticity of $e_{K},\vec{l}_{v_{I}^{\prime
}\tilde{v}_{I}}$ implies that $\vec{l}_{v_{I}^{\prime}\tilde{v}_{I}}$ is
contained in the line which joins $v$ to $v_{I}^{\prime}$ along the direction
${\vec{\dot{e}}}_{I}(v)$. From (\ref{posnvprimei}), no such line exists. If
${\vec{\dot{e}}}_{K\perp}(v)\neq0$ then $||\vec{\dot{e}}_{K\perp}%
||/||\vec{\dot{e}}_{K}||$ is of $O(1)$. On the other hand a Taylor series
expansion along the edge $e_{I}$ locates ${\tilde{v}}_{I}$ to $O(\delta^{2})$
from the line passing through $v$ in the direction of ${\vec{\dot{e}}}_{I}|_{v}%
$, which, together with equation (\ref{deltaqeperp}) implies that $||\vec
{l}_{\tilde{v}_{I}v_{I}^{\prime}\perp}||/||\vec{l}_{\tilde{v}_{I}v_{I}%
^{\prime}}||,$ is of $O(\delta^{})$, which, once again, rules out overlap.
The finite number of intersections with $\gamma(c)$ can be removed by
appropriate bumping which preserves the location of $\vec{l}_{v_{I}^{\prime
}\tilde{v}_{I}}$ above the plane $P$ of Section 4.4.2. Finally, the edge
tangents at the end point $v_{I}^{\prime}$ can be aligned with ${\vec{\dot{e}%
}}_{I}|_{v}$, and the end point $\tilde{v}_{I}$ transformed into a $C^{1}%
$-kink by  appropriate semianalytic diffeomorphisms which are compactly
supported in the vicinity of these end points and which are applied
\emph{only} to $\vec{l}_{v_{I}^{\prime}\tilde{v}_{I}}$.

Next, suppose that the above prescription leads to $v_{I}^{\prime}$ being a
non-GR vertex. Then we are done. If not, then proceed as follows. First note
that since the `bumping' is supported away from $v_{I}^{\prime}$, it follows
that in a small enough neighborhood of $v_{I}^{\prime}$, the edges $\tilde
{e}_{J\neq I}$ which connect $v_{I}^{\prime}$ to $\tilde{v}_{J}$ are straight
lines. Next, pick some $J\neq I$. Then it follows from the above discussion,
in conjunction with the GR property of $v_{I}^{\prime},$ that in a small
enough neighborhood of $v_{I}^{\prime},$ the plane which contains $\tilde
{e}_{J}$ and which is tangent to the direction $\vec{\dot{e}}_{I}(v)$ does not
intersect any other edge $\tilde{e}_{K\neq J\neq I}$. Now consider the vector
field which generates rotations about the axis passing through $v_{I}^{\prime
}$ in a direction normal to this plane. Multiplying this vector field with a
semianalytic function of small enough support about $v_{I}^{\prime}$ yields a
vector field of compact support which generates a diffeomorphism that rotates
the tangent $\vec{\dot{\tilde{e}}}_{J}(v_{I}^{\prime})$ to the edge $\tilde
{e}_{J}$ at $v_{I}^{\prime}$ into a direction exactly anti-parallel to that of
$\vec{\dot{e}}_{I}(v)$. We apply this diffeomorphism \emph{only} to the edge
$\tilde{e}_{J}$. As a result the vertex $v_{I}^{\prime}$ loses its GR property
since, now, any triplet of tangent vectors containing the tangents to the
$I^{\mathrm{th}}$ and $J^{\mathrm{th}}$ edges at $v_{I}^{\prime}$ lie in a
plane by virtue of the anti-collinearity of the (outward-pointing) tangents to
the $I^{\mathrm{th}}$ and the $J^{\mathrm{th}}$ edges.

\subsection{Relating Deformations by Diffeomorphisms}

\label{scrunchAppendix} \noindent\emph{1. Introductory Remarks:} For small
enough $\delta=\delta_{0}$ let the vertex $v_{I}^{\prime}$ be placed and
joined to the undeformed graph $\gamma(c)$ as described in Section 4.4.2 and
the first two sections of this appendix. This specifies the deformation at
triangulation fineness $\delta_{0}$. In the subsequent sections we generate
deformations for all $\delta$ such that $0<\delta<\delta_{0}$ by the
application of semianalytic diffeomorphisms to the deformation at $\delta_{0}%
$. Clearly, we need these diffeomorphisms to do the following:

\begin{enumerate}
\item[(a)] Leave the undeformed graph $\gamma(c)$ invariant;

\item[(b)] move the points $\tilde{v}_{J}$ down the edges $e_{J}$ to a
distance of $\delta^{q}$ from $v$ for $J\neq I$ and to a distance of $2\delta$
for $J=I$;

\item[(c)] move the immediate vicinity of the vertex $v_{I}^{\prime}$ to a
distance of approximately $\delta$ from $v$ in such a way that the tangents at
the new position, $v_{I}^{\prime}(\delta)$, satisfy Equations (\ref{doteI}),
(\ref{doteJ}).
\end{enumerate}

In order to implement (c) simultaneously with (a) and (b), we need to ensure
that the diffeomorphism which implements (c) is identity in the vicinity of
$\gamma(c)$. We find it simplest to proceed as follows. First we define the
position of the displaced vertex at parameter $\delta$ through Equation
(\ref{posnvprimei}). Thus the set of points $v_{I}^{\prime}\equiv
v_{I}^{\prime}(\delta)$ (for all positive $\delta$ less than $\delta_{0}$) are
contained in a plane tangent to $\vec{\dot{e}}_{I}(v),\vec{\hat{n}}_{I}$. Our
strategy is to choose $\vec{\hat{n}}_{I}$ such that this plane does not
intersect $\gamma(c)$ except at $v$ (and, at most, in a small vicinity of the
straight line passing through $v$ in the direction of $\vec{\dot{e}}_{I}(v)$).
More precisely, we show that this plane is contained in a small angle
%(of
%$O(\delta)$)
\textquotedblleft wedge\textquotedblright\ with axis along the straight line
passing through $v$ in the direction of $\vec{\dot{e}}_{I}(v)$, and, that this
wedge intersects $\gamma(c)$ at most along (a very small neighborhood of) its
axis. This enables the construction of an appropriate diffeomorphism which is
identity outside this wedge and which implements (c).

In order to show the existence of $\vec{\hat{n}}_{I}$ which allows the
construction of such a wedge, it is necessary to confine the edges which are
in the vicinity of the straight line passing through $v$ in the direction of
${\vec{\dot{e}}}_{I}(v)$ to manageable neighborhoods so that ${\vec{\hat{n}}%
}_{I}$ can be chosen to point away from them. In the GR case only the
$I^{\mathrm{th}}$ edge is of this type, whereas in the non-GR case there may
be several edges with tangent at $v$ along ${\vec{\dot{e}}}_{I}(v)$. It turns
out that in both cases these edges can themselves be confined to appropriately
small neighborhoods.

Given the importance of the `wedge neighborhoods', it is useful to develop
some nomenclature to refer to their construction. We do so in Part 2 below. In
Part 3, we show how to choose $\vec{\hat{n}}_{I}$ when $v$ is GR and in Part
4, when $v$ is not GR. Having chosen $\vec{\hat{n}}_{I}$ appropriately, we
construct, in Part 5, a diffeomorphism which implements (c) while respecting
(a). In Part 6 we construct diffeomorphisms which implement (b) while
respecting (a) in such a way that they are identity in the vicinity of
$v_{I}^{\prime}(\delta)$ so as not to affect the (prior) implementation of (c).

In Parts 3 and 4 we do not fix $\delta=\delta_{0}$. Rather the considerations
in these parts assures us of the existence of a small enough $\delta$ which
can be set equal to $\delta_{0}$ in Parts 5 and 6. Accordingly, from C.1, our
considerations in Parts 3,4 are restricted to the ball $B_{4\delta}(v)$ and,
in Parts 5,6 to $B_{4\delta_{0}}(v)$.

\bigskip

\noindent\emph{2. Some useful nomenclature:} Consider a pair of
linearly-independent vectors $\vec{a},\vec{b}$. Consider the set of points
\begin{equation}
\vec{x}=\alpha\vec{a}+\beta\vec{b} \label{plane}%
\end{equation}
for all $\alpha\in%
%TCIMACRO{\U{211d} }%
%BeginExpansion
\mathbb{R}
%EndExpansion
$ and all $\beta\geq0$ such that $\vec{x}\in B_{4\delta}(v)$. Clearly, the set
of these points comprises a \textquotedblleft half plane\textquotedblright%
\ which is bounded by the line passing through $v$ in the direction of
$\vec{a}$. We refer to this set of points as \emph{the half plane tangent to
$(\vec{a},\vec{b})$ with boundary through $v$ along }$\vec{a}$. Let us denote
this half plane as $P$. Rotate $P$ about its boundary through $v$ along
$\vec{a}$ by $\pm\theta$ to obtain a pair of half planes which bound a wedge
of angle $2\theta$. We shall refer to this wedge as \emph{the wedge of angle
$2\theta$ associated with $P$}.

\bigskip

\noindent\emph{3. Detailed choice of }$\vec{\hat{n}}$\emph{$_{I}$ for the GR
case:} Let the coordinates of the edge $e_{I}$ at parameter value $t$ be
$\vec{x}_{I}(t)$. Since $e_{I}$ is $C^{k}$, we may use the Taylor expansion:
%Analyticity near $v$ implies that
%%%%OCT 2 2012%%%
\begin{equation}
\vec{x}_{I}(t)=\sum_{n=1}^{k-1}\vec{v}_{n}^{I}t^{n} + O(t^{k}) \label{analei}%
\end{equation}
with $\vec{v}_{1}^{I}=\vec{\dot{e}}_{I}(v)$. For simplicity we rescale the
parameter $t$ so that $\vec{v}_{1}^{I}=\vec{\hat{e}}_{I}(v)$, where as in the
main text, $\vec{\hat{e}}_{I}(v)$ is unit in the $\{x\}$ coordinate metric.
Let $m$ be the smallest integer less than $k$ such that the pair $\vec{v}%
_{m}^{I},\vec{v}_{1}^{I}$ are not linearly-dependent. If no such $m$ exists
then we set $m=k-1$ so that $\vec{v}_{m\perp}^{I}=0$.

If $\vec{v}_{m \perp}^{I}\neq0$ then we proceed as follows. Let $P_{m}^{I}$ be
the half plane tangent to $(\vec{v}_{1}^{I},\vec{v}_{m}^{I})$ with boundary
through $v$ along $\vec{v}_{1}^{I}$. Then equation (\ref{analei}) implies that
for small enough $\delta$, the edge $e_{I}$ is confined to the wedge
$W_{m}^{I}(\theta)$ with $\theta$ of $O(\delta)$. Hence there is a
\textquotedblleft$2\pi-2\theta$\textquotedblright\ worth of possible choices
for ${\vec{\hat{n}}}_{I}$ such that $v_{I}^{\prime}$ does not lie on $e_{I}$.
We choose $\vec{\hat{n}}_{I}$ such that it lies an angle of $O(1)$ away from
the set of vectors $\{\vec{v}_{m\perp}^{I},\vec{e}_{J\perp}\},$ $J\neq I$.
Clearly, for small enough $\delta$, $v_{I}^{\prime}$ also does not lie on the
undeformed graph $\gamma(c)$.

If $\vec{v}_{m=k-1\perp}^{I}=0$ then we have that all $\vec{v}_{m\perp}%
^{I}=0$ for $m$ such that $1<m\leq k-1$. It follows that the edge $e_{I}$ is confined to a very
small neighborhood $S_{k}$ of the line through $v$ along the direction
${\dot{e}}_{I}(v)$. To define $S_{k}$, it is useful to rotate the coordinates
$\{x\}=(x,y,z)$ so that the $z$-axis points along ${\dot{e}}_{I}(v)$, $v$
being at the origin.
%and translate them so that $v$ is at the origin.
Then we define $S_{k}$ through:
%$S_k$ consists of the points $(x,y,z)$ such that%
\begin{equation}
S_{k}=\{(x,y,z)\}\;\mathrm{such\;that}\;x^{2}+y^{2}\leq z^{2k-2}%
,\;\;z\geq0.\label{defsk}%
\end{equation}
Since $p\ll k$, it follows from (\ref{posnvprimei}) that for small enough
$\delta$, $v_{I}^{\prime}$ lies outside $S_{k}$ for any choice of $\vec
{\hat{n}}_{I}$. We choose $\vec{\hat{n}}_{I}$ so that it it lies at an angle
of $O(1)$ away from the set of vectors $\{\vec{e}_{J\perp}\},$ $J\neq I$.

\bigskip

\noindent\emph{4. Detailed choice of }$\vec{\hat{n}}$\emph{$_{I}$ for the
non-GR case:} If there are no edges at $v$ other than $e_{I}$ with tangent
proportional to $\vec{\dot{e}}_{I}(v)$, we place $v_{I}^{\prime}$ as for the
GR case by choosing $\vec{\hat{n}}_{I}$ to be at an angular separation of
$O(1)$ from the set $\{\vec{v}_{m\perp}^{I},\vec{\dot{e}}_{J\perp}\}$ for the
case that $\vec{v}_{m\perp}^{I}\neq0$ and from the set $\{\vec{\dot{e}%
}_{J\perp}\}$ when $m=k-1$, ${\vec v}_{k-1\perp}=0$.

If there are $s$ edges $e_{J_{i}\neq I},$ $i=1,\dots,s$ such that $\vec
{\dot{e}}_{J_{i}}(v)$ is proportional to $\vec{\dot{e}}_{I}(v)$, then using
%analyticity of
the $C^{k}$ nature of these edges, we expand the coordinates $\vec{x}^{J_{i}%
}(t_{i})$ of $e_{J_{i}}$ as a Taylor series in the parameter $t_{i}$ so that:
\begin{equation}
\vec{x}^{J_{i}}(t_{i})=\sum_{n=1}^{k-1}\vec{v}_{n}^{J_{i}}(t_{i})^{n}%
+O(t_{i}^{k}) \label{analej}%
\end{equation}
with ${\vec{v}}_{1}^{J_{i}}$ proportional to ${\vec{\dot{e}}}_{I}(v)$. As in
step 3, for simplicity we rescale the parameters $t_{i}$ so that $\vec{v}%
_{1}^{J_{i}}=\vec{\hat{e}}_{I}(v)$ For each $i$, let $m_{i}$ be the smallest
integer less than $k$ such that $\vec{v}_{m_{i}}^{J_{i}}$ is not proportional
to $\vec{\dot{e}}_{I}(v)$.
%\footnote{If no such
%$m_{i}$ exists, then there is no restriction on the choice of $\vec{\hat{n}%
%}_{I}$ relative to $e_{J_{i}}$.}
If $\vec{v}_{m_{i}\perp}^{J_{i}}=0\;\forall m_{i}=1,2,\dots,k-1$ then set
$m_{i}=k-1$ so that $\vec{v}_{m_{i}\perp}^{J_{i}}=0$.

If $\vec{v}_{m_{i}\perp}^{J_{i}}\neq0$, let $P^{J_{i}}$ be the half plane
tangent to $(\vec{\dot{e}}_{I}(v),\vec{v}_{m_{i}}^{J_{i}})$ with boundary
through $v$ along $\vec{\dot{e}}_{I}(v)$. Let $W^{J_{i}}(\theta_{i})$ be the
wedge of angle $2\theta_{i}$ associated to this half plane. Using Equation
(\ref{analej}), we choose $\theta_{i}$ of $O(\delta)$ such that the edge
$e_{J_{i}}$ is confined to the wedge $W^{J_{i}}(\theta_{i})$. We choose
$\vec{\hat{n}}_{I}$ to be such that its angular separation is of $O(1)$ from
the wedges $W^{J_{i}}(\theta_{i}),$ $i=1,..,k$ as well as from the directions
along the vectors
%the perpendicular components
$\vec{\dot{e}}_{J\perp}(v),$ $J\notin\{I,J_{1},\dots,J_{k}\}$ (recall that
${\vec{\dot{e}}}_{J\perp}(v),$ $J\notin\{I,J_{1},\dots,J_{k}\}$ are the
perpendicular components of the tangents to the remaining edges ${{e}}_{J},$
$J\notin\{I,J_{1},\dots,J_{k}\}$ at $v$). Clearly, this, together with $p\ll
k$ ensures that for small enough $\delta$, $v_{I}^{\prime}$ does not lie on
$\gamma(c)$.

\bigskip

\noindent\emph{5. Moving the displaced vertex and its vicinity}: Let $P_{I}$
be the half-plane tangent to $(\vec{\dot{e}}_{I}(v),\vec{\hat{n}}_{I})$ with
boundary through $v$ along $\vec{\dot{e}}_{I}(v)$. For the purposes of this
part, we rotate the coordinate system $\{x\}=\{x,y,z\}$ so that $\vec{\dot{e}%
}_{I}(v)$ is along the $z$-direction and $\vec{\hat{n}}_{I}$ is along the
$y$-direction. Thus $P_{I}$ is a part of the $y$-$z$ plane. The choice of
$\vec{\hat{n}}_{I}$ implies that there
%%@@ OCT 2 2012%%
exists small enough $\delta= \delta_{0}$ and $\theta= \theta_{0}$ such that
wedge of angle $2\theta_{0}$ associated with $P_{I}$ does not intersect
$\gamma(c)$ except, at most, inside $S_{k}$.
%
%the origin, as well as, at most, along the $z$-axis.
Denote this wedge by $W_{I}(\theta_{0})$.

Clearly, at deformation parameter $\delta_{0}$, the point $v_{I}^{\prime
}\equiv v_{I}^{\prime}(\delta_{0})$ has coordinates $(y,z)=(\delta_{0}%
^{p},\delta_{0})$. Let the displaced vertex at parameter $\delta<\delta_{0}$
be denoted by $v_{I}^{\prime}(\delta)$. We place $v_{I}^{\prime}(\delta)$ on
$P_{I}$ with coordinates $(y(\delta),z(\delta))$ given by:
\begin{equation}
y({\delta})=\delta^{p},\qquad z({\delta})=\delta.
\end{equation}
Let the straight line joining $v_{I}^{\prime}({\delta_{0}})$ to $v_{I}%
^{\prime}({\delta})$ be $l_{\delta_{0},\delta}$. By virtue of the existence of
$W_{I}(\theta_{0})$ and the fact that $p<<k$, there exists a neighborhood of
this line which lies within $W_{I}(\theta_{0} )$ but outside $S_{k}$, and
hence does not intersect $\gamma(c)$. Hence, by multiplying the translational
vector field along the direction $\vec{l}_{v_{I}^{\prime}({\delta_{0}}%
),v_{I}^{\prime}({\delta})}$ by a suitable function of compact support, a
vector field can be constructed that generates a diffeomorphism which rigidly
translates a small enough neighborhood of $v_{I}^{\prime}(\delta_{0})$ to a
corresponding neighborhood of $v_{I}^{\prime}(\delta)$ while being identity in
a small enough neighborhood of $\gamma(c)$.

The rigid translation property ensures that the edge tangents at
$v_{I}^{\prime}(\delta_{0})$ and $v_{I}^{\prime}(\delta)$ are identical. It
remains to `scrunch' the edge tangents of all edges except the $I^{\mathrm{th}%
}$ together. Let the coordinates of $v_{I}^{\prime}(\delta)$ be $(x(v_{I}%
^{\prime}(\delta)),y(v_{I}^{\prime}(\delta)),z(v_{I}^{\prime}(\delta)))$ and
consider the following linear `anisotropic' scaling transformation $G$ near
$v_{I}^{\prime}(\delta)$:
\begin{align}
G(x-x(v_{I}^{\prime}(\delta)))  &  =\delta^{q-1}(x-x(v_{I}^{\prime}%
(\delta)))\nonumber\\
G(y-y(v_{I}^{\prime}(\delta)))  &  =\delta^{q-1}(y-y(v_{I}^{\prime}%
(\delta)))\nonumber\\
G(z-z(v_{I}^{\prime}(\delta)))  &  =(z-z(v_{I}^{\prime}(\delta))).
\label{scrunch}%
\end{align}
It can easily be verified that this transformation scrunches together the
tangent vectors at $v_{I}^{\prime}({\delta})$ as required. The transformation
$G$ is generated by the vector field $v_{G}^{a}=x(\frac{\partial}{\partial
x})^{a}+y(\frac{\partial}{\partial y})^{a}$. Once again, multiplying $\vec
{v}{_{G}}$ by an semianalytic function of compact support yields a vector
field which generates a diffeomorphism that generates the transformation
(\ref{scrunch}) at $v_{I}^{\prime}(\delta)$ and is identity in a small enough
neighborhood of $\gamma(c)$.

\bigskip

\noindent\emph{6. Moving the points }$\tilde{v}_{J}$: Since the edges $e_{J}$
are semianalytic
%in $B_{4\delta_{0}}(v)$,
the points $\tilde{v}_{J}$ can be independently translated along $e_{J}$ to
their desired position by appropriate semianalytic diffeomorphisms as follows.
At parameter value $\delta_{0}$ the point $\tilde{v}_{J}\equiv\tilde{v}%
_{J}(\delta_{0})$ is at a distance of $\delta_{0}^{q}$ from $v$ for $J\neq I$
and at a distance of $2\delta_{0}$ from $v$ for $J=I$. We seek to move
${\tilde{v}}_{J}(\delta_{0})$ to ${\tilde{v}}_{J}(\delta)$ along $e_{J}$ where
${\tilde{v}}_{J}(\delta)$ is at a distance of $\delta^{q}$ from $v$ for $J\neq
I$ and at a distance of $2\delta$ from $v$ for $J=I$.

Fix some edge $e_{J}$.
%Consider an analyic vector field which coincides with the tangent to $e_J$ along $e_J$.
Let the part of the edge $e_{J}$ between $\tilde{v}_{J}(\delta_{0})$ and
$\tilde{v}_{J}(\delta)$ be $e_{J}(\delta_{0},\delta)$. Let $U_{e_{J}%
(\delta_{0},\delta)}$ be a small enough neighborhood of $e_{J}(\delta
_{0},\delta)$ such that $U_{e_{J}(\delta_{0},\delta)}\cap\gamma(c)=e_{J}%
(\delta_{0},\delta)$ and such that there exists a small enough neighborhood of
$v^{\prime}(\delta)$ which does not intersect $U_{e_{J}(\delta_{0},\delta)}$.
Let $F_{J}$ be a semianalytic function which vanishes outside $U_{e_{J}%
(\delta_{0},\delta)}$ and which is unity on $e_{J}(\delta_{0},\delta)$. Let
${\vec{g}}_{J}$ be a semianalytic vector field which, when restricted to
$e_{J}$, coincides with the tangent vector to $e_{J}$. Then, clearly, the
semianalytic vector field $F_{J}g_{J}$ generates a diffeomorphism which moves
$\tilde{v}_{J}(\delta_{0})$ to $\tilde{v}_{J}(\delta)$ while preserving
$\gamma(c)$ and the vicinity of $v^{\prime}(\delta)$.

We note that the generation of deformations at $\delta<\delta_{0}$ as
described above preserves the following properties and/or equations which are
sufficient for the analysis of Sections 4-6:

\begin{enumerate}
\item[(i)] Equations (\ref{posnvprimei}), (\ref{doteI}) and (\ref{doteJ}).

\item[(ii)] The $C^{1}$ or $C^{0}$ nature of kinks.

\item[(iii)] The GR or non-GR nature of the displaced vertex.
\end{enumerate}

%\noindent{\em Note}: Above, the location of the points
%$\tilde{v}_{J\neq I}(\delta)$, $\tilde{v}_{I}(\delta)$,
%$v^{\prime}(\delta)$
%are `approximately' at distances $\delta^q, 2\delta, \delta$
%from $v$. By ``approximately'' we mean upto terms of $O(\delta^{2q},O(\delta^2), O(\delta^2)$.
%for the purposes of this work, we have ensured that they are located
%{\em exactly} at these distances from $v$ for the purposes of our forthcoming work
%\cite{meinprep}.

\bibliographystyle{unsrt}
\bibliography{u131b}

\end{document}